# Emerging material systems for integrated optical Kerr frequency combs

ANDRE KOVACH*,[1] DONGYU CHEN*,[2] JINGHAN HE,[3] HYUNGWOO CHOI,[1] ADIL HAN DOGAN,[4] MOHAMMADREZA GHASEMKHANI,[4] HOSSEIN TAHERI,[4,†] AND ANDREA M. ARMANI[1,2,3, ‡]

[1]*Mork Family Department of Chemical Engineering and Materials Science, University of Southern California, 1002 Childs Way, Los Angeles, CA 90089, USA*
[2]*Ming Hsieh Department of Electrical and Computer Engineering, University of Southern California, 1002 Childs Way, Los Angeles, CA 90089, USA*
[3]*Department of Chemistry, University of Southern California, 1002 Childs Way, Los Angeles, CA 90089, USA*
[4]*Department of Electrical and Computer Engineering, University of California Riverside, 900 University Ave, Riverside, CA 92521, USA*

*\*Both authors contributed equally to this work.*
[‡]*armani@usc.edu*
[†]*hossein.taheri@ucr.edu*

**Abstract:** The experimental realization of a Kerr frequency comb represented the convergence of research in materials, physics, and engineering, and this symbiotic relationship continues to underpin efforts in comb innovation today. While the initial focus developing cavity-based frequency combs relied on existing microresonator architectures and classic optical materials, in recent years, this trend has been disrupted. This paper reviews the latest achievements in frequency comb generation using resonant cavities, placing them within the broader historical context of the field. After presenting well-established material systems and device designs, the emerging materials and device architectures are examined. Specifically, the unconventional material systems as well as atypical device designs that have enabled tailored dispersion profiles and improved comb performance are compared to the current state of art. The remaining challenges and future outlook for the field of cavity-based frequency combs is evaluated.

# 1 Introduction

Frequency combs are one of the few inventions awarded the Nobel prize whose potential broad-ranging, societal impact was recognized in the first publications on the work [1–3]. However, it would be decades before this impact could begin to be realized in system, and it required a close partnership between scientists and engineers, making iterative discoveries. We are now poised on the edge of developing useful systems leveraging these technologies.

At a fundamental level, Kerr frequency combs rely on four wave mixing (FWM) processes for frequency generation, and therefore, they require both a nonlinear medium as well as a high power laser source [4]. The specific nonlinear process is determined by the nonlinearities present in or excluded from the material and the device geometry as well as the total optical pump power available to the system.

The initial strategy for generating combs approached the pump source and the nonlinear media as discrete components and optimized each component independently. However, this method also resulted in very large systems that had to be aligned and stabilized, limiting the utility of this technology. One proposed solution was to directly integrate the FWM media with the pump source.

Initial successful efforts included fiber-based strategies which did simplify system alignment and improve vibration tolerance [1,5]. However, the overall footprint was still large. With the advent of the on-chip optical ring waveguide or resonant cavity, the ability to directly fabricate an on-chip waveguide from a nonlinear medium became possible. This work enabled demonstrations in a wide range of fields including atomic clocks, spectroscopy, and lidar; however, the need for even more portability is still a driving research challenge. [6–17] As a result, research efforts developing highly nonlinear optical materials that are also compatible with on-chip fabrication methods as well as non-conventional device architectures has increased in importance.

This review will first present the basics of resonant cavities and the fundamentals of the FWM process that gives rise to comb generation. After establishing this framework, the focus of the review is dedicated to discussing the strengths and weaknesses of several material platforms for designing integrated frequency comb systems based on dielectric and on conducting materials as well as different device design strategies. When appropriate, non-integrated material systems and devices will also be discussed to provide appropriate historical context and to establish the basic theoretical framework.

# 2 Resonant cavity platforms

Whether it is acoustic, optical, or electrical energy, the ability to store and to amplify input energy is a defining principle of a resonant cavity. In the optical domain, optical resonators can be found throughout nature in butterfly wings and rain drops. Depending on how they interact with an incoming optical field, they are categorized as either a traveling wave or standing wave cavity. The present review is primarily focused on traveling wave optical cavities.

The first traveling wave, or whispering gallery mode, optical cavity with a high-quality factor was a freely suspended, spherical liquid droplet. First demonstrated in the mid-1980's, it was enabled by the development of the optical trap. [18–21] The extremely smooth, liquid-air interface formed an ideal surface to support the circulating optical cavity modes. After developing methods to stabilize droplets against rapid evaporation while still maintaining optical transparency, researchers quickly discovered the ability of the droplet to amplify input optical light. Using this platform, a rapid cascade of theoretical works and experimental discoveries occurred, and the field of droplet microcavity physics is still active today (Fig. 1). [22–29] However, given that the whispering gallery mode cavity was an alternative approach to the well-established standing wave resonator, such as the Fabry-Perot cavity, it was necessary to introduce quantitative metrics to allow for direct comparison, not only between different whispering gallery mode devices but also across resonant cavity types.

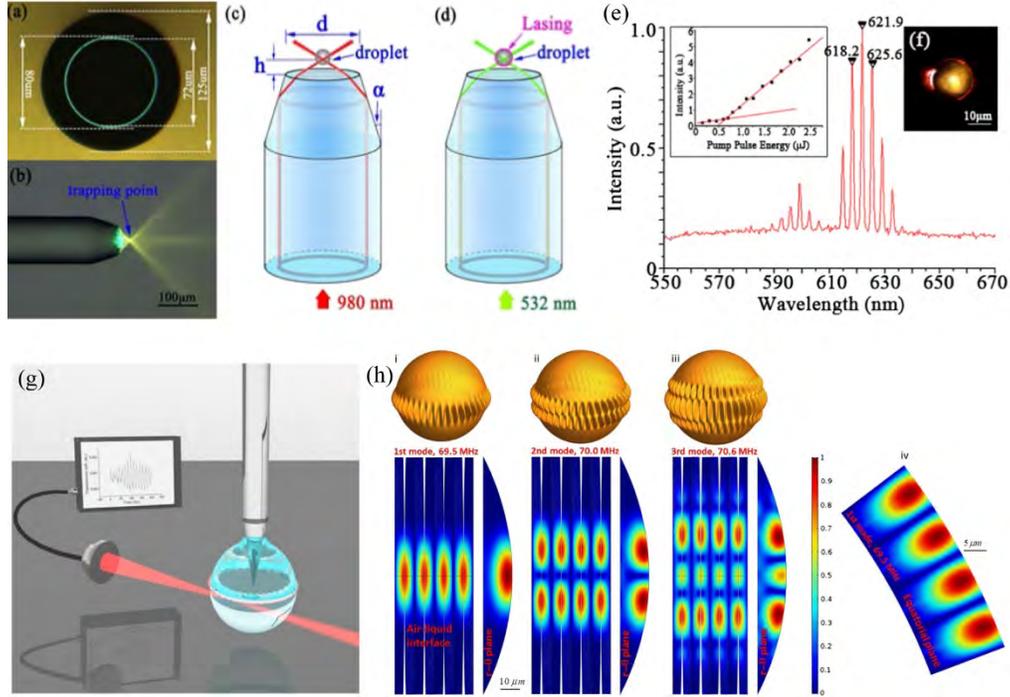

Fig. 1. Droplet resonators. (a)-(d) An oil droplet resonator doped with laser dye is suspended in an optical fiber trap. (e) -(f) Lasing is observed from the droplet resonator with <uJ pulse energy. Reprinted with permission from [30]. Copyright 2016 by the American Physical Society. (g) A suspended oil droplet resonator is formed, and opto-acoustic vibrations are excited. (h) Finite element method modeling of the different acoustic modes of the liquid cavity. Modes are shown with exaggerated depth (top) and absolute displacement values (bottom). Adapted from [31].

To characterize a resonant cavity, several properties must be analyzed. Two of the most basic are the linewidth of the resonant wavelength ($\delta\lambda$), which is inversely related to photon lifetime in the cavity, and the free spectral range (FSR) or the spacing between adjacent resonant wavelengths. [32] These parameters govern numerous other metrics and determine the suitability of a cavity for a specific application. For example, a narrow linewidth cavity confines light of a very precise wavelength, which can be useful in a simple add-drop filter application where a single resonator is coupled to a pair of waveguides [33]. However, a complementary device is comprised of multiple cavities coupled together. [34,35] In this coupled cavity system, also called an optical buffer or delay line, an ultra-narrow linewidth makes the device coupling more complex as any small offset in resonant wavelengths between cavities will result device failure.

Therefore, given the wide range in potential cavity applications, there is no single perfect cavity design. To address the wide range in technology demands, numerous device designs, varying both the cavity geometry and the cavity material, have been explored since the first suspended liquid droplets. Given this diversity, researchers developed several universal metrics that went beyond FSR and linewidth to compare device architectures and to evaluate their suitability for different applications. [32]

## 2.1 Characterization metrics of optical resonant cavities

The most common is the cavity quality factor (Q) or the closely related parameter, cavity finesse (F). [32] The intrinsic cavity quality factor ($Q_o$) describes the photon lifetime or storage time within the cavity. Therefore, a higher Q indicates a longer storage time or lower optical loss. Intrinsic cavity losses can arise from a plethora of sources. Therefore, the total cavity Q is described as a series of loss mechanisms: [36]

$$Q_{tot}^{-1} = Q_o^{-1} + Q_{ext}^{-1} = Q_{mat}^{-1} + Q_{scat}^{-1} + Q_{cont}^{-1} + Q_{rad}^{-1} + Q_{car}^{-1} + Q_{coupl}^{-1} \quad (1)$$

where the intrinsic losses include the material losses ($1/Q_{mat}$), scattering losses ($1/Q_{scat}$), contamination losses ($1/Q_{cont}$), radiation losses ($1/Q_{rad}$), and carrier losses ($1/Q_{car}$). These loss mechanisms are governed by the material and geometric properties, including refractive index and optical loss, the operating conditions, and the device geometry. [25,37,38] Therefore, these losses can be mitigated to a certain extent with careful planning. For example, with judicious choice of material system, device geometry, wavelength, and fabrication process, Q factors in excess of $10^{10}$ have been achieved. [39,40] These ultra-high-Q factors have enabled numerous applications ranging from fundamental science to applied technology. [6,25,41–83] As a result, maintaining the cavity Q is of critical importance to many researchers.

Moreover, it is important to note that, in addition to intrinsic cavity loss, there are extrinsic cavity losses. [32] These losses, such as coupling or photon injection losses ($1/Q_{coupl}$), can affect the cavity performance in a given application. [84–87] As a result, numerous approaches for coupling light into optical cavities have been developed with differing degrees of loss. [34,85,86,88–93] Currently, the lowest loss methods are the angle-polished fiber [94], a tapered fiber [95], a coupling prism [96,97], and an integrated bus waveguide [98] (Fig. 2). The first three nearly lossless couplers can be integrated with on-chip devices using thermal or UV-curable polymers or "pick-and-place" techniques. [89,99,100] However, because these methods are unable to be lithographically patterned, the total density of devices is fundamentally limited. The drive to increase density inspired research into alternative coupling methods, such as the bus waveguide, as well as innovation in device design.

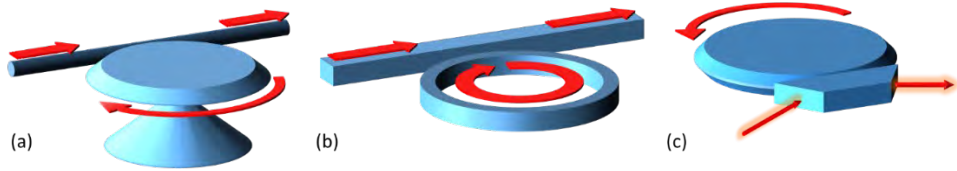

Fig. 2. Example microcomb generation resonators and coupling schemes. (a) Wedge resonator and tapered fiber coupling; (b) Integrated ring resonator and ridge waveguide coupling; (c) Chamfered crystalline resonator with prism coupling. The Lugiato-Lefever equation, discussed in the next section, can be used to model microcomb formation in all of these platforms.

Fortunately, this push for photonic integration occurred in parallel with advances in lithography and nanofabrication instrumentation. This synergy allowed researchers to balance the competing demands of high-Q and on-chip compatibility. As a result, the past two decades have witnessed rapid growth in integrated optical resonant cavity device platform designs. While the circular-nature of the cavity is maintained in nearly all cavities, the precise three-dimensional morphology varies greatly. The geometries shown in Fig. 3 are the most commonly used in nonlinear optics studies; however, the diversity of device morphology is much broader, including hollow bubbles, goblets, and nanowires among many others. [84,88,101–138]

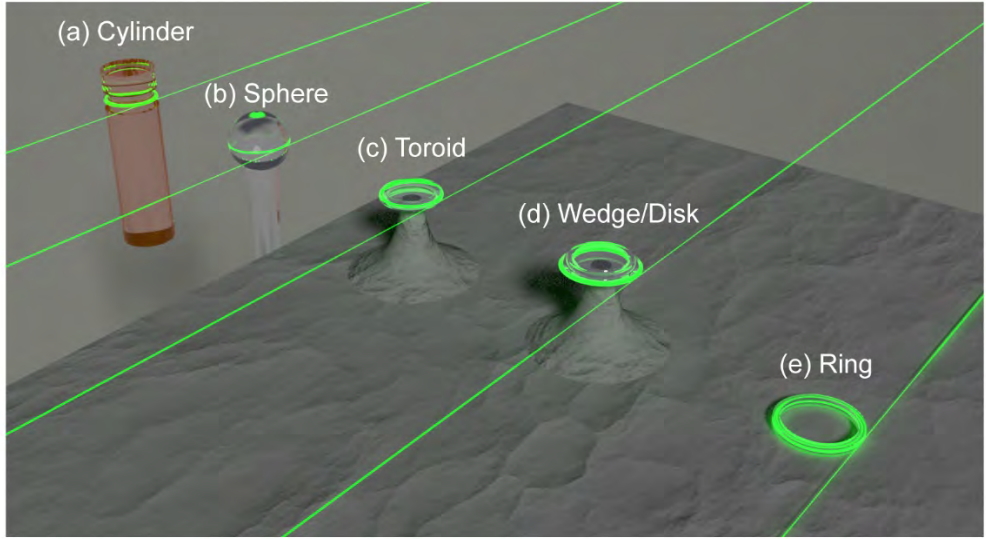

Fig. 3. Summary of common resonant cavity geometries used in frequency comb generation. While the free-standing and suspended geometries require complex optical coupling techniques such as prisms and tapered fiber waveguides, the on-chip ring can be directly integrated with a waveguide. The devices are shown in order of decreasing Q, from the fluoride cavities which have achieved Q factors above $10^{11}$ to the ring Q's which have achieved Q's above $10^7$. However, in addition to Q, optical mode volume is also a consideration, and the mode volume of ring structures is typically smaller than that of the other structures.

Additionally, to overcome extrinsic loss, numerous strategies for integrating waveguides on-chip as well as on-/off chip coupling have been explored. Initial methods focused on using similar materials for both the waveguide and cavity system [103,106,107,135,139,140], but more recent designs have expanded beyond this limitation, greatly improving performance [88,120,140,141]. Specifically, by achieving a high optical mode overlap or matching the effective index of the coupler and the mode for materials with differing refractive indices, coupling efficiencies have been improved. Therefore, while Q does provide insight into a device's performance, optical mode profile is another key parameter which must be considered in device design. [87,142]

The optical mode area is very sensitive to the device material and geometry, and even small changes can have a significant effect (Fig. 4). In addition, the TE and TM modes can have drastically different optical modes profiles. To date, the majority of Finite Element Method (FEM) and Finite-difference time-domain method (FDTD) modeling efforts, such as those shown in Fig. 4, have focused on modeling devices with intuitive geometries, inspired by ray optics. With advances in computational power, there has been a surge of research activity in inverse design of devices. This research has resulted in a wide range of optical structures with extremely unique architectures. This emerging field can only be expected to grow in the future.

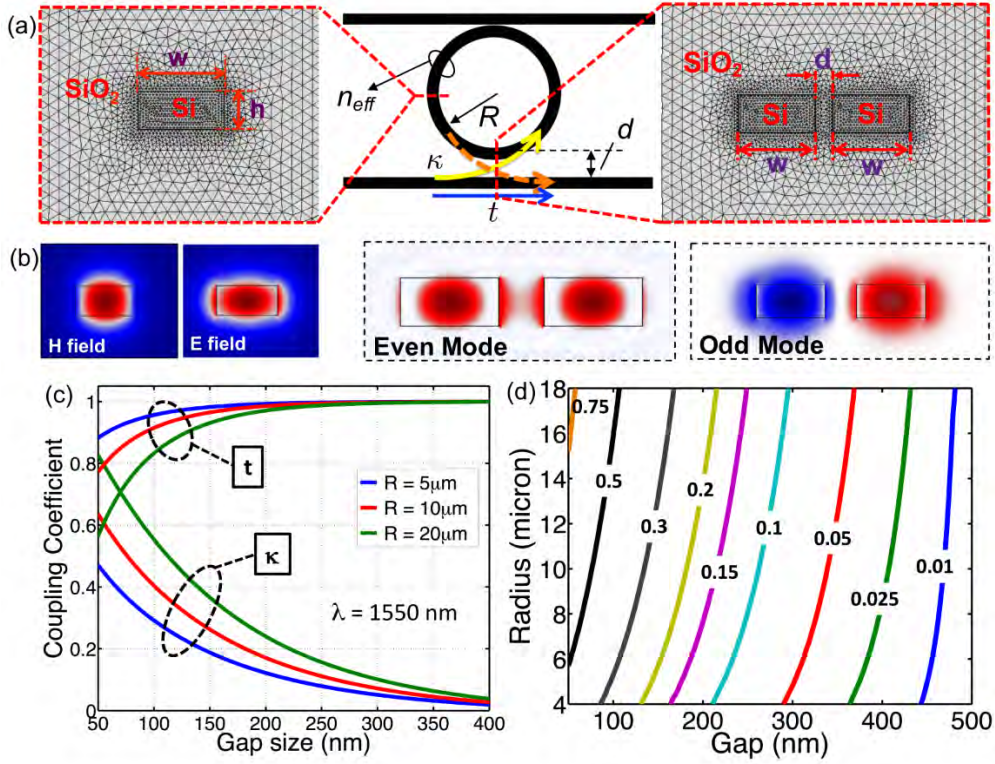

Fig. 4. FEM modeling of coupling from waveguides into microring resonators. (a) Schematic of model showing mesh scaling and dimensions. (b) Example mode profile results for 450x220 nm waveguides. (c) Coupling and transmission coefficients as a function of coupling gap and device radius at 1550 nm. (d) Dependence of coupling on device diameter and gap distance. © 2018 IEEE. Reprinted, with permission, from [143].

To generate an optical nonlinear behavior, it is critical to have enough photons as a source. The optical mode area plays a critical role in determining the optical field power or circulating optical intensity in the device. Specifically, circulating optical power ($P_{circ}$) and circulating optical intensity ($I_{circ}$) in a whispering gallery mode cavity are: [32,144]

$$\frac{P_{circ}}{P_{in}} = \frac{\lambda Q_o}{\pi^2 nR} \frac{K}{(1+K)^2} \quad (2)$$

$$I_{circ} = \frac{P_{circ}}{A_{eff}} \quad (3)$$

where $P_{in}$ is the input power, $Q_o$ is the intrinsic cavity Q, $\lambda$ is the resonant wavelength, R is the device radius, n is the refractive index of the optical mode, $A_{eff}$ is the optical mode area, and $K \equiv Q_o/Q_{coupl}$.

Thus, even if two devices have the same Q, the circulating powers or circulating intensities could be dramatically different. For example, assuming a microtoroid cavity with a Q of 100 million operating at 1550 nm with 1 mW of input power, the $P_{circ}$ can change dramatically with changes to the refractive index and the device major diameter, especially when the major diameters are small (Fig. 5). It is important to note that in devices with complex geometries, such as the microtoroid modeled in Fig. 5, the diameter contributes to these values through the $A_{eff}$ term and the effective refractive index term (which is can have a significant impact on the circulating intensity). Therefore, while Q is an important criterion, when designing a device for applications in nonlinear optics, particularly in frequency generation, one needs to look beyond the cavity Q and consider $I_{circ}$ or $P_{circ}$ in the final cavity design.

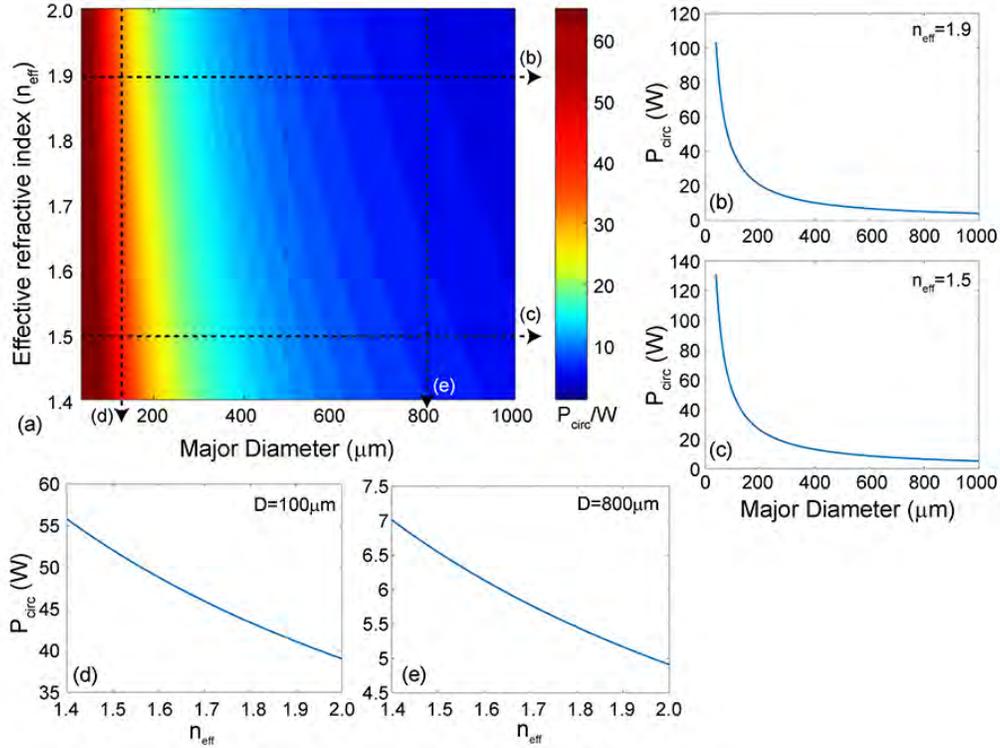

Fig. 5. Finite element method modeling to determine the dependence of the circulating intensity on the effective refractive index and device diameter. The geometry modeled is a toroidal resonator operating at 1550 nm. (a) A summary of all circulating intensity values obtained while simultaneously varying D and $n_{eff}$. (b)/(c) Plots of the horizontal dashed lines indicated in part (a) highlighting the dependence of $P_{circ}$ on diameter. (d)/(e) Plots of the vertical dashed lines indicated in part (a) highlighting the dependence of $P_{circ}$ on $n_{eff}$.

Given that the threshold for many nonlinear phenomena scales inversely with the square of the cavity quality factor, maintaining the device Q has been a focus of research. While this review is focused on frequency comb generation via four wave mixing processes, many of the initial experiments performed with whispering gallery mode cavities were focused on investigating nonlinear optical phenomena such as Raman and Brillouin scattering using ultra-high Q cavities. This work laid the foundation for the subsequent comb generation efforts, and these synergistic research efforts have continued to present day with investigations into optomechanics, optical cooling, and other nonlinear behaviors. [18,24,26,46,59,60,63,65,67–71,73,75–80,145–161]

*2.2 Refractive index modulation*

Before discussing different strategies for combining nonlinear optics for frequency generation and microcavities, it is important to highlight the role of the cavity refractive index. As can be seen in Fig. 5 in section 2.1, changes in the index can impact the circulating optical power. [162] Given the relationship between circulating power and nonlinear threshold, index tuning is a powerful strategy to manipulate a device's ability to generate nonlinear phenomenon. In addition to tuning the refractive index in a static manner, index modulation can be performed in a dynamic manner, modulating between two index values. Because the resonant frequency is defined, in part, by the cavity's index, this approach provides a path to tune a cavity on and off-resonance. Therefore, index modulation provides a simple path to control nonlinear effects. However, it also makes the cavity susceptible to instabilities that must be considered during cavity design.

In whispering gallery mode microcavities, the two most commonly used approaches for changing or modulating the refractive index are the thermo-optic effect and the electro-optic (EO) effect (Pockels effect) [144,163–176]; however, other mechanisms like pressure (or acoustics wave-induced compression) have been used to a lesser extent. [53,177,178] In most materials used in integrated optics, the thermo-optic effect is orders of magnitude slower in response than the electro-optic effect (Fig. 6). As such, they have very different applications.

It is important to note that the thermo-optic (TO) effect can be either a linear or a nonlinear effect depending on the heat source. When the thermal source is applied externally, such as an integrated resistive heater, the effect on the cavity is linear. This approach for changing the resonant wavelength ($\Delta\lambda$) is theoretically described by: [166]

$$\Delta\lambda = \lambda_o \Delta T [\varepsilon_{eff} + (dn_{eff}/dT)/n_{eff}] \quad (4)$$

where $\lambda_o$, $\varepsilon_{eff}$, $dn_{eff}/dT$, $n_{eff}$, and $\Delta T$ represent the cold cavity resonant wavelength, the effective expansion, the thermo-optic coefficient, the effective refractive index, and the change in temperature. While this method has been used to tune the refractive index thereby tuning the resonant wavelength, it is very energy intensive. Additionally, the time constant of the thermal effect is very slow (microseconds). Therefore, while great success has been achieved when employing this method to create environmental temperature sensors [47,49,50,129,167], it is not a popular method for designing high-speed optical modulators due to the timescales involved as well as intrinsic sensitivity to environmental noise. The linear TO effect is also responsible for the creation of thermal noise which impacts resonant wavelength stability.

If the circulating optical field inside the cavity exceeds a critical threshold, this field can directly heat the cavity. This thermal source is a nonlinear optical effect. [179] To calculate the $\Delta T$, or temperature change, that can be induced by the optical field, several parameters must be evaluated. First, only the portion of the circulating optical field which is absorbed by the cavity material is converted to heat. Therefore, the $\Delta T$ is directly related to the previously defined circulating optical power ($P_{circ}$), where the thermal energy is limited to the lost or dissipated power ($P_d$) in the cavity [166]:

$$P_d = \sigma_o^2 P_{circ} t_r \quad (5)$$

where $\sigma_o$ is the loss coefficient and $t_r$ is the round-trip circulation time. To accurately capture losses in converting optical energy to thermal energy, $P_d$ must be determined using modeling, such as COMSOL Multiphysics.

Second, because the resonant wavelength is determined by the refractive index, the cavity exhibits thermal bistability, oscillating between heating and cooling. [167,180] Because $dn/dT$ is a ubiquitous parameter for all materials, this behavior plays a role in nearly all nonlinear optical devices, as some optical power will be lost to heat generation, reducing the efficiency of the system. [181] As such, it must be considered when designing a resonant cavity-based frequency comb generator. However, this power can be used to create a "thermal lock", increasing the stability of the system. [182]

The electro-optic or Pockels effect is a second-order nonlinearity. It can be used to tune the wavelength by applying an electrical field to modify the refractive index according to: [183]

$$n_{eff} = n_{eff,o} - \frac{1}{2} n^3 r \frac{V_o}{G} \Gamma_1 \Gamma_2 \quad (6)$$

where $n_{eff,o}$ is the initial effective refractive index, n is the waveguide index, r is the electro-optic coefficient, $V_o$ is the applied voltage, G is the gap between electrodes, and $\Gamma_1$ and $\Gamma_2$ are the overlap integral between the electrical and optical field and the dynamic reduction factor in this overlap as the index changes. The electro-optic coefficient (r) is dependent on which optical mode is being excited. [183]

However, unlike the TO coefficient, not all materials exhibit a second order nonlinearity. In fact, the most common optical materials, like silica and hydex, have zero second order nonlinearity. The potential for high speed modulation ability motivated research into alternative material systems, including lithium niobate, III-V, and polymeric materials among others. Although these integrated microcavity devices often require additional processing steps, like poling or structuring to maximize the amplitude modulation ability, extremely fast optical modulators have been demonstrated and commercialized (Fig. 6). [164,169,171,184,185]

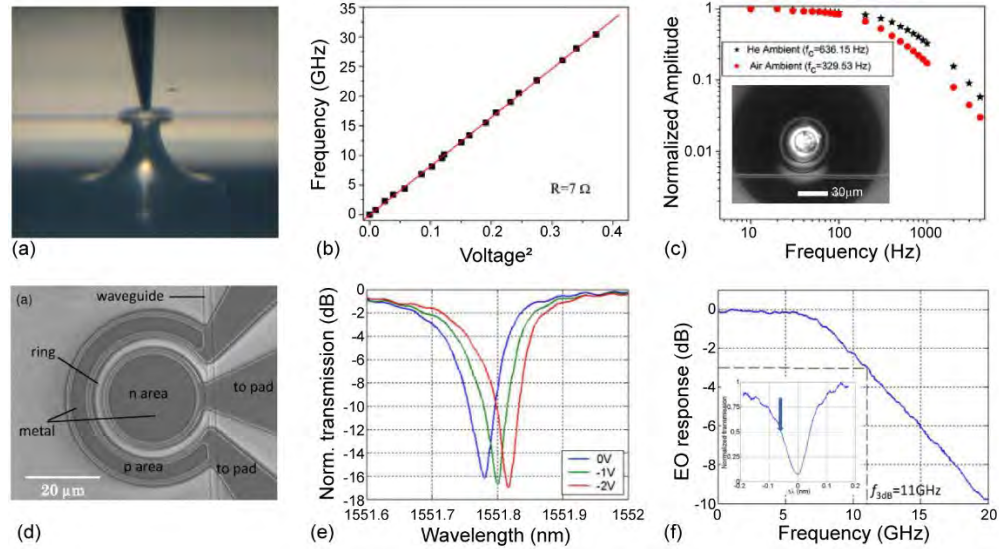

Fig. 6. Modulating resonant wavelength with refractive index. (a)-(c) Silica toroidal cavity is thermally tuned resistively heating the oxide. Reprinted with permission from [168]. Copyright 2016 by the American Physical Society. (d)-(f) Silicon ring resonator is electro-optically tuned. Adapted from [186]. While the silica cavities, such as that shown in part (a) can achieve higher Q's, the TO tuning rates are significantly slower as compared to the lower Q Si cavities.

## 3  Theory of resonant cavity-comb generation

### 3.1  Brief historical review

The theoretical and numerical modeling of microresonator-based optical frequency comb generation are deeply rooted in numerous methods that have been developed for complementary nonlinear optical systems. The goal of the following sections is not to provide an exhaustive literature survey or historical account, but instead to put in perspective the evolution of this rich area of nonlinear optics. A conspicuous theme throughout this section is the iterative nature of the advances enabled by the symbiotic collaborations between theoretical and experimental findings. Frequency microcomb literature is rife with examples where predictions of theory inspired experimental demonstrations or, vice versa, where new experimental observations initiated theoretical and numerical investigations.

The focus here is on optical frequency combs in microresonators fabricated in centro-symmetric materials, i.e., those with inversion symmetry, in which the dominant nonlinearity is cubic. Other material platforms, particularly quadratic materials, have also attracted attention for comb generation in recent years [187–192]. However, in principle, inversion symmetry prohibits contribution from the $\chi^{(2)}$ nonlinear susceptibility [193,194], and the nonlinear effects in centro-symmetric material platforms arise from the $\chi^{(3)}$ coefficient, making the comb generation process more efficient. We will refer to such materials as Kerr media.

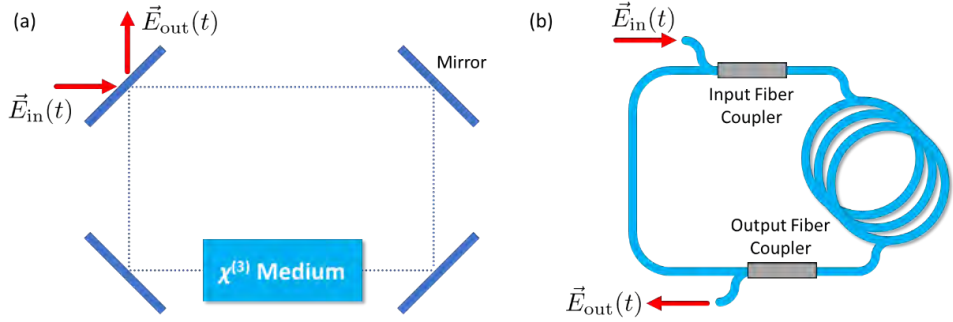

Fig. 7. Simple schematic of (a) spatial and (b) fiber loop cavities.

The fundamental physical effect responsible for frequency comb generation in Kerr media is four-wave mixing (FWM), a process in which four optical waves (with generally different frequencies) interact, subject to the conservation of energy and momentum [193]. When some of the interacting frequencies are degenerate, FWM reduces to self- or cross-phase modulation, which are also involved in the microcomb dynamics [162]. FWM was studied in low-loss fibers soon after they became available and long before such effects were studied in high-quality-factor (high-Q) microresonators. This effect can be described using a set of coupled nonlinear ordinary differential equations (ODEs). The description is essentially based on modal expansion of the electromagnetic field and polarization vectors and includes as many equations as the frequencies considered. Each equation follows the temporal evolution of the amplitude and phase of one of the frequencies involved in the process.

During the 1980s and 1990s, the excitement in the field was focused on Kerr-nonlinear spatial and fiber cavities (Fig. 7) with research efforts including both theory and experimental discoveries (see [195] and references therein). In particular, the possibility of observing stationary spatial dissipative structures was predicted by Lugiato and Lefever using a damped and driven nonlinear Schrodinger equation (NLSE) [196] which is now commonly referred to as the Lugiato-Lefever equation (LLE), particularly among the microcomb community. It was also shown that a nonlinear dynamical map (the Ikeda map) can successfully describe instabilities of continuously pumped Kerr-nonlinear cavities [197,198]. Additionally, a mean field approximation of the Ikeda map, similar to the LLE, was shown to accurately describe a wealth of phenomena, especially in experimental fiber loop systems [199]. While exact soliton solutions of the NLSE had been known since two decades earlier [200], the existence and stability of soliton waveforms in damped and driven variants of this equation were studied in areas of physics other than optics under the name externally (or ac-) driven, damped NLSE [201,202]; see also [203].

In the context of nonlinear optics in high-Q optical resonators, after initial experimental demonstrations, a small number of nonlinear coupled wave equations (CWEs) were quickly developed to understand effects such as side band generation threshold [204,205]. CWEs were later adopted for an (in principle) arbitrarily larger number of frequencies mixing in the nonlinear cavity [206,207]. Importantly, the damped and driven NLSE was recognized as the governing master equation and used to theoretically demonstrate the possibility of generating coherent Kerr microcomb pulses. Broadband mode-locked microresonator-based frequency combs were observed experimentally and shown to be of the dissipative soliton type in the following few years [208–212]. It was also shown that the nonlinear CWE set can be converted into a single partial differential equation, i.e., the LLE, using Fourier summation of the modes and changing to a moving reference frame [213]. By this time, the nonlinear spatial cavity and fiber-based resonator community were attracted to microcomb research, and it was shown that the mean field approximation of the Ikeda map, previously heavily utilized in the study of spatial cavities and fiber loops, can be exploited in nonlinear optical microresonators as well [214]. Although length scales and quality factors, and hence experimental procedures and challenges, are very different when moving from spatial or fiber-based resonators to high-Q microresonators, the physical effects are analogous in many ways. Diffraction in spatial cavities is replaced by dispersion in fiber loops or microresonators, and instead of spatial patterns, temporal pulses are observed. It is worth noting that relying on the LLE is not always sufficient for explaining the microcomb physical phenomena, and resorting to the more fundamental Ikeda map is sometimes warranted [14,215–217].

Kerr microcomb research has been a very active and prolific area of optics in the past decade, and various effects and multiple modified variants of the LLE have been introduced to study different pump-, dispersion-, and detuning-controlled, and thermal effects. Modeling of such physical effects usually requires adding terms to modify the LLE, or coupling the LLE to other dynamical equations and solving them simultaneously. Examples included avoided mode crossing [218–220], higher-order group velocity dispersion (GVD) and Cherenkov dispersive wave (DW) emission [221–224], phase- and intensity-modulated pumping [225–227], pumping by two pumps [228–232], synchronous pumping [233], Raman self-frequency shift [74,234,235], soliton crystals [230,236], thermal effects [237,238], and laser cavity-soliton microcombs. [239] Some of these effects, such as DW emission [240], have counterparts in the spatial or fiber loop cavities and some, like thermal resonance shift, are more specifically relevant to the high-Q microresonator platforms. The following sections focus on the most fundamental effects in Kerr microcomb generation.

### 3.2 Basic principles
### 3.2.1 The simple structure of a frequency comb

In this Section, we will introduce frequency combs as a mathematical object without concerning the reader about the source of generating them. This discussion will introduce the two main frequencies which define the comb, i.e., the repetition rate $f_{\text{rep}}$ and the carrier envelope offset frequency $f_{\text{CEO}}$, and hint at the link between them. This will lay the foundation for the next section on comb stabilization and self-referencing and will help the reader better appreciate some of the main goals pursued among the microcomb community, such as the generation of octave-spanning frequency combs. In what follows, we will first consider the simpler special case in which the carrier envelope offset frequency is zero. The influence of a non-zero $f_{\text{CEO}}$ is then included in a second step [241].

**An offset-free optical frequency comb**

Recall that the Fourier transform of a train of Dirac delta functions is itself a train of Dirac delta functions [242]. We will denote this train function with $\text{III}_T(t)$,

$$\text{III}_T(t) = \sum_{m=-\infty}^{+\infty} \delta(t - mT), \tag{7}$$

where $T$ is the time period such that $\text{III}_T(t + T) = \text{III}_T(t)$, and $\delta(\cdot)$ is the Dirac delta function. This function, called the impulse train, the sampling function, the Shah function, or the Dirac comb, is periodic and can therefore be expanded as a Fourier series,

$$\text{III}_T(t) = \frac{1}{T} \sum_{n=-\infty}^{+\infty} \exp(in\frac{2\pi}{T}t). \tag{8}$$

whose Fourier transform is another Dirac comb, namely,

$$\mathcal{F}\{\text{III}_T(t)\} = \frac{1}{T}\text{III}_{1/T}(f) = \sum_{n=-\infty}^{+\infty} \exp(-in2\pi Tf).$$

(9)

An optical pulse $p(t)$ has a carrier frequency and an envelope. The electromagnetic wave of the pulse can mathematically be written as the product of the carrier wave and the envelope: $E(\text{t}) = p(\text{t}) \times \cos(\omega_c t)$, where $p(t)$ is the pulse envelope and $\omega_c$ is the carrier frequency. Using the convolution theorem, the influence of the carrier frequency in the frequency domain is to shift the center frequency of the pulse envelope transform from zero to the carrier frequency. Consequently, the carrier frequency, no matter how large it may be for an optical pulse, can be ignored so far as one is concerned with the pulse shape (envelope). Therefore, in the figures throughout this section, and in almost all the theoretical work on the frequency comb literature, the spectrum of a comb is commonly drawn around zero frequency (or, zero mode number), and the existence of a carrier frequency (or, its corresponding mode number in the optical resonator) is to be understood.

In general, the Fourier transform of a single temporal isolated pulse constitutes a continuum of frequencies. When, however, one looks at a train of pulses (i.e., a periodic waveform of infinite length), the frequency spectrum will change into a discrete set of frequencies. The mathematical structure of such a train of pulses can be described in both the time and frequency domain using the Dirac comb. In particular, an infinite train $p_T^{\text{train}}(t)$ made of $T$-shifted copies of a pulse envelope $p(t)$ can be written as

$$p_T^{\text{train}}(t) = \sum_{m=-\infty}^{+\infty} p(t - mT) = \text{III}_T(t) \star p(t), \qquad (10)$$

where the $\star$ denotes convolution. Exploiting the convolution theorem, the Fourier transform of $p_T^{\text{train}}(t)$ is equal to the product of the Fourier transforms of the impulse train and the pulse envelope; see Fig. 8. But the Fourier transform of the impulse train is another impulse train. As a result, an infinite train of equally-spaced pulses in the time domain is an equally-spaced train of Dirac delta functions in the frequency domain in which the 'height' of each spike (the power of each comb line) is determined by the Fourier transform of the pulse shape (envelope) at the discrete spike locations.

It is worth noting that although we have talked about a 'pulse', no notion of phase synchronization and mode locking is necessary for the forgoing discussion to hold. Although a pulse will not form when the comb teeth are not oscillating synchronously, the output waveform which includes the superposition of all the comb teeth powers and phases will still be periodic, and hence the presented analysis will still be applicable.

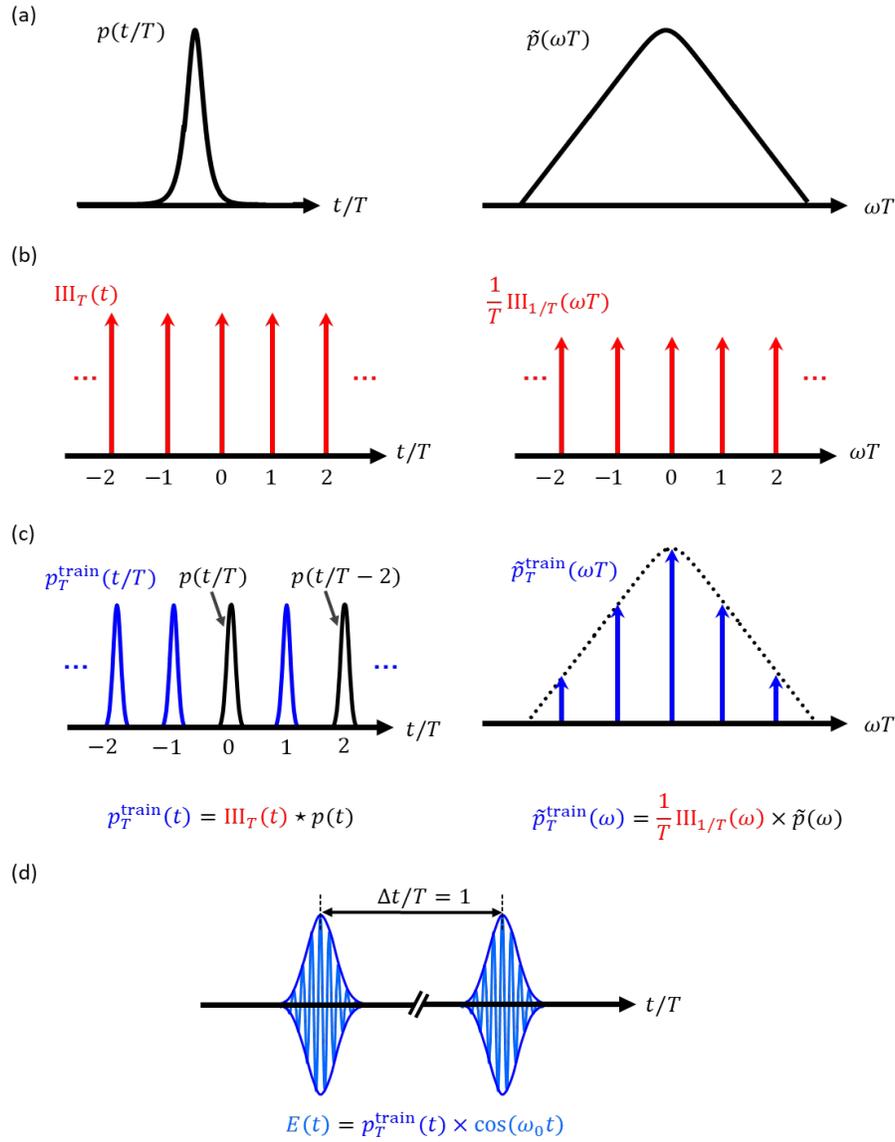

Fig. 8. The structure of a frequency comb as a mathematical object. (a) The Fourier transform of a single pulse *p(t)* (left) includes a continuum of frequencies (right). The parameter *T* is an arbitrary constant with units of time while *ω* is frequency. (b) The Fourier transform of a periodic impulse train is yet another impulse train in the frequency domain. Here, *T* is the temporal period of the impulse train. (c) An infinite pulse train can be considered the result of the convolution of the pulse envelop (left panel in (a)) and the temporal impulse train (left panel in (b)). In the frequency domain, the continuous spectrum of the pulse envelope (right panel in (a)) should multiply the frequency impulse train (right panel in (b)). (d) In (a)-(c), the carrier frequency was ignored. The effect of the carrier frequency is to shift the center frequency of the comb spectrum (position of the spectral peak in the right panel in (c)) from zero to the carrier frequency.

## Carrier envelope offset frequency

We will briefly discuss carrier envelope phase, an important frequency comb stability parameter which, as the reader will note later in this Section, does not interestingly appear in the theoretical formulation of microcombs using the LLE. We assumed in our dry mathematical introduction of frequency combs above that the peak of the pulse train envelope is always aligned with the peak of the carrier sinusoid during one of the periods of the carrier, as in Fig. 8(d). But this was a simplifying assumption and does not generally hold, as seen in Fig. 9(a). To account for the phase offset between the envelope and carrier, one can write the electric field of the pulse train $E(t)$ in terms of the pulse envelope $p(t)$ considering the carrier frequency $\omega_c$, i.e., $p(t)\exp(i\omega_c t)$, and a phase $\phi_{CEO} = \phi_0 + n\Delta\phi_{CEO}$ for the $n$-th pulse in the train. This generalization will consider the phase difference mentioned earlier. Hence,

$$E(t) = \sum_{n=-\infty}^{+\infty} p(t-nT)\exp[i\omega_c(t-nT)]\exp[i(\phi_0 + n\Delta\phi_{CEO})]. \quad (11)$$

If we define the (continuous-time) Fourier transform of a generic function $g(t)$ by $\tilde{g}(\omega) = \int_{-\infty}^{\infty} g(t)\exp(-i\omega t)\,dt$, then taking the Fourier transform of the previous expression for $E(t)$ leads to

$$\tilde{E}(\omega) = \exp(i\phi_0)\,\tilde{p}(\omega - \omega_c) \sum_{-n}^{+n} \exp[in(\Delta\phi_{CEO} - \omega T)] \quad (12)$$

$$= \exp(i\phi_0)\,\tilde{p}(\omega - \omega_c) \sum_{-n}^{+n} \delta(\omega T - \Delta\phi_{CEO} - 2\pi n), \quad (13)$$

where we have used Eq. (9) with $f = \omega/2\pi - \Delta\phi_{CEO}/2\pi T$. The latter expression means the spectrum of the pulse train will comprise a frequency comb at frequencies where the argument of the Delta function under the summation vanishes, i.e.,

$$\omega_n = \frac{\Delta\phi_{CEO}}{T} + n\frac{2\pi}{T}. \quad (14)$$

Hence, the carrier envelope phase leads to an offset frequency, from the equidistant grid of frequencies separated by $1/T$ and starting from zero, equal to $\Delta\phi_{CEO}/T$ for each frequency of the pulse train. Because successive pulses have a phase difference of $\Delta\phi_{CEO}$ and since the electric field is a continuous function, going from one comb envelope peak to the next, there should be an integer number of carrier frequency cycles, i.e.,

$$\omega_c T - \Delta\phi_{CEO} = 2\pi m, \quad (15)$$

where $m$ is an integer. Hence,

$$\omega_c = \frac{\Delta\phi_{CEO}}{T} + m\frac{2\pi}{T}. \quad (16)$$

Comparison of this equation with Eq. (14) shows that the carrier frequency itself should be one tooth of the frequency comb.

To summarize, the frequencies of an OFC can be written as

$$f_n = f_{CEO} + nf_{rep}, \quad (17)$$

where

$$f_{CEO} = \frac{\Delta\phi_{CEO}}{2\pi T} = \frac{\Delta\phi_{CEO}}{2\pi} f_{rep} \quad (18)$$

is the offset frequency of the comb from zero of the frequency axis and $f_{rep} = 1/T$ is the repetition rate. While a frequency comb comprises frequencies in the optical regime (i.e., hundreds of THz), each of its frequencies can be defined in terms of only two frequencies $f_{CEO}$ and $f_{rep}$, which are smaller frequencies. Typical mode-locked solid-state lasers emit with pulse repetition rates roughly between 50 MHz and a few GHz and microcomb repetition rates in the few tens of GHz down to a few GHz range have been demonstrated [17,243,244].

Microresonator-based frequency combs repetition rates can be larger, up to tens of THz, depending on the resonator size.

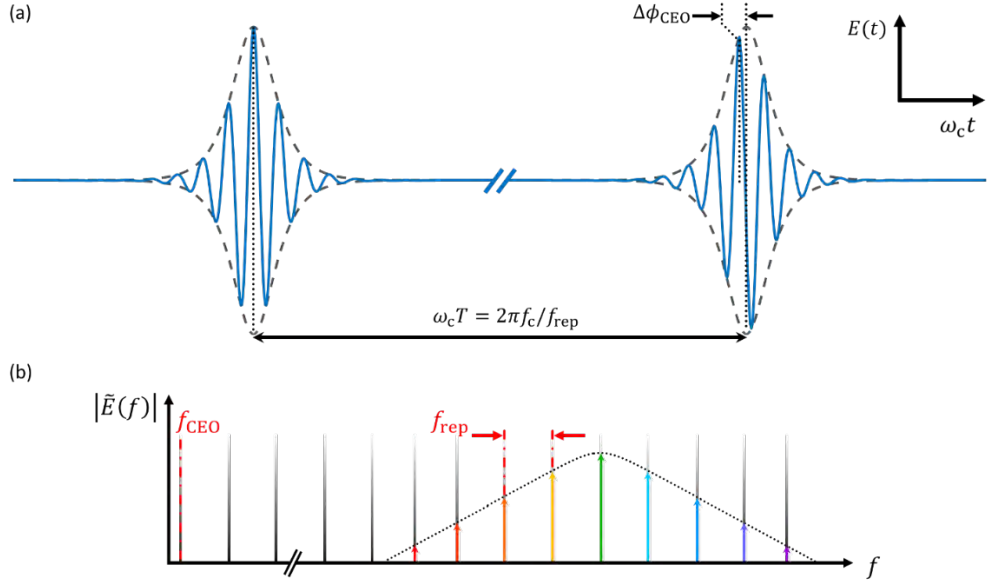

Fig. 9. Carrier envelope phase offset and carrier envelope frequency. (a) The carrier wave, in general, has a phase offset with respect to the pulse envelope; if the peak of the pulse envelope is aligned with the maximum value of the carrier sinusoid for one pulse, it will not, generally, keep this synchrony in the following pulse. The carrier envelope phase offset originates from the difference between group velocity and phase velocity. (b) The carrier envelope offset phase can easily be included in the mathematical description of a frequency comb (see text) and shown to be proportional to the offset frequency of the frequency comb from zero: if the comb is virtually extended toward the zero frequency, there will, generally, be an offset from the origin, smaller than the repetition rate and given by Eq. (18).

### 3.2.2  *Self-stabilization of an optical frequency comb*

Any frequency of an optical frequency comb can be identified in terms of the two frequencies $f_{\text{CEO}}$ and $f_{\text{rep}}$, as defined in Eq. (17). If these two frequencies are measured, then all frequencies of the comb are known. The measurement of the repetition rate is straightforward: it is equal to the beating frequency of any two adjacent frequencies in the comb. Repetition rates up to a few tens of GHz can directly be measured with commercially available spectrum analyzers and photodetectors. The carrier envelope offset frequency, on the other hand, is not so easy to measure. Let's recall that this frequency is in fact the offset from zero frequency of a fictitious extension of the comb down toward DC (very small) frequencies. To measure $f_{\text{rep}}$, a technique known as the *f-2f* is used, Fig. 10. A frequency $f_m = f_{\text{CEO}} + m f_{\text{rep}}$ on the low-frequency side of the comb spectrum skirt is passed through a nonlinear $\chi^{(2)}$ media and a tone equal to $2 f_m = 2 f_{\text{CEO}} + 2 m f_{\text{rep}}$ is generated. Another frequency $f_{2m} = f_{\text{CEO}} + 2 m f_{\text{rep}}$ on the high-frequency side of the comb spectrum is then sent with the latter doubled frequency through a photodetector. The beating of these two frequencies reveals the missing carrier offset frequency:

$$2 f_m - f_{2m} = 2 f_{\text{CEO}} + 2 m f_{\text{rep}} - \left( f_{\text{CEO}} + 2 m f_{\text{rep}} \right) = f_{\text{CEO}}. \tag{19}$$

If the two defining frequencies of a frequency comb, $f_{\text{CE}}$ and $f_{\text{rep}}$, are measured through this technique (or generally, not using another reference frequency), the comb is said to be *self-* *referenced* [245]. The *f-2f* technique assumes that the comb is octave-spanning, i.e., so wide that it covers, with detectable power, from one frequency up to another twice as large. This assumption poses technical challenges, as the width of a physical comb is always limited; beyond certain frequencies, on both ends, comb teeth are soaked in noise. One can think of other similar techniques which require a smaller comb span. For instance, if a frequency $f_m$ on the lower-side of the comb is tripled (using a cubic nonlinear medium) and another $f_n$ is doubled (in a quadratic medium), and if $3m = 2n$, then $f_{CEO}$ can again be found, and instead of an octave-spanning comb, one spanning three quarters of an octave could be self-referenced [246]. However, such techniques cannot always be easily implemented, as, e.g., nonlinear effects based on cubic (or higher) wave mixing processes need exceedingly high powers. As we will highlight in our discussion of dispersion later in this Section, microresonator dispersion engineering for broadband comb generation has been a long-term goal of the microcomb research community.

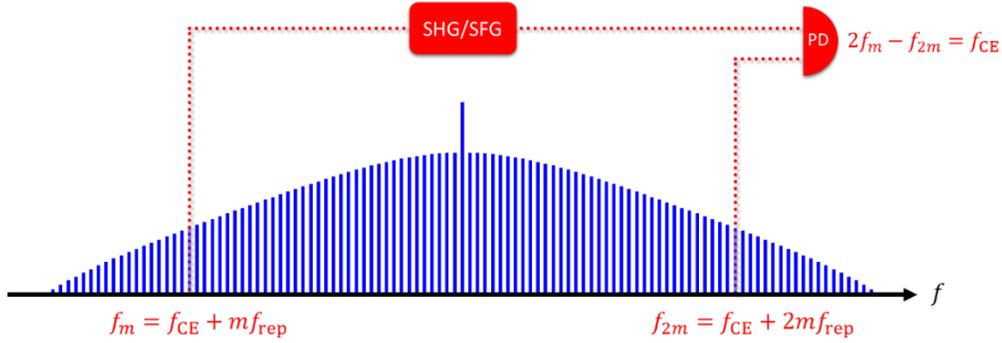

Fig. 10. Self-referencing of an optical frequency comb using the *f-2f* technique. If the optical frequency comb is wide enough that there is detectable power in two comb teeth in its skirt separated by an octave, the smaller frequency $f_m$ can be doubled and passed through a photodetector with the larger frequency $f_{2m}$. This method will resolve the unknown carrier envelope offset frequency. Similarly, $f_{CEO}$ might be found using the *2f-3f* technique, as described in the text. This surrogate method will require more power in the lower frequency which is to be tripled using a cubic nonlinear material, but the comb span can be smaller than octave-spanning.

### 3.2.3 *Conservation of momentum and energy in Kerr nonlinear wave mixing processes in microresonator*

Typical passive mode-locked lasers consist of a cavity, gain medium, and saturable absorber [247], see Fig. 11(a). The nonlinear FWM process in a high-Q resonator driven by a pump essentially combines the laser cavity and gain medium Fig. 11(b). Instead of gain provided by population, parametric gain will lead to the generation of new frequencies from vacuum fluctuations [248,4]. The comb generation process starts with degenerate FWM, where two photons of a strong enough pump are annihilated to give rise to one photon of higher and another of lower energy and frequency (traditionally called anti-Stokes and Stokes sidebands or, borrowing from the microwave nomenclature, idler and signal, respectively), Fig. 12. In an efficient process, the generated signal and idler waves are sufficiently strong and can interact with each other as well as with the pump, through what is referred to as non-degenerate FWM, to generate new sidebands. The cascaded frequency mixing process can lead to the generation of yet other frequencies, ultimately forming a frequency comb.

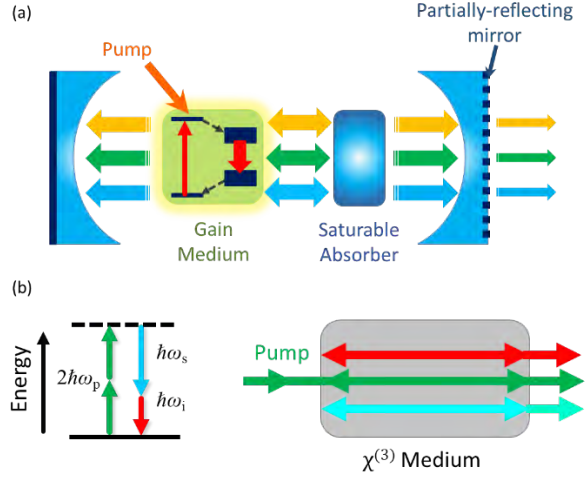

Fig. 11 Frequency comb generation in pulsed lasers and optical microresonators. (a) A mode-locked pulsed laser with a laser cavity, a gain medium which is pumped to create population inversion, and a saturable absorber. A saturable absorber makes mode-locked laser oscillation and pulse generation possible. (b) Frequency comb generation based on parametric frequency conversion in a $\chi^{(3)}$-material. The cubic nonlinear material leads to the generation of new frequencies from the pump frequency. In this process, signal and idler photons are generated at the expense of two pump photons. The generated signal and idler frequencies can then interact with each other and with the pump through the nonlinear media to generate new frequencies given their intensities are high enough and they can phase match.

Mode locking in passive mode-locked lasers is made possible by a saturable absorber. The saturable absorber could be replaced with a small block of Kerr-nonlinear material and an aperture, leading to Kerr-lens mode-locking (KLM). The Kerr medium provides an intensity-dependent refractive index such that the central part of the beam inside the laser cavity feels a larger refractive index than the external parts. This effect focuses the beam, which is then filtered by the aperture leading to larger loss for the outer lower-intensity parts of the beam. As a result, a saturable absorber is emulated. The advantage compared to mode locking through a saturable absorber is that the response time of a Kerr medium is very fast. Pulse formation based on microcombs, on the other hand, is said to occur through *spontaneous self-synchronization*, i.e., mode locking here does not rely on a saturable absorber [249–251]. Microcombs are therefore efficient pulsed laser sources [216].

Parametric frequency generation is subject to both energy and momentum conservation. Consider the schematic in Fig. 12, where $\eta \in \{0, \pm 1, \pm 2, \pm 3, \ldots\}$ is the *relative* azimuthal resonator mode number, which also labels the frequency comb teeth with respect to the pumped mode. The pumped mode (the one closest in frequency to the pump laser frequency) satisfies the resonance condition $j_0 \times \lambda_{j_0} = 2\pi R \times n_e(\lambda_{j_0})$, where $j_0$ is the mode number (an integer, usually a few hundred to a few thousand, based on resonator size), $\lambda_{j_0}$ is the resonance wavelength, $R$ is the resonator radius, and $n_e(\lambda_{j_0})$ is the modal effective refractive index at wavelength $\lambda_{j_0}$. Other modal frequencies satisfy a similar resonance condition, $j \times \lambda_j = 2\pi R \times n_e(\lambda_j)$. The relative mode number is therefore equal to $\eta = j - j_0$, such that the pumped mode is labeled by $\eta = 0$ and the rest of the comb teeth are numbered with respect to it.

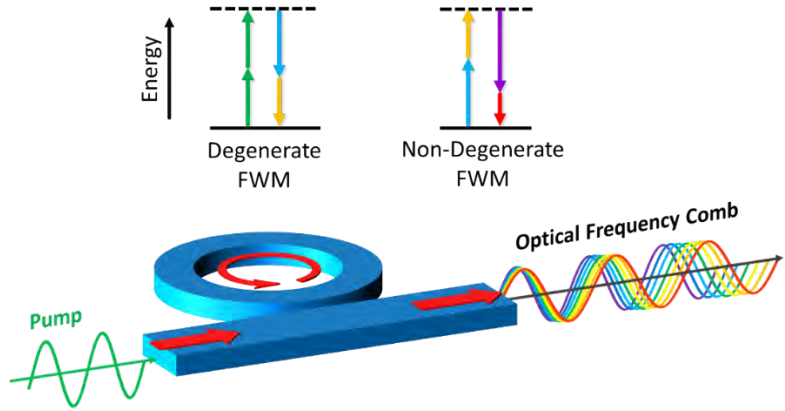

Fig. 12 Schematic showing frequency comb generation in a microring resonator driven by a CW laser. Field enhancement inside a high-Q microresonator results in nonlinear optical interactions at reduced powers. A set of new frequencies are generated as a result of cascaded FWM: seeded by vacuum fluctuations, pump-degenerate FWM generates photons at two new side-band frequencies. The generated photons can mix with the pump photons and with each other to create photons at other frequencies through non-degenerate FWM. Frequency generation in this process is both energy- and momentum-conserved. $\eta$ is the azimuthal microresonator mode number relative to the mode number of the pumped resonance.

Consider four resonator modes $\eta = l$, $m$, $n$, and $\mu$. Efficient interaction of these modes is subject to the phase matching condition

$$k_{j_0+l} - k_{j_0+m} + k_{j_0+n} = k_{j_0+\mu}.$$

(20)

Modes of a circular resonator can be considered angular momentum eigenstates with discrete propagation constants proportional to the mode number,

$$k_j = \frac{2\pi n_e(\lambda_j)}{\lambda_j} = \frac{j}{R}.$$

(21)

Substituting Eq. (21) in Eq. (20) and recalling that $j = j_0 + \eta$, shows that the phase matching condition, Eq. (20), will be satisfied if the mode numbers of the four interacting modes satisfy the constraint $l - m + n = \mu$. Consequently, modes of a resonator can harbor light frequencies interacting through Kerr nonlinearity and the interaction will automatically be phase-matched. Conservation of energy,

$$\hbar\omega_l - \hbar\omega_m + \hbar\omega_n = \hbar\omega_\mu,$$

(22)

where $\hbar$ is the reduced Planck constant, and,

$$\omega_\eta = \frac{2\pi c}{\lambda_{j_0+\eta}} = \frac{j_0 + \eta}{R} \cdot \frac{c}{n_e},$$

(23)

(for $\eta \in \{l, m, n, \mu\}$) is the resonance frequency in radians per second, on the other hand, places stringent requirements on the resonator material and geometric dispersion. Equation (22) means the generated frequencies must maintain a constant spacing, i.e., $\omega_l - \omega_m = \omega_\mu - \omega_n$. However, the effective index $n_e$ is not a constant and is affected by both change of frequency and resonator shape. (To be more accurate, even $R$ in the resonance condition and in Eq. (23) is not constant with frequency, because, strictly speaking, it is the effective radius of the resonator for the respective mode, measured roughly from the peak of the modal intensity profile to the cavity center.) In other words, the modes of a resonator are not necessarily equally spaced and the resonator free spectral range (FSR) varies with frequency. This simplified argument hints at the important fact that conservation of energy will limit the efficiency of cascaded FWM in a Kerr-nonlinear resonator and only sufficiently close to the pump, the accumulated FSR variations are small enough such that energy conservation will be respected. Influence of modal dispersion and energy conservation is, then, limiting the comb span: even if abundant laser power is available, comb span will be limited and power in the comb teeth will not be uniform. As we noted before, dispersion engineering, therefore bears a significant weight in microcomb generation research.

## 3.3 Comb Dynamics

Our discussion in the previous subsection was concerned with the basic physical concepts of microcomb generation. In this subsection, we will discuss the theory of microcomb generation in more detail and will focus particularly on the Lugiato-Lefever equation (LLE), introduced earlier in the historical overview.

### 3.3.1 Lugiato-Lefever formulation of Kerr frequency comb generation

The externally driven damped NLSE, or the LLE, is a partial differential equation with periodic boundary conditions which can describe nonlinear frequency mixing in optical microresonators with Kerr nonlinearity. The wave mixing process could also be described in the frequency domain in terms of a set of coupled ODEs [206]. It is straightforward to move from the LLE to the CWE picture and vice versa using the discrete Fourier transform. While the two descriptions are equivalent, each of them has their own advantages from the perspective of ease of numerical implementation, integration, and inclusion of other physical effects.

The goal of this section is not to rederive the LLE. For a detailed derivation, the reader can refer to the available literature [199,206,213,252]. The goal here is to outline and review the derivation steps briefly, and to highlight steps which are usually omitted from the literature. This understanding will enable the reader to derive variants of the LLE when necessary. We will start by the coupled-mode equations and then find the LLE based on them. The approximations involved in this derivation are underscored and their significance is discussed.

Consider a single mode of a cavity in the linear regime. If the cavity (or the mode) is not driven, the temporal dynamics of the complex mode amplitude $A$ (magnitude and phase) of this mode will be governed by

$$\frac{\mathrm{d}A}{\mathrm{d}t} = \mathrm{i}\omega A - \frac{\Delta\omega^{(\mathrm{int})}}{2} A, \tag{24}$$

where $\omega$ is the resonance or oscillation frequency of the mode and $\Delta\omega^{(\mathrm{int})}$ is the mode linewidth originating from intrinsic losses of the cavity (the cavity is not coupled to any sources or access waveguide) [253]. The above equation simply states that the electromagnetic field in this mode oscillates with frequency $\omega$ and its energy $|A|^2$ decays at a rate $\Delta\omega^{(\mathrm{int})}$. For an optical mode, the frequency $\omega$ is very large and can simply be removed from this equation by the substitution $A \rightarrow A \exp(-\mathrm{i}\omega t)$, i.e., focusing on the field envelope. If the cavity mode is driven, an excitation term should be added to this equation, and the decay rate should also be modified to account for external coupling losses. In the presence of nonlinearity, the cavity mode can be coupled to other cavity modes, and so other terms must be added to the right-hand side of the above equation depending on the nonlinearity type. Since different modes are now involved, we must label each complex amplitude and frequency, for instance, by their resonant mode number $j$. The nonlinear coupled-wave equations for the temporal evolution of the (slowly varying) intra-cavity complex amplitudes $A_j$ in a Kerr-nonlinear resonator hence take the following form,

$$\frac{\partial A_j}{\partial t} = -\frac{\Delta\omega_j}{2} A_j + \text{Terms due to Kerr nonlinearity } + \text{Drive term(s)}$$
(25)

We assume that modal energies are normalized such that $|A_j|^2$ is the number of photons in the mode labeled $j$. The integer $j$ is the azimuthal mode number of the resonator mode closest to this comb tooth, which satisfies $j \times \lambda_j = 2\pi R \times n_e(\lambda_j)$. In the latter expression, $\lambda_j = 2\pi c/\omega_j$ is the free-space resonance wavelength, $c$ being the vacuum speed of light, $R$ is the resonator radius, and $n_e(\lambda_j)$ is the effective refractive index at the resonance wavelength. $\Delta\omega_j$ is the resonance linewidth, defined by $\Delta\omega_j = 2\pi c/(\lambda_j Q_{L,j})$, where and $Q_{L,j}$ is the loaded optical quality factor of the resonance labeled $j$. The coupling between the modal complex amplitudes results from the Kerr nonlinearity and is reflected in the nonlinear terms. Therefore, they are proportional to the third-order nonlinear susceptibility $\chi^{(3)}$ of the cavity material and appear as triple products of the comb teeth complex amplitudes. It is through this term that the nonlinearity of the cavity material plays a direct role in the performance of the Kerr frequency comb. If the resonator is driven by a single CW laser, only one of the coupled-wave equations will have an explicit driving term. The general equation describing the temporal evolution of the complex-valued slowly-varying amplitude of the comb tooth closest to resonance with mode number j will hence read

$$\frac{\partial A_j}{\partial t} = -\frac{\Delta\omega_j}{2} A_j + ig \sum_{l,m,n} A_l A_m^* A_n \exp[-i(\omega_l - \omega_m + \omega_n - \omega_j)t] \delta_{j'j}$$
$$+ \sqrt{\Delta\omega_{j_0}^{(\text{ext})}} \mathcal{F}_0 \exp[-i(\omega_P - \omega_j)t] \delta_{j_0 j}.$$

(26)

where $j' = l - m + n$, and $l$, $m$, and $n$ are integers. The parameter $\mathcal{F}_o$ is the driving force at the pump frequency $\omega_P$ which is related to the pump laser power $P_{\text{in}}$ through

$$\mathcal{F}_0 = i\sqrt{\frac{P_{\text{in}}}{\hbar\omega_{j_0}}} e^{i\phi_{\text{in}}},$$

(27)

where $\phi_{\text{in}}$ is the phase of the pump laser, $\Delta\omega_j = \Delta\omega_j^{(\text{ext})} + \Delta\omega_j^{(\text{int})}$, $\Delta\omega_j^{(\text{ext})}$ is the linewidth attributed to external coupling, and $\Delta\omega_j^{(\text{int})}$ is the linewidth attributed to intrinsic losses. The parameter $g$ is the four-wave mixing gain [254] and is given by

$$g = \frac{n_2 c}{n_0^2} \frac{\hbar\omega_{j_0}^2}{V_{j_0}}.$$

(28)

In this expression, $n_2$ is the nonlinear (or second-order) index of refraction [194], related to $\chi^{(3)}$ through $n_2 = 3\chi^{(3)}/(4n_0^2 \epsilon_0 c)$, $\epsilon_0$ being the permittivity of free space and $n_0$ the (usual weak-field) index of refraction, $\omega_{j_0}$ is the resonance frequency of the pumped mode, and $V_{j_0}$ is the effective nonlinear mode volume for the pumped mode. The latter parameter naturally arises in the derivation [206] and for a resonator of volume $V_{\text{cavity}}$ is given by

$$V_{j_0} = \frac{\left[\int_\infty dV \ |\vec{E}(\vec{r})|^2\right]^2}{\int_{V_{\text{cavity}}} dV \ |\vec{E}(\vec{r})|^4}.$$

(29)

Here, $\vec{E}(\vec{r})$ describes the spatial distribution of the electromagnetic modal field. For a resonator with periphery $L$ and cross-section $A_{\text{cavity}}$, this expression can be simplified if one first evaluates surface integrals in the plane perpendicular to the direction of wave propagation in the resonator and considers that it does not change for a mode over resonator perimeter, namely,

$$V_{j_0} = LA_{j_0} = L\frac{\left[\int_\infty dA \ |\vec{E_t}(\vec{r})|^2\right]^2}{\int_{A_{\text{cavity}}} dA \ |\vec{E_t}(\vec{r})|^4}.$$

(30)

where $A_{j_0}$ is the effective nonlinear modal area and $\vec{E}_t(\vec{r})$ is the transverse part of the electromagnetic modal field. For a resonator of circular periphery with radius $R$, $L = 2\pi R$.

Cold cavity resonator modes are identified by expanding modal frequencies around resonance labeled $j_0$ with frequency $\omega_{j_0}$

$$\omega_j = \omega_{j_0} + D_1(j - j_0) + \frac{1}{2!}D_2(j - j_0)^2 + \frac{1}{3!}D_3(j - j_0)^3 + \cdots.$$

(31)

The coefficients $D_k$ in this equation are the dispersion parameters and will be further discussed in Section 3.4. The spatiotemporal intra-cavity waveform can be found from summing up the contributions of all the comb teeth

$$A(\theta, t) = \sum_j A_j(t) \exp\bigl[-i(\omega_j - \omega_{j_0})t + i(j - j_0)\theta\bigr].$$

(32)

Here $\theta$ is the azimuthal angle around the microresonator, which is related to the fast time parameter $T$ through $\theta = v_g T/R$ (modulo $2\pi$), being the group velocity of the waveform $A(\theta, t)$. (The fast time $T$ is a variable and should not be confused with the repetition rate $T$ used in Section 3.2.) Using the temporal derivative of the above expression,

$$\frac{\partial A}{\partial t} = \sum_j \left[\frac{\partial A_j}{\partial t} - i(\omega_j - \omega_{j_0})A_j(t)\right] \exp\bigl[-i(\omega_j - \omega_{j_0})t + i(j - j_0)\theta\bigr],$$

(33)

and its spatial $n$-th order derivative,

$$(-i)^n \frac{\partial^n A}{\partial \theta^n} = \sum_j (j - j_0)^n A_j(t) \exp\bigl[-i(\omega_j - \omega_{j_0})t + i(j - j_0)\theta\bigr],$$

(34)

together with Eqs. (11), (16), and (17), the spatiotemporal equation describing frequency comb generation can be found,

$$\frac{\partial A}{\partial t} = \left(-\frac{\Delta\omega_{j_0}}{2} + ig|A|^2 + \sum_{n=1}^{N\geq 2}(-i)^{n+1}\frac{D_n}{n!}\frac{\partial^n}{\partial \theta^n}\right)A + \sqrt{\Delta\omega_{j_0}^{(\text{ext})}}\mathcal{F}_0 \exp(-i\delta_0 t).$$

(35)

Here, $N$ is an integer describing the number of coefficients in Eq. (31) necessary to properly describe the particular resonator dispersion and $\delta_0 = \omega_P - \omega_{j_0}$ is the detuning of the pump from its nearest cavity resonance. In deriving the above equation, the frequency dependence of resonator linewidth has been ignored, i.e., we have assumed that $\Delta\omega_j = \Delta\omega_{j_0}$, for any $j$.

Equation (35) can be written in a rotating reference frame with angular velocity equal to $D_1$. This change of reference frame will remove the first-order spatial derivative of the intra-cavity field on the right-hand-side of Equation (35) and is possible through the following change of variable $\theta - D_1 t = \theta'$. Considering also $t = t'$, it is easy to show that

$$\frac{\partial}{\partial \theta} = \frac{\partial}{\partial \theta'} \tag{36}$$

$$\frac{\partial}{\partial t} = -D_1 \frac{\partial}{\partial \theta'} + \frac{\partial}{\partial t'}. \tag{37}$$

Equation (35) can then be written in a simplified form,

$$\frac{\partial A}{\partial t} = \left(-\frac{\Delta\omega_{j_0}}{2} + i\delta_0 + ig|A|^2 + \sum_{n=2}^{N\geq 2}(-i)^{n+1}\frac{D_n}{n!}\frac{\partial^n}{\partial \theta^n}\right)A + \sqrt{\Delta\omega_{j_0}^{(\text{ext})}}\mathcal{F}_0.$$

(38)

In this final equation, the primes over $\theta'$ and $t'$ have been removed to simplify notation. Note that the $D_1 \partial A/\partial \theta$ term has disappeared and the summation now starts from $n = 2$. In Eq. (38) detuning also has been moved from the exponent of the driving force to the coefficient of the field envelope through $A \to A \exp(-i\delta_0 t)$. This equation can describe many of the observed Kerr microcomb phenomena and for $N = 2$ reduces to the standard LLE in dimensional form:

$$\frac{\partial A}{\partial t} = \left(-\frac{\Delta\omega_{j_0}}{2} + i\delta_0 + ig|A|^2 + i\frac{D_2}{2}\frac{\partial^2}{\partial\theta^2}\right)A + \sqrt{\Delta\omega_{j_0}^{(\text{ext})}}\mathcal{F}_0. \tag{39}$$

Equation (39) does not have an analytical solution [203,255,256] but the simpler NLSE ($\Delta\omega_{j_0} = 0$, and $F = 0$) including detuning has steady-state *bright* soliton solutions of the form $A(\theta) = A_0 \text{sech}(B\theta)$ [200]. Substituting this solution in the steady-state NLSE leads to

$$\delta_0 + gA_0^2 \text{sech}^2(B\theta) + \frac{D_2}{2}B^2[1 - 2\text{sech}^2(B\theta)] = 0. \tag{40}$$

Or two simple equations for $A_0$ and B, namely, $gA_0^2 = D_2 B^2$ and $B^2 = -2\delta_0/D_2$. These expressions indicate that the bright soliton solutions of the NLSE (and approximate bright soliton solutions of the LLE) exist only for $D_2 > 0$ (recall that $g$ is a positive value) and $\delta_0 < 0$, i.e., for anomalous dispersion and the red-detuned pumping regime ($\omega_P < \omega_{j_0}$). While this finding is correct, as we will discuss in the next section, LLE has other types of stable, unstable, and breathing solutions too, such as hyperparametric oscillations or Turing patterns, breathing solitons, and chaotic combs [199,257–265].

We have introduced here the standard LLE. To model other effects, such as Raman scattering [266], thermal resonance shift, avoided mode crossing, or comb generation in material platforms or frequency regimes with both $\chi^{(2)}$ and $\chi^{(3)}$ nonlinearity other variations of this equation ought to be used. For instance to understand Raman self-frequency shift other nonlinear terms should be added to the left-hand side of Eq. (38) [74,234,267–269], to study thermal effects this equation should be coupled to other ODEs describing the thermal resonance shift of the resonance mode [237,270,271], or for simultaneous $\chi^{(2)}$-$\chi^{(3)}$ comb generation it should be coupled to another partial differential equation for comb generation around the second harmonic of the pump [272].

### 3.3.2 Numerical integration of the LLE

Before proceeding to numerical integration of Eq. (38), we will cast it in non-dimensional form,

$$\frac{\partial \psi}{\partial \tau} = \left(-1 - i\alpha + i|\psi|^2 - \sum_{n=2}^{N \geq 2} (-i)^{n+1} \frac{d_n}{n!} \frac{\partial^n}{\partial\theta^n}\right)\psi + F. \tag{41}$$

In this equation, we have normalized the waveform and the drive term through

$$\psi = A\sqrt{\frac{2g}{\Delta\omega_{j_0}}}, \tag{42}$$

$$F = \left[8g\Delta\omega_{j_0}^{(\text{ext})}/\left(\Delta\omega_{j_0}\right)^3\right]^{1/2} \mathcal{F}_0, \tag{43}$$

and time and dispersion through $\tau = t\Delta\omega_{j_0}/2$, $\alpha = -2\delta_0/\Delta\omega_{j_0}$, and $d_n = -2D_n/\Delta\omega_{j_0}$. Physically, the waveform $\psi$ is normalized to the sideband generation threshold, time is normalized to twice the photon lifetime of the cavity (at the pumped resonance), and pump-resonance detuning and dispersion coefficients are normalized to the half-width at half maximum (HWHM) of the pumped resonance. Use of the minus sign in defining $\alpha$ and $d_n$ is just a matter of choice and convenience. For the normalized dispersion coefficient $d_n$, in particular with this minus sign included, normal and anomalous dispersion regimes will be characterized by $\beta_2$ and $d_2$ of the same sign, i.e., in the normal dispersion regime $\beta_2 > 0$ and $d_2$

> 0, and in the anomalous dispersion regime $\beta_2 < 0$ and $d_2 < 0$. Equation (41) can be integrated numerically using the split-step Fourier transform (SSFT) method [162,273,274].

As noted earlier in this section, it is also possible to numerically model microcomb generation using coupled-wave equations (CWEs) of Eq. (26). We can use the Fourier transform of Eq. (41) to find the suitable normalized coupled ODEs which could be integrated using standard methods, such as the Runge-Kutta technique. Defining the discrete-time Fourier transform pair $\psi(\theta, \tau) \leftrightarrow \tilde{a}_\eta(\tau)$, with respect to the variables $\theta$, the azimuthal angle, and $\eta$, the mode number, by

$$\psi(\theta, \tau) = \sum_{\eta=-\infty}^{\infty} \tilde{a}_\eta(\tau) \exp(i\eta\theta), \tag{44}$$

$$\tilde{a}_\eta(\tau) = \frac{1}{\pi} \int_a^b \psi(\theta, \tau) \exp(-i\eta\theta) \, d\theta, \tag{45}$$

the desired normalized coupled ODE set can be found,

$$\frac{d\tilde{a}_\eta}{d\tau} = -(1 + i\alpha_\eta)\tilde{a}_\eta + i\sum_{l,m,n} \tilde{a}_l \tilde{a}^*_m \tilde{a}_n \, \delta_{\eta'\eta} + \tilde{F}_\eta. \tag{46}$$

In this equation, $\eta' = l - m + n$, and $\tilde{F}_\eta$ is the discrete-time Fourier transform of the drive term, which for the simple case of CW pumping reduces to $\tilde{F}_\eta = F\delta_{0\eta}$, i.e., only one equation in the ODE set (corresponding to the pumped resonance for which $\eta = 0$) has an explicit driving term. For more complex pumping schemes, such as a modulated pump or dichromatic or synchronous pumping, this driving term will be more elaborate [225–227,230,233,275]. The parameter $\alpha_\eta$, given by

$$\alpha_\eta = \alpha_0 - \sum_{n=2}^{N \geq 2} \frac{d_n}{n!} \eta^n, \tag{47}$$

is the cold cavity detuning of the comb harmonic at resonance $\eta$.

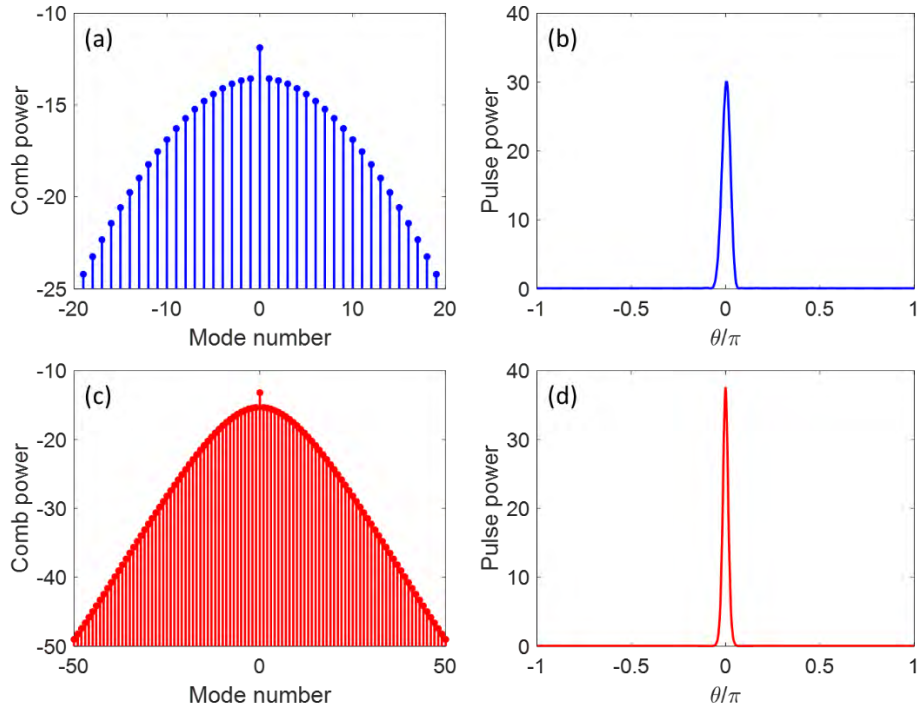

Fig. 13 A microcomb found by the numerical integration of the LLE using the coupled-wave equations (a, b) and the split-step Fourier transform (c, d). (a, c) show the comb power spectrum while (b, d) depict the pulse power. In the coupled-wave integration, 41 modes and in the split-step integration 128 modes are considered. It is seen that if the total number of modes considered in the two techniques are notably different, the final comb spectrum and pulse profile cannot be the same. Parameter values are $F = \sqrt{20}$, $\alpha = 19$, and $d_2 = -0.1$.

Fig. 13 shows results of integrating the LLE using the fourth-order Runge-Kutta (RK4) method, Fig. 13 (a, b), and also using SSFT, Fig. 13 (c, d). For the RK4 integration 41 modes and for SSFT 128 modes have been considered (only 100 of these modes are shown in the latter case). It is seen that if the total number of modes considered in the two techniques are notably different, the final comb spectrum and pulse profile cannot be the same. Generally, the direct integration of Eq. (46) is very time consuming and the integration time increases rapidly with increasing number of modes considered [206,212]. It is possible to use the RK4 technique in combination with fast Fourier transform (FFT) to increase integration speed [276].

### 3.4 Role of microresonator dispersion

In Section 3.2, we talked about the fundamental role of microresonator modal dispersion in frequency comb generation. Microresonator dispersion and quality factor essentially control the comb span and dynamic range of parameters such as pump power and pump-resonance detuning where stable combs can be generated. Engineering resonator dispersion has particularly been a major effort in the microcomb community, particularly with the objective of generating octave-spanning microcombs suitable for self-referencing. Dispersion requirements for broadband comb generation have been studied meticulously [218,222,224,277], and extensive work has been done on dispersion management in different resonator material platforms and geometries [218,222,223,277–292]. Below we will explore the main concepts and ideas related to microcomb dispersion. We will first discuss the nonlinearity-dispersion balance in Kerr nonlinear media which enables stable soliton formation [193], and then review higher-order dispersion, avoided mode crossing, DW emission, and their impact on microcomb repetition rate and stability.

### 3.4.1 Nonlinearity-dispersion balance in Kerr nonlinear media

Physically, GVD results in pulse broadening in a nonlinear media because different frequencies which comprise the pulse travel in the host media with different velocities. However, broadening caused by dispersion can be compensated by nonlinearity in a Kerr media. Because of the fundamental significance of this effect, we will describe it here by focusing on the standard NLSE. In microresonator-based frequency comb generation, the nonlinearity-dispersion balance is accompanied by the gain-loss balance, where cavity losses are offset by external pumping [217].

We can revert to the NLSE by ignoring the detuning, loss, and drive terms in Eq. (39),

$$\frac{\partial A}{\partial t} = \left(\mathrm{i}g|A|^2 + \mathrm{i}\frac{D_2}{2}\frac{\partial^2}{\partial \theta^2}\right) A.$$

(48)

In the absence of dispersion, Eq. (25) simplifies and includes only nonlinear self-phase modulation, i.e.,

$$\frac{\partial A}{\partial t} = \mathrm{i}g|A|^2 A. \tag{49}$$

The solution of this equation is $A(\theta, t) = A(\theta, 0) \exp[\mathrm{i}g|A(\theta,0)|^2 t]$, which is a pulse whose intensity remains unaltered during propagation but whose phase experiences a shift with time $\phi = g|A(\theta,0)|^2 t$, and therefore an instantaneous frequency drift $\delta\omega$ = -d$\phi$/dT, where T is the fast time related to $\theta$ by $\theta = v_\mathrm{g} T/R$, see Eq. (32). As noted earlier in the discussion following Eq. (39), $A(\theta, 0) = A_0 \mathrm{sech}(B\theta) = A_0 \mathrm{sech}(T/T_0)$ is a solution of Eq. (48), and $T_0$ being a measure of temporal pulse duration. For this pulse shape,

$$\delta\omega = -\frac{\mathrm{d}\phi}{\mathrm{d}T} = -g|A_0|^2 t \frac{\mathrm{d}}{\mathrm{d}T}\mathrm{sech}^2(T/T_0). \tag{50}$$

Fig. 14 depicts the instantaneous frequency shift $\delta\omega$ for a focusing cubic medium (for which $n_2$ > 0 and hence $g$ > 0) and shows that the instantaneous frequencies in the leading half of the pulse will decrease while those in the trailing half of the pulse will increase. If GVD is present, group velocity $v_\mathrm{g} = (\mathrm{d}\beta/\mathrm{d}\omega)^{-1}$ will change with frequency. If d$v_\mathrm{g}$/d$\omega$ > 0, group velocity increases with frequency. As a result, components of the pulse with higher instantaneous frequency (those in the trailing half of the pulse) will travel faster than those with lower instantaneous frequency (in the leading half of the pulse). This effect will lead to pulse narrowing and counter the self-frequency shift induced by self-phase modulation. The effect is reversed for d$v_\mathrm{g}$/d$\omega$ < 0, in which pulse broadening will be enhanced. The case d$v_\mathrm{g}$/d$\omega$ > 0 corresponds to anomalous GVD, since d$^2\beta$/d$\omega^2$ = d(1/$v_\mathrm{g}$)/d$\omega$ < 0. As a result, anomalous GVD can balance nonlinearity in focusing Kerr nonlinear media. We note that, while this analysis was based on a bright soliton and showed nonlinearity-dispersion balance in Kerr nonlinear media with anomalous dispersion, similar analysis can justify the same balance for a dark soliton in normal dispersion.

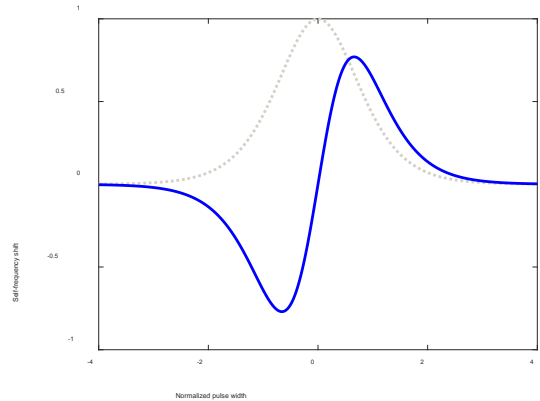

Fig. 14 Instantaneous self frequency shift (blue) of a soliton pulse in a Kerr nonlinear medium because of self-phase modulation as a function of pulse duration. The dotted grey curve in the background shows the pulse intensity. For the focusing Kerr medium considered above, the instantaneous frequencies in the leading half of the pulse will decrease while those in the trailing half of the pulse will increase. This frequency shift can be compensated by anomalous group velocity dispersion; see text.

### *3.4.2 Higher-order dispersion effects*

Below we will briefly review the main effects arising from higher-order microresonator dispersion.

**Dispersive wave emission and soliton recoil**

We talked earlier about the need for broadband combs for self-referencing [4]. As the comb bandwidth increases, considering higher-order dispersion effects will be inevitable [217]. This means resonator intermodal spacing (FSR) and group velocity will not uniformly increase or decrease with frequency, as in the case of simply anomalous or normal dispersion but will have a more complex behavior. One immediate consequence is that it is possible for modes farther away from the pump to have FSRs very close to that of the resonator at the pumped resonance and hence for phase matching between these weaker comb harmonics and the stronger harmonics near the pump. This means that in the case of soliton formation in the cavity, soliton group velocity and repetition rate, essentially dictated by the resonator FSR near soliton spectral peak, will match the phase velocity of some spectrally distant comb harmonics so that some of the soliton energy will be resonantly dispersed to these linear waves. The resulting higher-power comb sidebands are termed dispersive waves (DWs), see Fig. 15. This phenomenon is akin to Cherenkov radiation in particle physics which occurs when a charged particle travels in a dielectric medium with phase velocity larger than that in this dielectric and emits electromagnetic radiation. As a result, DW emission is also referred to as Cherenkov DW radiation in the microcomb literature [240]. DW emission has been studied theoretically and observed experimentally in fiber cavities and microcombs [98,221–223,230,281,293,294].

DW emission results in larger power on the spectral wings of the comb and therefore enhanced comb spectral coverage. Microresonator dispersion engineering has therefore been proven as an effective tool for broadband comb generation. Engineering resonator geometry to achieve desired dispersion profiles and generating one or two energetic dispersive wave signatures on the weaker comb spectrum wing has been demonstrated [221,282,292].

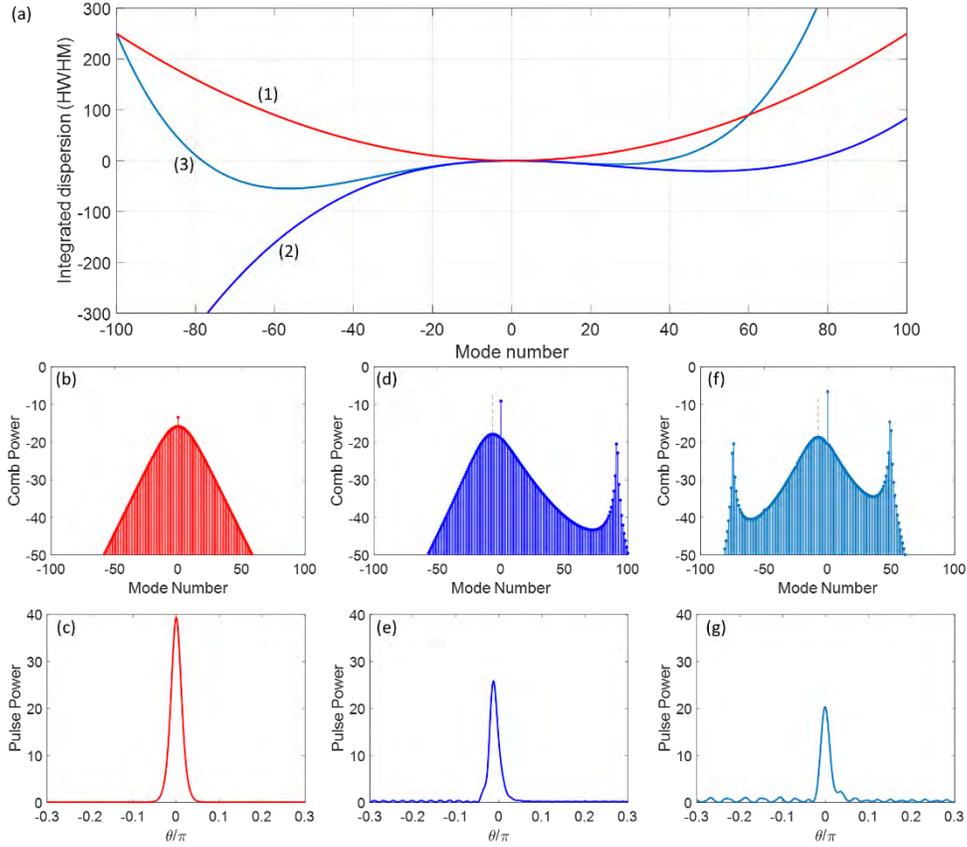

Fig. 15 Dispersive wave emission in microcombs resulting from higher-order dispersion. (a) Three different dispersion profiles, with significant contribution from (1) lowest order GVD coefficient $D_2$, (2) $D_2$ and $D_3$, and (3) $D_2$-$D_4$. (b, d, f) Comb spectra corresponding to the dispersion profiles (1, 2, 3) of (a). (c, e, g) Pulse powers profiles for the comb spectra depicted in (b, d, f). The vertical grey dashed lines in panels (d) and (f) signify soliton spectral peaks which do not coincide with the pump frequencies because of DW-induced soliton recoil.

In analogy with quantum mechanics, a momentum parameter can be defined for a frequency comb [225,240,295]. It has been shown that comb momentum plays the role of the comb spectrum center of mass [222,296], Fig. 16. This center of mass is always at zero for a steady-state CW-pumped comb even in the presence of higher-order dispersion [230]. As a result, DW emission is accompanied by 'soliton recoil' (Fig. 16): DW pushes the soliton spectral peak in the opposite direction such that it no longer coincides with the pump frequency, thus preserving soliton 'momentum'. Comparing panel (b) in Fig. 15, where third- and higher-order dispersion terms can be neglected, with panels Fig. 15 (d, f), in which higher-order dispersion coefficients ought to be considered, shows that in the presence of higher-order dispersion, soliton spectral peak (dashed line) can acquire an offset with respect to the pump. Soliton recoil, in turn, results in the change of comb repetition rate and in the increased sensitivity of the repetition rate to pump power fluctuations. Repetition rate fluctuation appears as increased phase noise of the RF signal demodulated on a photodetector and has significant practical ramifications [297–300]. The signature of DW emission in the time domain is modulation of the pulse CW background, Fig. 15 (e, g). In Fig. 15 (a), the vertical axis is the integrated dispersion,

$$D_{\text{int}}(j - j_0) = \omega_j - \omega_{j_0} - D_1(j - j_0) = \frac{1}{2!}D_2(j - j_0)^2 + \frac{1}{3!}D_3(j - j_0)^3 + \cdots, \quad (51)$$

expressed in units of pumped cavity half-width at half maximum (HWHM) linewidth ($\Delta\omega_{j_0}/2$) which appears in the LLE, Eq. (38). DWs appear roughly at the spectral point where integrated dispersion crosses zero.

We should note that the interplay of soliton Raman self-frequency shift can also shift the soliton spectral peak and comb repetition rate, and may compensate DW-induced soliton recoil [74,234,235].

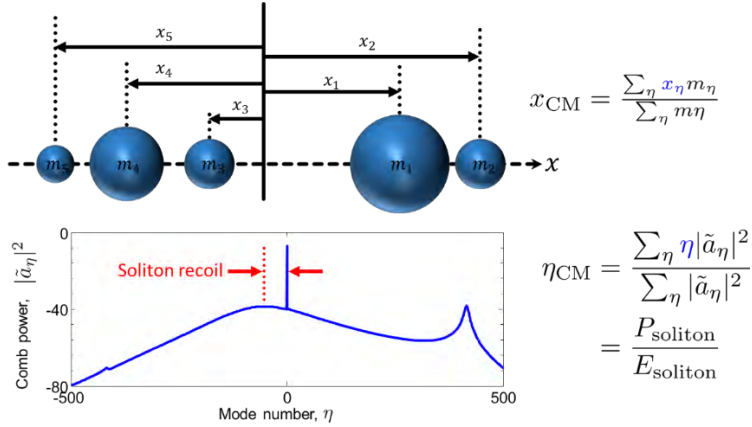

Fig. 16 Dispersive-wave-induced soliton recoil. Like center of mass (CM) defined for a group of point masses, an effective CM mode number can be defined for a microcomb as the ratio of comb momentum $P$ and energy $E$. For a steady-state comb, the comb momentum and therefore effective CM mode number should be zero, even in the presence of higher-order dispersion. This means appearance of DW on one side of the pump pushes soliton spectral peak to the other side.

**Avoided mode crossing**

Besides higher-order dispersion arising from the frequency dependence of one family mode, the inevitable effect of the interaction between the comb generating mode family and other mode families in a resonator cannot be ignored, especially in resonators of larger radius, which are of interest for RF and microwave signal generation, or at shorter pumping wavelengths, which are desirable for visible microcombs. Degenerate modal frequencies in two different mode families result in avoided mode crossings (AMCs, also called mode anti-crossings) which disrupt the dispersion profile of both mode families and have a significant impact on comb generation [219,283]. For instance, it has been shown that strong mode anti-crossings near the pump will inhibit soliton formation [218]. The significant role of AMCs in soliton crystal formation has also been studied both theoretically and experimentally [230,236,301].

AMC modeling is possible by a two-parameter model for the lifted degeneracy of the interacting mode families [218]. A more complete picture showing the transfer of power from the main pumped mode family to the crossing one has also been introduced [219,302]. Fig. 17 shows an example using the former modification of the LLE. An AMC around mode number 50, Fig. 17 (a), results from the interaction of the two mode families (red dotted line) and is reflected as a sudden disruption in the microcomb spectrum near the same mode number, Fig. 17 (b), and as the modulation of the soliton CW background in the microcomb temporal profile, Fig. 17 (c).

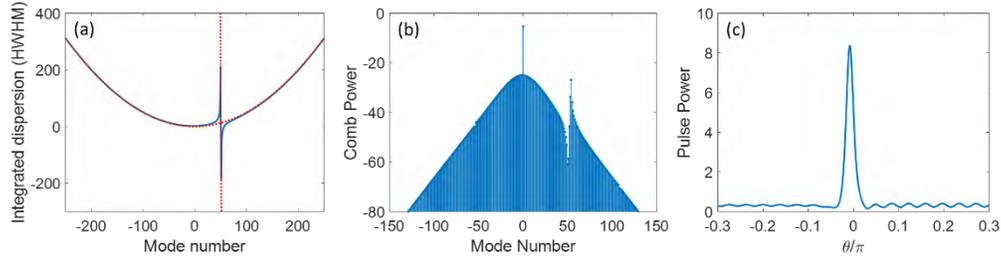

Fig. 17 Mode anti-crossing in the integrated dispersion profile of a resonator (a) and its impact on the microcomb spectrum (b) and pulse profile (c). An AMC around mode number 50 results from the interaction of the two mode families (red dotted line) (a) and is reflected as a sudden disruption in the microcomb spectrum near the same mode number (b), and as the modulation of the soliton CW background in the microcomb temporal profile (c).

Before closing this Section, we emphasize that while we have focused our discussion here on anomalous-dispersion microcombs, frequency comb formation in microresonators with normal GVD has been studied in detail as well [303–312]. An example is shown in Fig. 18. While normal dispersion combs are usually dark pulses, e.g., as in Fig. 18 (c), it has been shown that dispersion of the cavity quality factor could facilitate the generation of bright pulses in the normal dispersion regime [313]. Additionally, it has been shown that normal-dispersion microcombs can generally result in higher nonlinear conversion efficiency (NCE). [314] NCE is the ratio of the power in the comb sidebands at the resonator output port divided by the pump power at its input port, and, in resonators with dominant quadratic dispersion, increases with cavity finesse, GVD, and the coupling between the external waveguide and resonator. [315] Recently, higher conversion efficiency in resonators with dominant anomalous quartic (fourth-order) dispersion has been reported. [224,316]

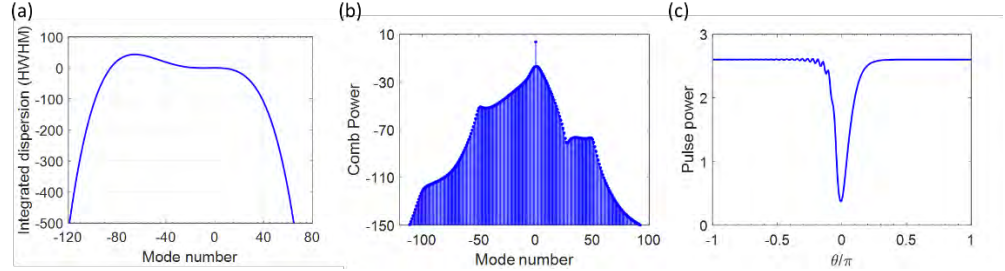

Fig. 18 An example normal dispersion frequency comb. (a) Integrated dispersion profile; (b) Comb power spectrum; (c) Pulse power profile.

### 3.4.3 Dispersion coefficients

For a waveguide, it is customary to define dispersion in terms of the coefficients in a Taylor series expansion of the propagation coefficient $\beta(\omega)$, sometimes also referred to as the wavenumber and denoted by $k(\omega)$, in the frequency $\omega$ around the pump or carrier frequency $\omega_0$. The expansion is usually written as

$$\beta(\omega) = \beta_0 + \beta_1(\omega - \omega_0) + \frac{1}{2!}\beta_2(\omega - \omega_0)^2 + \frac{1}{3!}\beta_3(\omega - \omega_0)^3 + \cdots .$$

(52)

The wavenumber $\beta$ has dimensions of [L]$^{-1}$ (SI units of m$^{-1}$). The coefficients in Eq. (53) are defined through

$$\beta_k = \frac{d^k \beta}{d\omega^k},$$

(53)

where all the derivatives are evaluated at $\omega_0$.

A ring (or more generally racetrack) resonator could be thought of as formed when a waveguide is bent onto itself and its two ends are connected to each other. This recipe will lead to periodic boundary conditions and as a result the continuum of waveguide modes will change into a set of discrete resonator modal frequencies. The frequency $\omega$ will then have to bear a label $j$ found from the resonance condition $j \times \lambda_j = 2\pi R \times n_e(\lambda_j)$, with $\lambda_j = 2\pi c/\omega_j$, $c$ being the vacuum speed of light, $R$ the ring radius, and $n_e(\lambda_j)$ the modal effective refractive index at the wavelength of the mode of interest. If the radius of curvature of the bends in this resonator are not very small (i.e., some tens of microns for typical materials like silicon), the dispersion parameters of the waveguide could be used for the resonator as a good approximation.

The modal dispersion in a resonator could also be identified by looking at the frequency dependence of the modal frequencies, rather than that of the propagation coefficient (as is the case for waveguides). This approach is more practically relevant when a resonator supporting whispering-gallery modes is considered, particularly because it is common to experimentally measure the spectral position of different modes involved in comb generation when characterizing a microresonator. When these modal frequencies are known, they can be fit to a polynomial in the modal number $j$ around a pumped mode number $j_0$, as written in Eq. (31) and repeated for convenience below.

$$\omega_j - \omega_{j_0} = D_1(j-j_0) + \frac{1}{2!}D_2(j-j_0)^2 + \frac{1}{3!}D_3(j-j_0)^3 + \cdots .$$

(54)

The coefficients $D_k$ are defined in a similar way to the $\beta_k$ coefficients, Eq. (53), with the caveat that their definition is based on discrete differences rather than continuous variable derivatives. In practice, they can be found by polynomial curve fitting.

It is straightforward, in principle, to find the $D_k$ coefficients with a knowledge of the $\beta_k$ parameters, or vice versa. To find their relationship, one notes that the resonance condition for a mode of frequency $\omega_j$ can be written as

$$\frac{j}{R} = \frac{2\pi}{\lambda_j} n_e(\omega_j) = \frac{\omega_j}{c} n_e(\omega_j) = \beta(\omega_j).$$

(55)

Hence,

$$\frac{j-j_0}{R} = \beta(\omega_j) - \beta(\omega_{j_0}).$$

(56)

Using Eq. (52) to expand the propagation coefficient $\beta(\omega_j)$ at $\omega_0 = \omega_{j_0}$, and substituting the result in the above equation leads to

$$\frac{j-j_0}{R} = \beta_1(\omega_j - \omega_{j_0}) + \frac{1}{2!}\beta_2(\omega_j - \omega_{j_0})^2 + \frac{1}{3!}\beta_3(\omega_j - \omega_{j_0})^3 + \cdots .$$

(57)

This expression gives us $j - j_0$ in terms of $\omega_j - \omega_{j_0}$. It is then possible to replace the different powers of $j - j_0$ in Eq. (54) with the above summation in terms of $\omega_j - \omega_{j_0}$.

$$\omega_j - \omega_{j_0} = D_1\left[R\beta_1(\omega_j - \omega_{j_0}) + \frac{R}{2!}\beta_2(\omega_j - \omega_{j_0})^2 + \frac{R}{3!}\beta_3(\omega_j - \omega_{j_0})^3 + \cdots\right]$$
$$+ \frac{1}{2!}D_2\left[R\beta_1(\omega_j - \omega_{j_0}) + \frac{R}{2!}\beta_2(\omega_j - \omega_{j_0})^2 + \frac{R}{3!}\beta_3(\omega_j - \omega_{j_0})^3 + \cdots\right]^2$$
$$+ \frac{1}{3!}D_3\left[R\beta_1(\omega_j - \omega_{j_0}) + \frac{R}{2!}\beta_2(\omega_j - \omega_{j_0})^2 + \frac{R}{3!}\beta_3(\omega_j - \omega_{j_0})^3 + \cdots\right]^3 + \cdots .$$

(58)

Comparing the coefficients of different powers of $\omega_j - \omega_{j_0}$ on both sides of the above equation leads to expressions linking $D_k$s to $\beta_k$s. For instance,

$$1 = D_1 R \beta_1 \rightarrow D_1 = \frac{1}{R\beta_1}$$

(59)

$$0 = \frac{1}{2}D_1 R\beta_2 + \frac{1}{2}D_2 R^2 \beta_1^2 \rightarrow D_2 = \frac{-\beta_2}{\beta_1^3 R^2}. \tag{60}$$

This algorithm can, in principle, give the relationship between any $D_k$ coefficient with the $\beta_k$s, although collecting all powers of $\omega_j - \omega_{j_0}$ on the right-hand-side of Eq. (58) rapidly becomes algebraically elaborate as the index $k$ increases. For $k = 3, 4$ one finds

$$D_3 = \frac{-\beta_3}{R^3 \beta_1^4} + 3\frac{\beta_2^2}{R^3 \beta_1^5}, \tag{61}$$

$$D_4 = \frac{-\beta_4}{R^4 \beta_1^5} - 33\frac{\beta_2^3}{R^4 \beta_1^7} + 16\frac{\beta_2 \beta_3}{R^4 \beta_1^6}. \tag{62}$$

It is noted that although we have derived equations for $D_k$s strictly in terms of $\beta_k$s only (and not $D_k$s with smaller index value $k$), in a numerical routine, the $D_k$s can be found recursively (i.e., first $D_1$, then $D_2$, then $D_3$, and so on) and not all at once. Hence, polishing the algebraic equations as was done above is not necessary. Obviously, same holds for finding $\beta_k$s given $D_k$s.

## 4 Material systems for frequency combs

### 4.1 Overview

Based on the analytical expressions in Section 3, there are two aspects that must be considered when designing an integrated microcavity-based comb: 1) the device optical performance and 2) the optical properties of the material. However, material properties and device performance are related through metrics like the optical absorption and the refractive index. The interdependence of geometry and material properties on cavity performance can be clearly seen $Q_{mat}$ and $Q_{ss}$: [36,121]

$$Q_{mat} = \frac{2\pi n}{\alpha \lambda} \tag{63}$$

$$Q_{ss} = \frac{\lambda^2 D}{2\pi^2 \sigma^2 B} \tag{64}$$

where $\alpha$ is the optical loss, D is the diameter, $\sigma$ is the rms size, and B is the correlation length of surface inhomogeneities. Therefore, the optimization of the material system and device design cannot be performed independently. In addition, factors like compatibility with nanofabrication processes and on-chip integration with optical couplers must be considered. As a result, designing a frequency comb is a multi-parameter, complex challenge.

Given the complexity of this optimization problem, the initial research efforts into microcavity-based frequency combs chose to pursue non-integrated device designs. [4,39,60,176,207,269,317–328] This approach removed restrictions related with on-chip integration and nanofabrication compatibility during these investigations. However, to translate from the lab environment, researchers were motivated to expand beyond the conventional integrated photonics toolbox and to explore alternative material systems and integration strategies.

After providing a brief history on a few of the initial demonstrations using dielectric materials, a more in-depth discussion will present the emerging dielectrics and semiconducting materials being explored for comb generation as well as discuss strategies based on hybrid device structures that combine multiple material types.

### 4.2 Dielectric material systems

The first whispering gallery mode optical cavities were fabricated from dielectric materials, starting with liquid droplets and moving quickly to polymer and silica spheres. [19,21,24,25,29,36,329] In the past couple decades, this range of materials and device geometries has greatly expanded to include crystalline materials as well as a broad range of on-chip geometries. A brief overview of shown in Fig. 19. Much of the foundational understanding of optical resonant cavity physics was developed based on the initial spherical devices. [25]

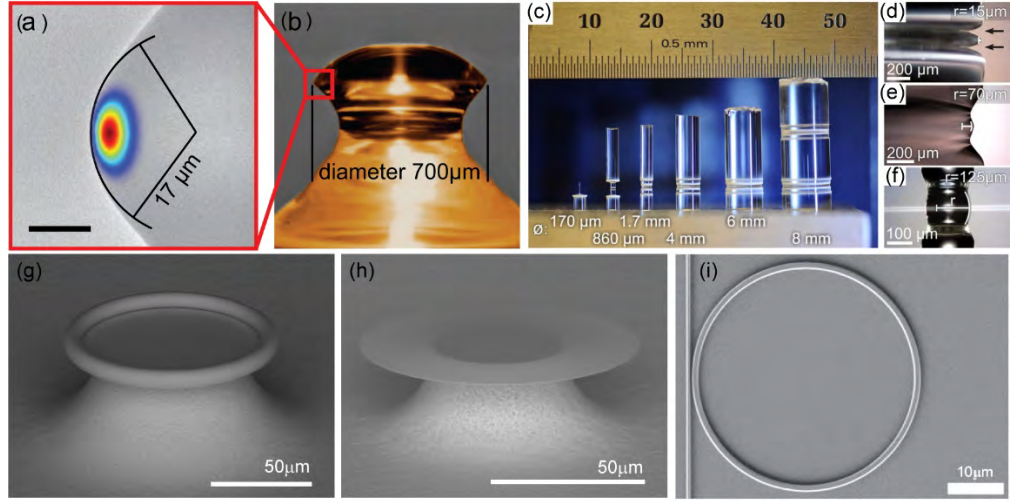

Fig. 19. Example materials and geometries of dielectric resonant cavities. (a) SEM image with FEM model indicating optical mode location and (b) optical image of $MgF_2$ cavity. Reprinted by permission from Nature [322]. © 2013 (c)-(f) Optical images of a range of laser machined quartz optical cavities. Reprinted with permission from [330]. Copyright 2013 by the American Physical Society. SEM images of (g) SiO2 toroidal cavity, (h) $SiO_xN_y$ microdisk cavity, and (i) $Si_3N_4$ microring. Reprinted by permission from Nature [109]. © 2009

The first cavity-enhanced frequency combs relied on dielectric cavities; specifically, $MgF_2$, $CaF_2$, and silica toroidal cavities. [4,317] When combined with low optical loss coupling methods, such as tapered optical fibers and prism couplers, the silica spheres and toroidal cavities could easily achieve quality factors above 100 million, and the Q's of the fluoride cavities were above 10 billion. These ultra-high-Q factors allowed frequency combs to be easily generated despite the low nonlinear coefficients and low refractive index values (**Table 1**).

**Table 1. Summary of nonlinear coefficients of dielectric materials and device behavior*.**

| Material | Major Diameter r (μm) | Minor Diameter (Width x Height) (μm) | Core Index | Cladding Index | Q (x$10^6$) | $n_2$ (x$10^{-20}$ $m^2$/W) | $P_{circ}$(W)** |
|---|---|---|---|---|---|---|---|
| $MgF_2$ [285] | 3800 | 120 | 1.37 | 1 | 200 | 0.7-0.9 | 3 |
| $CaF_2$ [40] | 5000 | 200 | 1.43 | 1 | 6000 | 1.2 | 66 |
| $BaF_2$ [331] | 1680 | 150 | 1.47 | 1 | 500 | 2.85 | 16 |
| $SrF_2$ [332] | 10600 | 920 | 1.43 | 1 | 1000 | 1.76 | 5 |
| Silica | | | 1.45 | 1 | | 3 | |
|   toroid [317] | 75 | ~4 | | | 100 | | 77 |
|   wedge [324] | 3000 | -- | | | 400 | | 7 |
| $SiO_xN_y$ [104] | 54 | 5.6 | 1.50 | 1 | 130 | unknown | 135 |
| $SiO_xC_y$ [333,334] | 270 | 1.45 x 1.5 | 1.7 | 1.45 | 1.2 | 11.5 | 0.22 |
| | | 0.725 x 1.650 | | 1.45 | 3 | 25 | |
| $Si_3N_4$ [335] | 200 | | 2.0 | | | | 2 |
| Diamond [128] | 50 | ~0.850 x 0.750 | 2.38 | 1.45 | 1 | 13 | 1 |

*at 1550 nm unless indicated. **The metal fluoride cavities are posts, two types of silica cavities are modeled, and the $Si_3N_4$ and diamond cavities are rings. All cavities are operating at critical coupling at 1550 nm with 1mW input power. All values and geometries are based on literature values as indicated.

Interestingly, while similar Q's are obtained in both material systems, the device fabrication methods are very different. The initial silica cavities relied on a thermal reflow process whereas the crystalline fluoride cavities used a polishing method.

The first silica cavities were microspheres fabricated from optical fiber melted using either a flame or a $CO_2$ laser. The increase in temperature induced a surface tension-induced reflow process, resulting in atomically smooth surfaces. Subsequent efforts developing suspended silica toroidal devices also relied on this same reflow technique, and more recently, chemical polishing methods have been used to create wedge-shaped devices. [278]

Subjecting a crystalline material to a thermal reflow step would induce a change in the crystal structure, stoichiometry, and even lead to an increase in surface roughness due to alternating degrees of purity along the crystal facets. Therefore, crystalline materials require alternative methods for the fabrication of whispering gallery mode resonators. Typically, these materials need to be mechanically ground and polished to obtain smooth surfaces with sub-nanometer surface roughness, like free-space optical components such as lenses or high-reflectance mirrors. The most common crystalline materials used in comb generation are the metal fluorides consisting mostly of calcium fluoride ($CaF_2$), magnesium fluoride ($MgF_2$), strontium fluoride ($SrF_2$), and barium fluoride ($BaF_2$).

Due to the low refractive index, fluoride and silica cavities poised a fundamental challenge with integration on a silicon substrate, namely, leakage of light into the higher index silicon substrate. As a result, the initial dielectric cavities were air clad (Fig. 19). However, this approach limited integration with additional on-chip components, like waveguides, detectors, and optical sources. Recently, a shift to nitrides and diamond as well as the implementation of low index oxides as buffer layers has enabled planar dielectric platforms to be designed and demonstrated. [102,104,106,107]

### 4.2.1  Metal Fluorides

The first metal fluoride fabricated into a whispering gallery mode resonator for generating Kerr frequency combs was $CaF_2$. [176] Single crystal $CaF_2$ possesses a wide transparency window from 130 nm to 10 µm, enabling high optical performance throughout this range. However, it has a relatively low non-linear index of refraction ($n_2$). Despite this low value, a tunable Kerr frequency comb was able to be generated with a span of 85 nm with an input pump power of 50 mW using a monolithic, mechanically polished $CaF_2$ whispering gallery mode resonator. This achievement was possible due to the device's quality factor, which was $2.5 \times 10^9$ (Fig. 20). The large intrinsic quality factor allows the resonator to build-up high circulating intensities, further enabling non-linear optical effects such as hyperparametric oscillations and stimulated Raman scattering (Stokes and Anti-Stokes), which are also visible in Fig. 20.

Soon after, $CaF_2$ resonators were fabricated with increased quality factors ($6 \times 10^9$) and Kerr combs, as well as Kerr-Raman combs, were observed. [39] Due to the increased quality factor, an increased span was observed (158 nm) when pumped at 1560 nm with 25 mW of input pump power. The improved span of the comb included both Stokes and anti-Stokes regions and any subsequent wave-mixing contributing to the line density of the comb. The increase in competition between stimulated Raman scattering (SRS) and Kerr comb generation was largely due to the change in resonator diameter and change in pump wavelength. The competition becomes fiercer at lower pump wavelengths as the group velocity dispersion (GVD) of the material remains normal and was near zero at the first expected Raman mode. This change made SRS a more efficient non-linear process than hyperparametric oscillation and therefore allowed for the observation of Stokes and anti-Stokes lines. Competition between hyperparametric oscillation and stimulated Brillouin scattering (SBS) also occured given the right geometry and pump wavelength. In over-moded $CaF_2$ resonators pumped at 1550 nm, Stokes and anti-Stokes SBS lines were generated, aided by the four-wave mixing processes already present. [283] In accordance with theory, the $CaF_2$ resonator exhibited a Brillouin shift of about 12.1 GHz.

Given requirements related to device fabrication as well as the occurrence of a host of other non-linear optical processes that may interfere, it is understood that whispering gallery mode resonators fabricated from $CaF_2$ present a challenge in terms of design and operating wavelength for the creation of spectrally pure Kerr frequency combs. $MgF_2$, however, has demonstrated low phase noise Kerr frequency combs and are currently used in commercial applications.

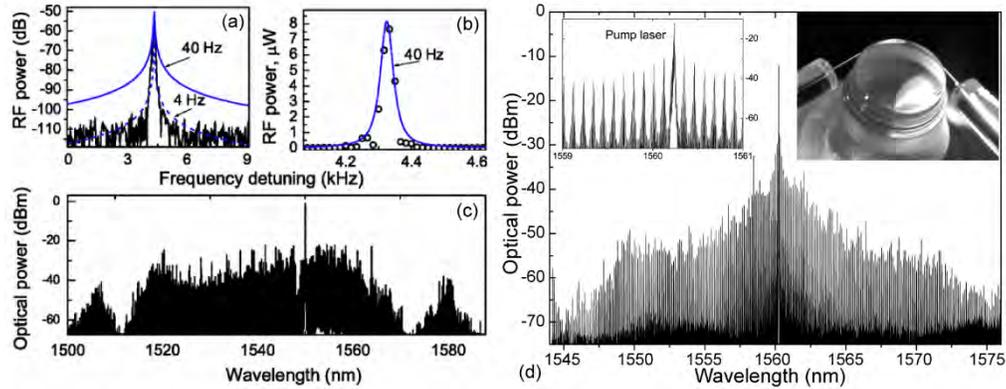

Fig. 20. Example combs generated by two different $CaF_2$ devices. (a)-(c) Microwave signal generated (in two different power scales) and the optical comb used. Fit of the microwave data on a logarithmic scale gives a more accurate determination of the signal bandwidth. Image adapted from [176]. (d) Comb generated by a $CaF_2$ resonator. Inset (left): zoom in around pump to observe the line spacing of 0.112 nm (which agrees with the cavity FSR of 13.81 GHz). Inset (right): Image of the 4.85 mm diameter device under test. Reprinted with permission from [39]. Copyright 2008 by the American Physical Society

$MgF_2$, like $CaF_2$, also possesses a very wide transparency window from 130 nm to 9 μm. The refractive index and non-linear index of refraction are however lower than $CaF_2$, at 1.38 and $0.7$-$0.9 \times 10^{-20}$ $m^2/W$ respectively. [336] $MgF_2$ is also crystalline and possesses a larger hardness (around 6 Mohs) than $CaF_2$, making it less sensitive to polishing and grinding techniques. Its mechanical robustness also makes it the preferred material for use in packaged devices such as opto-electronic oscillators and small form-factor ultra-narrow linewidth lasers. [337,338]

The first Kerr frequency comb generated from a monolithic $MgF_2$ disk whispering gallery mode resonator was demonstrated to be the most power-efficient comb produced at that time, as measured by its bandwidth to input power ratio. [320] The whispering gallery mode resonator was pumped with 2 mW and produced a 20 nm wide Kerr frequency comb centered around a pump wavelength of 1543 nm. The comb produced was phase locked as measured by an RF signal produced from the comb lines beating against each other on a fast photodiode. The intrinsic quality factor of the disk was reported to be $3 \times 10^9$. An improvement on $MgF_2$ combs was reported shortly thereafter with an improved span of about 200 nm (Fig. 21). [285] The improvement in the span is argued to be due to the carefully engineered spectrum of the whispering gallery mode resonator itself. By tuning the ratio between the linewidth and the dispersion of the resonator, a coherent frequency comb with initial sidebands of only one FSR distance away from the pump was generated. These coherent frequency combs produced stable and narrow RF signals, which is crucial for spectroscopic and metrological applications. Along with carefully tuned coupling conditions and pump powers, the excitation of a broadband Kerr frequency comb was observed as the number of sidebands non-adiabatically jumped from three to over 100 with a minimal power increase. [260]

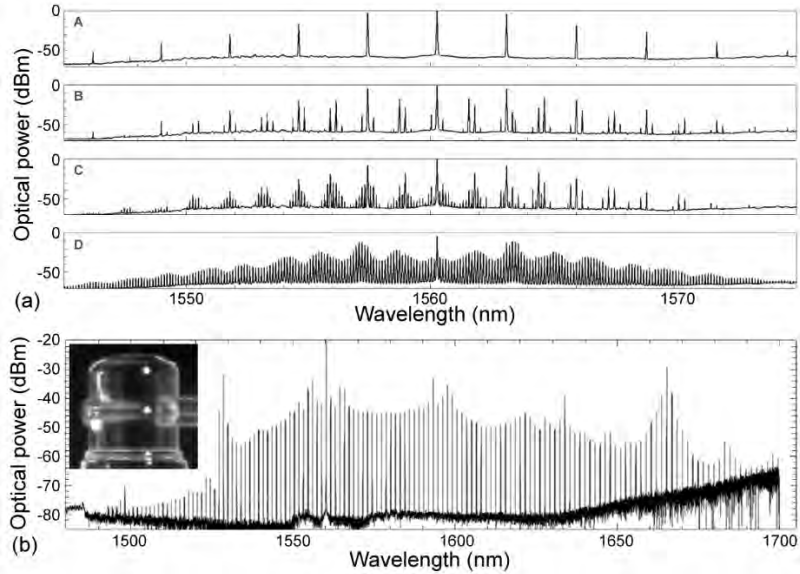

Fig. 21. Examples of frequency combs generated by a $MgF_2$ resonant cavities. (a) The comb lines in the 3.8 mm diameter cavity are initially spaced at 19 cavity FSRs (A) and they gradually fill the spectrum until steady state comb generation is reached. (b) In a device with an engineered geometry (and thus engineered dispersion), the comb spectrum is much broader. In this comb generated by a 0.403 mm diameter cavity, over a hundred lines spanning over 200 nm are generated with 50 mW of input power. Figures adapted from [285].

Other metal fluoride compounds which have been utilized for whispering gallery mode resonators and even Kerr comb generation are barium fluoride ($BaF_2$) and strontium fluoride ($SrF_2$). While lithium fluoride has been used for whispering gallery mode resonators and even Brillouin lasing, Kerr comb generation has not been reported for this material. [339] The first reports of a Kerr frequency comb generated from a $BaF_2$ whispering gallery mode resonator came out in the same year (2016), and both focused on $BaF_2$ as a universal non-linear scattering platform (FWM, Brillouin, Raman, etc.) at pump wavelengths near 1550 nm. [340] Typical quality factors of these resonators can reach upwards of one billion and are polished and machined in much the same way that the other metal fluorides are. $BaF_2$ possesses a low anomalous dispersion in the mid-IR and is therefore an attractive material for Kerr comb generation in that wavelength regime. However, at 1550 nm, it possesses a slightly higher refractive index than silica (1.466), therefore making prism coupling a more attractive option for efficient excitation of ultra-high Q modes. Although many non-linear phenomena were present in the $BaF_2$ resonator, the span of the Kerr frequency comb was relatively small, at only about 40 GHz. Another study reported spans of about 348 GHz, but also with competing non-linear effects such as Brillouin lasing, ultimately limiting Kerr comb generation through hyperparametric oscillation. [331] However, competing non-linear effects in whispering gallery mode resonators can be used to synergistically promote one, such as Kerr-assisted Raman lasing. Unsurprisingly, a Kerr-Raman comb is generated with $BaF_2$ with a larger span (~300 GHz) than the Kerr comb, centered around 1608 nm. Overall, the metal fluorides and their usage have been largely responsible for the detailed exploration of microcomb technology and the establishment of the governing theory. As such, they paved the way for compact, commercially available opto-electronic oscillators and all optical time keeping devices, whether fabricated from fluorides or other material systems.

While cavities based on fluoride materials can solve several critical challenges in developing a practical comb-system, namely, environmental stability, operation over a large wavelength range, and extremely high Q factors, they also suffer from one fundamental limitation. Fluorides are extremely inert and non-reactive. While this property does contribute to their environmental stability, it also means that fluorides are nearly impervious to all standard lithographic etching methods. As a result, fluoride cavities are typically fabricated using hand-polishing methods, resulting in free-standing devices. [40,187] This fabrication approach creates two issues.

First, the precision and accuracy of hand-polishing in achieving a pre-defined diameter is not as robust as nanolithography. Because the resonant wavelength, FSR, and dispersion are dependent on the diameter (as well as material properties), any variation between devices will give rise to variability between the combs that are generated. This dependence motivated recent research into developing motorized or automated mechanical polishing methods. [341] Second, it is necessary to have low loss photon injection to achieve high performing combs. To meet this criterion, low loss couplers, such as prisms or tapered fibers, are used. However, the alignment relies on high precision stages, and even slight perturbations can dramatically change the coupled power, as shown by the coupling coefficient K term in equation (2), thus changing the performance.

Significant efforts have been invested into overcoming the coupling challenge. One approach relies on pre-mounted micro-prism couplers which are aligned with the cavity while monitoring transmission. [322] Researchers have also investigated combining fluoride cavities with on-chip, lithographically patterned, slab waveguides [88]. Both approaches provide a path to solving the packaging challenge, and they have both had great success in achieving combs in the near and mid-IR. Additionally, the broad transparency window and dispersion profile of metal fluorides make them a very attractive material system for high performance comb applications where device to device reproducibility is not as high of a concern.

*4.2.2 Silicon Dioxide*

One approach to overcome this hurdle is to use integrated photonic devices. The additional control provided by lithography improves the reproducibility of the device fabrication. Motivated by this concept, the first on-chip devices were based on the silica toroidal platform (Fig. 22 (a)). Building on previous work in parametric oscillation and Raman [61,61,62,64,187,342], silica microtoroids were used for generating and studying broadband Kerr frequency combs in 2007. [317] Benefiting from the smooth surface of the devices, quality factors of the silica toroids can achieve over $10^8$ easily, which greatly enhances the circulating power inside the cavity and boosts comb generation.

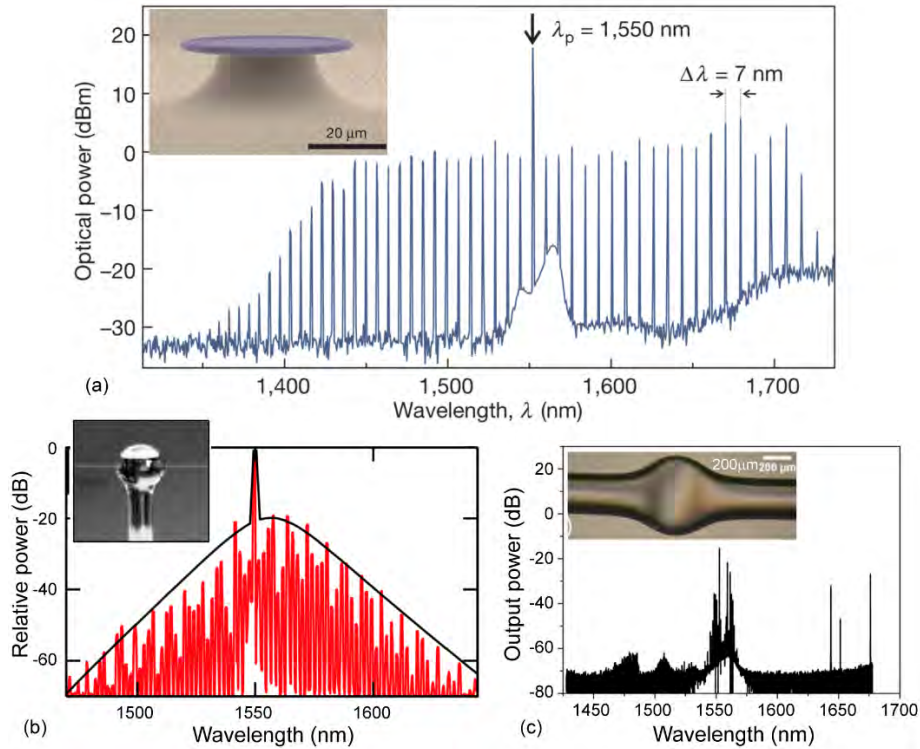

Fig. 22. Example combs generated by three different types of silica resonator pumped at 1550 nm. Insets: Images of the corresponding device geometries. (a) Silica microtoroid. Reprinted by permission from Nature [317] © 2007. (b) Silica microsphere. Adapted from [343]. and (c) Silica microbubble. Adapted from [344].

Besides silica microtoroids, silica microspheres and microbubbles have also demonstrated for frequency combs generation (Fig. 22 (b) and (c)). Silica microspheres can be easily made by melting the end of a standard optical fiber and ultra-high quality factors can also be easily achieved for this type of devices. Due to the high symmetry and spatial freedom of the microsphere structure, exceptionally dense spatial modes present in these devices, which can lead to avoided mode crossing and prevent the soliton formation. However, by careful alignment of the microsphere relative to the coupling fiber taper, it has been demonstrated that higher-order spatial modes can be suppressed and that cavity soliton formation can be achieved from silica microspheres. [343] Silica microbubbles are also studied for frequency combs generation. The silica microbubbles are fabricated by heating slightly pressurized silica capillaries uniformly. Quality factors of the silica microbubbles are usually above $10^7$ as a result of thermally smoothed surface. Comparing to silica microsphere, additional controls on the device structure are introduced, like the wall thickness. Dispersion of the silica microbubbles are engineered by modifying the geometry parameters and Kerr frequency combs are generated from these devices. [321,344–346]

However, while the toroidal cavities are based on a lithographically defined microdisk, the final fabrication step involves a laser-induced reflow process to remove any residual surface roughness. [121] This step enables ultra-high-Q factors to be readily achievable, but it makes the device diameter even harder to control than when hand-polished methods are used. This process also complicates integration with on-chip waveguides. Because the device diameter shrinks by several microns during the reflow step, the device will move away from any pre-patterned waveguide structures. As a result, there has been limited success creating an in-plane lithographically patterned toroid-waveguide system [129] with most efforts focusing on fabricating the waveguide and resonator separately and then packaging them. For silica microspheres and microbubbles, due to the similar reflow or thermal heating process, they have the same difficulty in precise control of the device size and integration with the coupler.

To overcome this limitation, silica wedge disk resonators were developed [278]. Inspired by the original microdisk resonator [138], the wedge resonator increased the silica thickness and leveraged a chemical polishing method to obtain Q factors over 875 million. Notably, the lithographically defined cavity diameters can be extremely large (millimeters), which is a requirement for achieving very small FSRs and controlling dispersion. While initial measurements were performed with a packaged taper coupler system, more recently, silicon nitride bus waveguides have been integrated on the chip to couple light to the disk resonators, creating a more packaged platform.

The ability to package the silica devices is a critical feature. Unlike the fluoride cavities which have a very stable surface, silica forms a layer of hydroxyl groups which have high optical loss in the near-IR. [36,104] Therefore, in order to maintain ultra-high-Q factors in silica devices for long periods of time in ambient environments, it is necessary to protect the surface from the air. Given this requirement, it is clear why alternative material systems that combined nanofabrication compatibility with environmental stability became of interest. These requirements inspired the initial silicon nitride platform development and the more recent work in diamond as well as work investigating high index doped oxides.

### *4.2.3 Doped-Silicon Oxides*

The two primary doped oxides that have been studied are silicon oxynitride ($SiO_xN_y$) and silicon oxycarbide ($SiO_xC_y$). In both materials, the dopant serves to increase the refractive index. These materials are typically deposited using chemical vapor deposition (CVD) based techniques, allowing for uniform, high-purity films to be created. Additional metal-doped oxide films, such as Zr-doped or Ti-doped, have also been demonstrated using solution-based processing and deposition. However, due to limitations associated with film cracking as well as impurities in reactants, these films are of lower quality. As a result, devices fabricated solely from metal-doped films have not demonstrated comb generation.

In both oxynitride and oxycarbide, the index can be precisely tuned by controlling the relative ratios of the three constituent elements. This flexibility allows higher index materials to be deposited and provides a method to control the zero point dispersion of a device, which is a key aspect of frequency comb design. [288] Importantly, because the material deposition is compatible with PECVD, extremely optically pure materials can be deposited over large areas, further improving device innovation.

Silicon oxycarbide is more commonly known as Hydex. This material was originally commercialized by Little Optics in 2000 and subsequently acquired by Infinera. As a result of the commercialization and incorporation into product lines, the relative composition of oxycarbides has been fairly stable. Due to the inclusion of carbon in the silicon dioxide, the refractive index of Hydex can reach 1.7, which is approximately 17% higher than the refractive index of silica. However, it maintains its reactivity to common CMOS processing methods. For example, using reactive ion etching, waveguide sidewalls can achieve exceptionally low roughness, enabling resonator quality factors to exceed 1 million.

This combination of properties enables Hydex-based components to be directly integrated on chip with a fused silica cladding instead of being suspended in air. Initial applications in the early 2000's focused on integrated waveguides, on-off chip couplers, and optical circuit elements. More recently, leveraging efficient four wave mixing and parametric oscillations, frequency comb generation and other nonlinear behaviors have been observed from the Hydex ring cavities. [334,347,348]

In the case of SiO$_x$N$_y$ devices, the film is deposited in a PECVD reactor using silane, ammonia, nitrogen, and nitrous oxide. By changing the ratio of oxygen and nitrogen in the deposition process, the refractive index of silicon oxynitride can be modified as well. This tunability in refractive index gives silicon oxynitride another degree of freedom for dispersion engineering, which is critical for the frequency comb generation. Also, due to the introduction of nitrogen to the material, the density of the hydroxyl groups on the surface of silicon oxynitride is decreased or eliminated in comparison to silica, which has dense surface hydroxyl group coverage and poor environmental stability as a result [104].

Using silicon oxynitride, microtoroidal resonators have been fabricated. Comparing to silica devices, due to the lack of hydroxyl groups on the material's surface, the performance of the silicon oxynitride toroids were stable even when the devices were stored and tested in an ambient environment. The quality factors of the silicon oxynitride toroids remained above $10^8$, which was comparable to the quality factors of the silica microtoroids.

Kerr frequency combs have also been demonstrated from an ultra-high quality factor silicon oxynitride microtoroid with a quality factor of 1.3 x $10^8$. Dispersion of the toroid resonators were tuned by changing the geometry of the cavities, and frequency combs with a span of more than 350 nm were observed from the silicon oxynitride microtoroidal cavity (Fig. 23). Additionally, the environmental stability of the generated combs was demonstrated by characterizing the comb span on the first day and the ninth day after fabrication without any treatment to the devices. The performance of the combs remained constant after the nine-day timeframe. [288] As an emerging material being used for whispering gallery mode microcavities and frequency comb generation, silicon oxynitride demonstrates the potential for high quality comb generation, and the tailorability of the index provides additional flexibility for cavity design and fabrication.

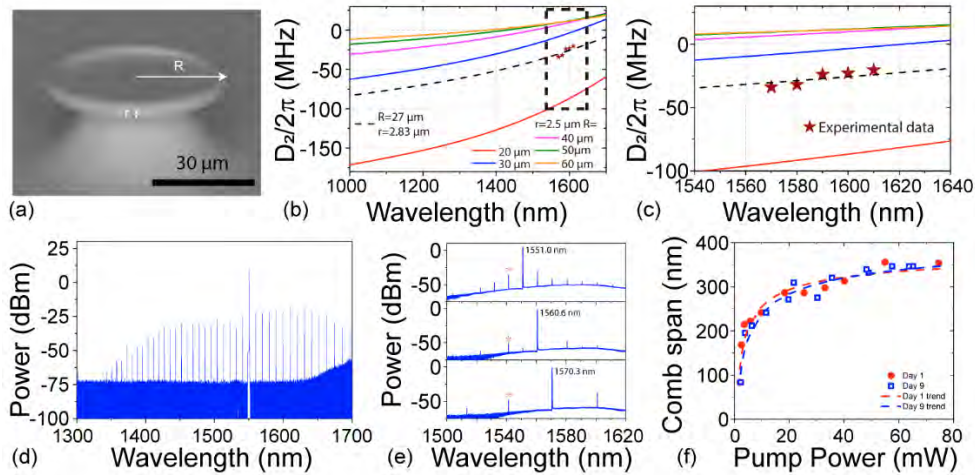

Fig. 23. Silicon oxynitride toroidal resonant cavity. (a) SEM image of a silicon oxynitride resonant cavity with major (R) and minor (r) radii indicated. (b) Calculated dispersion as a function of wavelength for a series of silicon oxynitride devices. The dashed black line corresponds to the precise geometry that was lithographically fabricated. (c) A zoom in of the indicated region in part (b). The stars are experimental measurements. The results align well with the theoretical predictions. (d) An example frequency comb generated by a silicon oxynitride cavity with 60 mW of input power. The total comb span is approximately 350 nm. (e) Comb generation at three different pump wavelengths with the initial sideband fixed at the location indicated by the red star. (f) An analysis of the comb span as a function of pump power immediately after device fabrication and after being stored for nine days. Within the error of the measurement, there is minimal change. Reprinted with permission from [288]. Copyright 2019 by the American Physical Society

*4.2.4 Silicon Nitride*

As can be seen in **Table 1**, silicon nitride ($Si_3N_4$) has several advantages from a fundamental materials perspective over both silicon dioxide and fluoride material systems. This difference enables a wide range of integrated photonics device designs.

The refractive index is approximately 2, which is nearly 40% higher than the other two materials, and it is also larger than the doped oxides (**Table 1**). This increase allows for completely different device geometries and architectures to be designed. Unlike either the silica or the fluoride devices which must be air clad in order to achieve optical mode confinement, the $Si_3N_4$ devices can be located directly on the substrate. [106,114,125,349] This approach enables a significant increase in device density. It also allows for direct integration with on-chip waveguides. Therefore, even before the resonant cavity field had been catalyzed by interest in frequency combs, significant research efforts had developed robust methods for fabricating these types of integrated systems.

In addition, silicon nitride has extremely favorable optical properties for comb generation over a large wavelength range. It has a wide transparency window from the visible to mid-infrared [350]. Additionally, due to the increase in refractive index, the optical mode area is smaller and more tightly confined within the device. Therefore, the dispersion of the cavity can be engineered by changing the geometrical parameters of the waveguide cross section [351] [335] [352]. However, these types of designs are only possible when a device geometry can be tightly controlled. In the case of $Si_3N_4$, this precision is possible because of its compatibility with nanofabrication methods.

Specifically, from a device fabrication perspective, the material stack used to create a $Si_3N_4$ waveguide is a classic semiconductor material system. As such, there are numerous existing nanofabrication etch and system integration etch recipes that can be leveraged to not only integrate the waveguide with the resonant cavity but also to integrate additional components, like lasers and detectors, on-chip. This type of fully integrated system is critical to translate the cavity from the benchtop to real-world applications.

Given the interest in fabricating $Si_3N_4$ devices for combs and other applications, several different approaches for depositing $Si_3N_4$ films have been developed. The two most commonly used were based on plasma enhance chemical vapor deposition (PECVD) or low-pressure chemical vapor deposition (LPCVD). These methods were preferred as they produced the lowest optical loss films. However, the stress in the film was also high, resulting in cracks in films that were thicker than around 250 nm [106]. However, the limited thickness increases the overlap of the optical field with the top surface which was rough. This interaction increases the scattering loss, decreasing the quality factor. Therefore, before $Si_3N_4$ devices could be used in comb generation, researchers had to develop strategies to mitigate stress-induced defects or to reduce surface scattering. Two different approaches were developed in parallel to address this challenge.

Thicker waveguides can be fabricated by introducing trenches on the wafer before the deposition of $Si_3N_4$, which prevent the propagation of the crack and create crack free regions on the wafer, where the waveguides can be patterned [353] [354] [130] . Another way for fabricating thick $Si_3N_4$ waveguides is first carefully defining the trenches on the silica substrate, which has the same pattern and size as the waveguides. Then a layer of $Si_3N_4$ is deposited on the whole wafer, as well as in the trenches. In this process, the predefined trenches not only prevent the crack from propagating, but also function as the mold of the waveguides [355] [356]. To further reduce the scattering loss and increase quality factor, several different strategies have been employed including optimizing the reactive ion etching (RIE) recipe, performing a reflow step to improve sidewall roughness, and using chemical mechanical polishing (CMP) to reduce top surface roughness. With all these methods, quality factors as high as 37 million for a $Si_3N_4$ ring with a 2.5 μm width have been achieved. While these Q factors are at least one order of magnitude smaller than the silica and fluoride cavities, the integration ability and dispersion engineering capability compensate for the slightly lower Q factor.

In addition, because the devices are fabricated lithographically, unconventional resonator structures, like concentric or tapered microresonators where the modes have asymmetric or nonuniform profiles, can be fabricated. The dispersion in these unique devices is modified as a result of the mode interaction, enabling improved comb performance. An example is such a device system is shown in Fig. 24 [357] . However, achieving $Si_3N_4$ frequency combs in the visible is a remaining challenge because of the difficulty in achieving anomalous dispersion in this wavelength range.

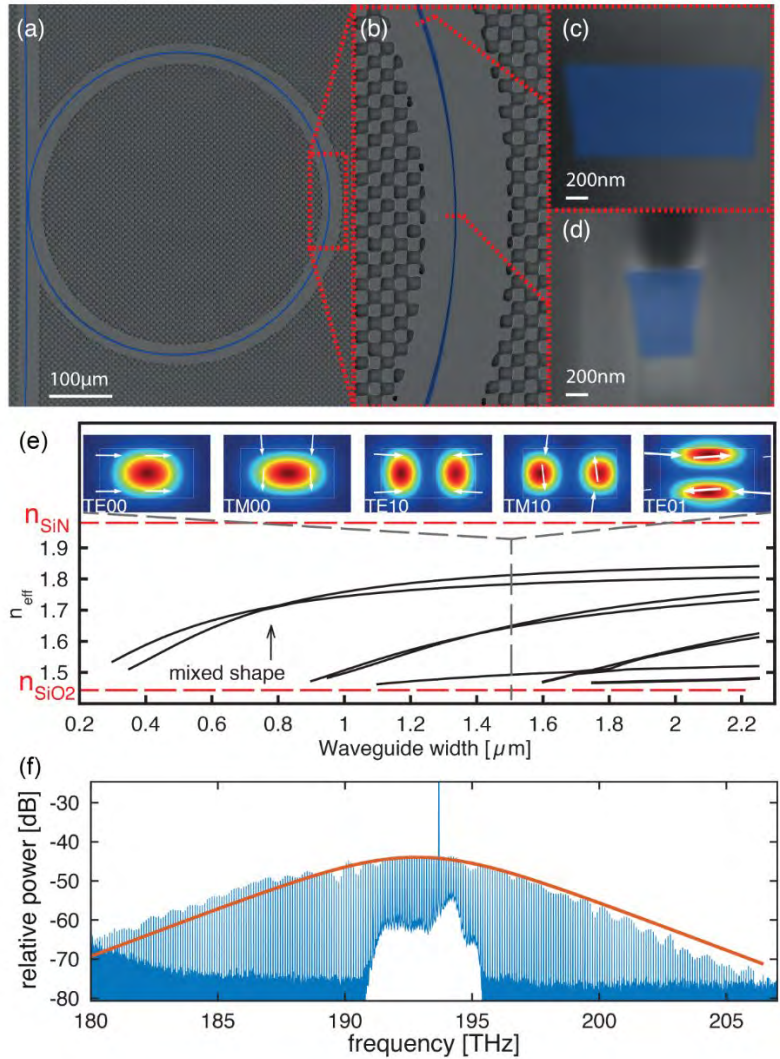

Fig. 24. Engineering dispersion in $Si_3N_4$ microring through geometry. (a) SEM image of entire $Si_3N_4$ device. A key feature includes (b) adiabatic tapered section which enables mode filtering. (c)/(d) Cross section of the ring and the tapered section. (e) Effective index as a function of waveguide width for several optical modes at 1550 nm. (f) Frequency comb with single dissipative Kerr soliton is generated by the engineered $Si_3N_4$ ring. Adapted from [289].

## 4.2.5 Diamond

Diamond is an emerging material system in nonlinear integrated photonics [122]. With high linear and nonlinear refractive indices and low absorption losses with a large wavelength window (from UV to far-IR region), it has many similar properties to $Si_3N_4$ (**Table 1**). Additionally, diamond combines the advantages of a high thermal conductivity and a low thermo-optic coefficient, which results in a reduced temperature sensitivity to high input powers. However, unlike many of the materials discussed so far, diamond has only recently been used in integrated photonics. Therefore, most investigations into diamond photonics have focused on developing the toolbox of lithography and fabrication methods needed to make resonant devices. The initial microring devices achieved Q factors in the 1 million range, which was moderate when compared with the other dielectric devices discussed thus far. [102,107,358] However, given diamond's higher linear and nonlinear refractive indices, the potential for generating frequency combs was clearly evident.

One recent landmark demonstration leveraged optical parametric oscillation (OPO) for frequency combs in the visible to near-IR range based on a fully integrated, monolithic, single crystal diamond microresonator [122]. Using a wedged microring with an optimized structure, the dispersion of the diamond resonator was engineered to achieve anomalous dispersion (GVD > 0) over a wide bandwidth, spanning 850 ~ 2,350 nm, according to numerical modeling calculations. Notably, the high refractive index of diamond allowed it to have anomalous dispersion even at visible wavelengths. The large anomalous dispersion indicated that the device's performance was only limited by the optical pump power propagating inside the resonator and the resonator's optical losses, overcoming the dispersion limitations of other dielectric materials.

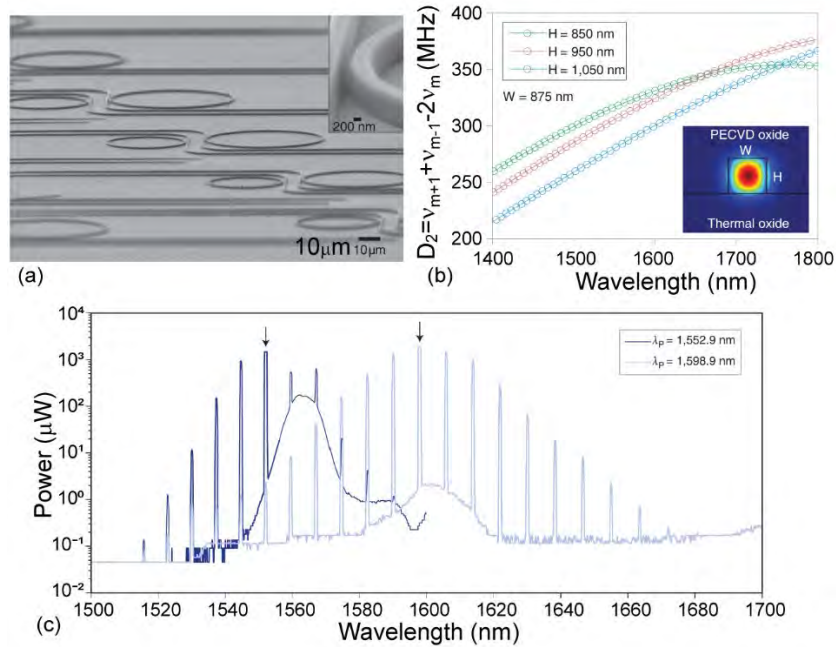

Fig. 25. Diamond cavity frequency combs. (a) SEM image of an array of on-chip diamond ring resonators. Inset: SEM image of the waveguide-cavity coupling region. (b) By changing the ring height, the total cavity dispersion can be engineered to achieve anomalous dispersion at a pre-determined wavelength. Inset: Optical mode profile in the diamond resonator (clad by silica). (c) Optical parametric oscillation spectra for two different pump wavelengths using the same cavity. The pump power is constant (80 mW). The change in the total comb is related to the change in dispersion as well as the change in coupling efficiency. Reprinted by permission from Nature [128]. (c) 2014

In the initial experimental demonstration, the loaded Q-factor of the device was as high as $Q_L \approx 1.14 \times 10^6$, and the authors inferred an intrinsic Q-factor of $1.35 \times 10^6$ with a waveguide propagation loss of 0.34 dB cm. [128] Due to the high Q-factor, the diamond device has an OPO threshold of only ~20 mW in the waveguide with a total power conversion efficiency of up to 5%. With the combination of diamond's unique properties, the device generates 20 lines, spanning a range of 165 nm using a 1598 nm pump wavelength with ~80 mW of pump power (Fig. 25). This demonstration marked the first fully integrated cavity-enhanced comb that spanned from the UV to the near-IR. Given the development stage of diamond, it is clearly an exciting material system with a long future ahead.

*4.2.6   Applications of dielectric materials*

As a result of the significant energy that has been invested into improving the performance of the dielectric devices, the requirement for input power has been dramatically decreased. Coupled with improved on/off chip coupling methods, the potential for fully integrated structures, including the pump laser source, as well as applications of combs have begun to emerge. A recent study demonstrated a battery-powered, integrated comb source. [359] This light-weight system marked a turning point in comb development. In addition, numerous demonstrations using larger footprint comb systems have been explored including dual comb spectroscopy [6,120,360], high speed telecommunications [361–365], ultrafast distance ranging [366,367], optical clocks [16,221,368,369], frequency synthesis [141], and microwave generation and pulse synthesis. [370–374] However, all of these applications are not possible using the exact same device platform, further underscoring the importance of having access to a material toolbox and not being limited to a single material system during fabrication.

Thus far, among the integrated dielectric devices, the silicon nitride rings and silica disks have been the most successful device platforms. One example is using the on-chip frequency comb source for dual comb spectroscopy. [6,360] Dual comb spectroscopy takes advantage of the high frequency resolution and accuracy as well as the broad bandwidth characteristics of the frequency combs, which allows the spectral response to be measured rapidly on a comb tooth-by-tooth basis. To further simply the system, two counter-propagating solitons that simultaneously orbit in one single resonator has been proposed for dual-comb spectroscopy. [375] However, thus far, this work has focused on spectroscopic detection in the near-IR. Spectroscopic measurements are typically performed in the mid-IR and far-IR, where the vibrational signatures of molecules are the strongest. Therefore, until materials compatible with integration and that operate in the mid-IR and far-IR are developed, being limited to the near-IR will be a fundamental limitation of using dielectric cavities in molecular spectroscopy.

Another important and promising application demonstrated from these integrated dielectric devices is wavelength-division multiplexing based telecommunications. Soliton frequency combs provide many stabilized single frequency sources to encode data. More than 50 terabits per second transmission speed was demonstrated on 179 individual comb lines in C and L telecommunication bands using this type of compact platform. [363] The development of a portable, real-time encoding system could be transformative for a wide range of application and situations.

Compact optical clocks are another important application for the microresonator based frequency combs. Comparing to the current atomic clock systems, a resonant-cavity optical clock can increase the accuracy of time-keeping by orders of magnitude, which can greatly benefit GPS systems, having wide-range societal impact. In these systems, the combs play two distinctly different roles, and two different types of combs are used. The first comb serves as the gears of the optical clock, linking the high oscillation frequencies of the atoms to the microwave frequencies that can be directly counted by the electronic systems. Meanwhile, a second comb is linked to a reference source which can count the offset of the combs. Theoretically, a single octave self-referenced frequency comb could be used, enabling a compact system with high accuracy, but so far, there are no integrated, octave frequency combs with electronically detectable FSRs. Instead, other approaches, such as locking the combs to rubidium resonances [16] and self-referencing with additional lasers [221], have been demonstrated.

Recently, using this strategy, a highly integrated optical atomic clock with a stability of 4.4 x $10^{-12}$ /$\tau^{1/2}$ was demonstrated. [368] In this work, the clock laser was referenced to rubidium, and two interlocked combs were generated. One comb was a $Si_3N_4$ comb with octave spanning for *f-2f* self-referencing but 1 THz repetition rate, and the other comb was a silica comb with narrow bandwidth but 22 GHz repetition rate for fine measurement of the repetition rate of the $Si_3N_4$ comb (Fig. 25).

Besides the applications reviewed above, many other applications are being explored. Meanwhile, in parallel, numerous device-centric research is still being pursued, focusing on topics like increasing the quality factors, broadening the span of the combs, and decreasing the power consumption of the combs, as well as further integrating the system, like combining the laser source and other photonics and electronic components on-chip.

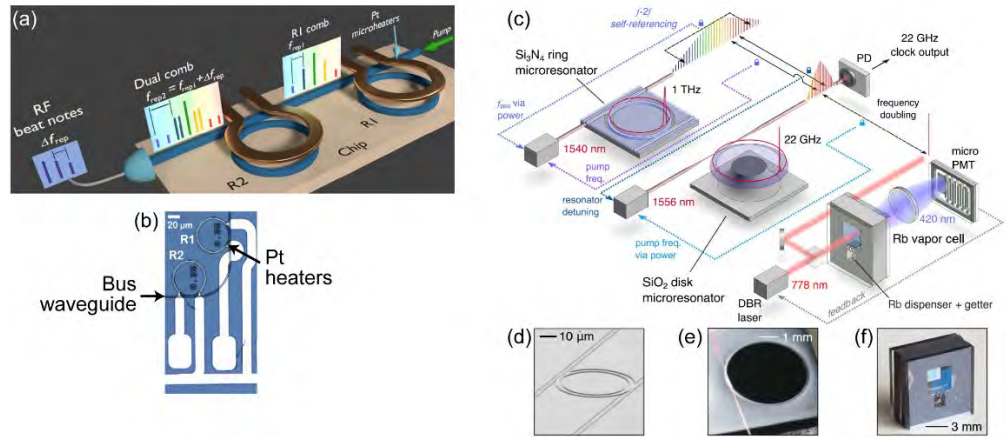

Fig. 26. Example applications of dielectric microcavity frequency combs. (a) Schematic of a dual $Si_3N_4$ ring frequency comb spectroscopy system. The pair of combs are beat against each other to improve sensitivity over a single comb detection approach. (b) Optical image of the dual comb system described in part (a). Reprinted/modified from [360]. © The Authors, some rights reserved; exclusive licensee American Association for the Advancement of Science. Distributed under a Creative Commons Attribution NonCommercial License 4.0 (CC BY-NC) http://creativecommons.org/licenses/by-nc/4.0/ (c) Schematic of an optical atomic clock comprised of three modules: (d) $Si_3N_4$ ring frequency comb, (e) silica disk frequency comb, and (f) Rb vapor cell. Adapted from [368].

*4.3 Conducting material systems*

Not surprisingly, the first microcavities with integrated waveguides were based on a silicon on insulator (SOI) platform. However, the initial Si microring devices had moderate Q factors that were limited by surface scattering, due to limitations with e-beam lithography and etching roughness at the time. [135,376] With advances in nanofabrication technologies, the Q factors and device complexity increased, and recent Si devices have obtained Q factors in excess of 1 million. [37,103,137,175,185,377,378] These intrinsic Q's are now primarily limited by carrier losses which are not present in dielectric cavities. Therefore, while these Qs are several orders of magnitude lower than fluoride devices, they are comparable to diamond with the advantage of compatibility with fabrication in foundries.

Additionally, Si and other semiconducting materials have linear and nonlinear refractive index values that are significantly higher than dielectric materials (**Table 2**). As discussed previously, this increase provides several critical advantages when designing a cavity-enhanced comb system, including the ability to achieve a high circulating intensity, providing a path for comb generation.

Table 2. Summary of nonlinear coefficients of conducting materials and device behavior*.

| Material | Major Diameter (μm) | Width x Height (μm) | Core Index | Cladding Index | Q (x$10^6$) | $n_2$ (x$10^{-20}$ m$^2$/W) | $P_{circ}$ (W) |
|---|---|---|---|---|---|---|---|
| Silicon [379] | 200 | 1.400 x 0.500 | 3.47 | 1.45 | 0.59 | ~100 at 2.5mm | 0.072 |
| Lithium Niobate [10] | 160 | 1.300 x 0.600 | 2.21 | 1; 1.45 on substrate | 1.1 | 18 | 0.280 |
| AlN [380] | 120 | 3.500 x 0.650 | 2.12 | 1.45 | 0.8 | 23 | 0.274 |
| GaP [381] | 100 | 0.500 x 0.300 | 3.05 | 1.45 | 0.3 | 600 | 0.107 |
| AlGaAs [382] | 810** | 0.320 x 0.620 | 3.33 | 1.45 | 0.2 | 2700 | 0.023 |

*at 1550 nm unless indicated. All cavities are operating at critical coupling at 1550nm with 1mW input power. All values and geometries are based on literature values, as indicated. **circumference, not diameter.

Given the advantages in terms of index values, nonlinear coefficients, and ease of integration, interest in developing cavity-enhanced combs based on semiconducting materials initially started with silicon. To compensate for the reduced Q values, the work quickly expanded to include other optical materials with higher optical nonlinearities, such as AlN and GaP. [380,382,383] Leveraging simulation methods originally developed for silica and silicon devices, optical mode modeling was rapidly applied to these devices to optimize their performance in comb generation, greatly accelerating their impact on the field. For example, as show in Fig. 27, predictive modeling of AlN devices can aid in fabrication design, and the dispersion is able to align with the telecommunications window [380].

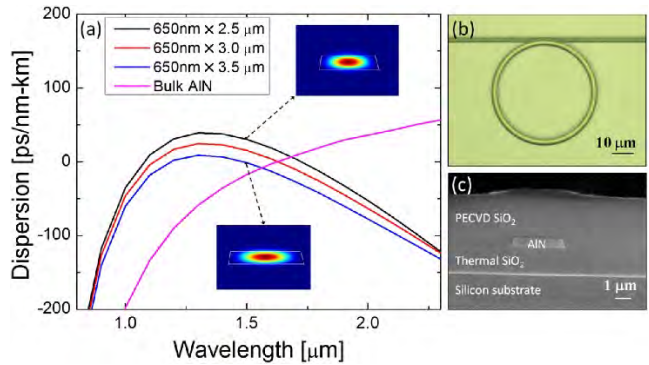

Fig. 27. Modeling to optimize dispersion in an AlN cavity. (a) The dispersion of device can be very different from the dispersion of the material, and even small changes in waveguide width can shift the zero-dispersion point. (b)/(c) Optical and SEM imaging of the proposed AlN device. Adapted from [380].

Additionally, as summarized in **Table 2**, the optical transparency windows of these materials are significantly different from dielectric materials, enabling unique applications from the dielectric cavities. The primary geometry investigated has been a waveguide integrated microring-based device. In some respects, this approach has enabled faster device development path and translation, as the basic device design and simulations for a bus waveguide coupled to cavity system were already created. However, it has also overlooked some important questions related to performance optimization, particularly considering recent work using silica on-chip wedge structures with $Si_3N_4$ buried waveguides.

### 4.3.1 Silicon

As mentioned, integrated silicon microrings were one of the first on-chip optical components, leading to the establishment of silicon photonics foundries. This positioning gave this platform a significant head start. As such, it is not surprising that silicon devices were one of the first on-chip frequency comb generators.

Due to its wide-transparency window from 1.2 μm to 8 μm coupled with its large third-order optical nonlinearity ($n_2$ = ~$10^{-18}$ $m^2$/W at 2.5 μm wavelength), silicon offers the opportunity for creating near-IR and mid-IR combs. In past work, an integrated silicon microring device with an intrinsic Q-factor of 5.9 x $10^5$ at 2.6 μm generated an optical frequency comb in the mid-IR region (2.1 to 3.5 μm) with 150 mW of power coupled into the waveguide [379]. To achieve this comb, it was necessary to engineer the device to exhibit anomalous group velocity dispersion in the mid-IR region [384,385]. The generated frequency comb exhibited a line spacing of 127 ± 2 GHz. The threshold of four-wave mixing was 3.1 ± 0.6 mW coupled power, which is in good agreement of a predicted threshold on the order of 10 mW.

The significant breakthroughs achieved with silicon photonic devices motivated researchers to explore other possible material systems that balanced optical loss and index, second and third order nonlinear coefficients, and thermal conductivity. Before expanding into entirely new material combinations, the first systems studied were based on optical materials that had previously been used in other integrated device platforms, such as lithium niobate.

### 4.3.2 Lithium Niobate

For a conducting material, $LiNbO_3$ has a moderate third-order nonlinear index (1.8x$10^{-19}$ $m^2$ $W^{-1}$) and $\chi^{(2)}$ ($r_{33}$ = 3×$10^{-11}$ m $V^{-1}$). $LiNbO_3$ microring resonators have previously been used to fabricate high performance and extremely stable optical modulators that were used in a wide range of technologies. [164] In the context of a frequency comb, the second-order nonlinear electro-optic response made it possible to manipulate the generated frequency combs by an external electric field. In parallel, $LiNbO_3$ also has an extremely large optical transparency window (0.35-5 μm). This low optical loss opened the door for generating frequency combs over a wide wavelength range, assuming challenges related to dispersion are mitigated. However, due to limitations in fabrication, integrated high-Q $LiNbO_3$ whispering gallery mode cavities were only possible in the past few years. Therefore, while this material and device performance combination is particularly interesting, it has only recently been demonstrated.

In recent work, an EO modulated frequency comb from $LiNbO_3$ was demonstrated ( Fig. 28). This work, which combines a pair of resonant cavities, was the culmination of a body of research in device fabrication, on/off chip coupling design, and systems integration. [10,140,386,387] The first ring was used to generate the optical comb. The dispersion of the ring was engineered by both changing the ring geometry and the cladding layer. With air cladding, the $LiNbO_3$ rings achieved anomalous dispersion and an intrinsic quality factor up to 1.1 million. With the engineered dispersion, frequency combs with a line spacing of ~250 GHz and span of ~700 nm were generated with a pump power of ~300 mW. The second ring was used to create an electrically programmable add-drop filter. This device selected a single comb line out of the spectrum, and the intensity of the selected line was further modulated through a high speed modulator via the $\chi^{(2)}$ effect. The whole process was performed on a monolithic integrated chip, taking advantage of the loss low and high second and third order nonlinearities of lithium niobate.

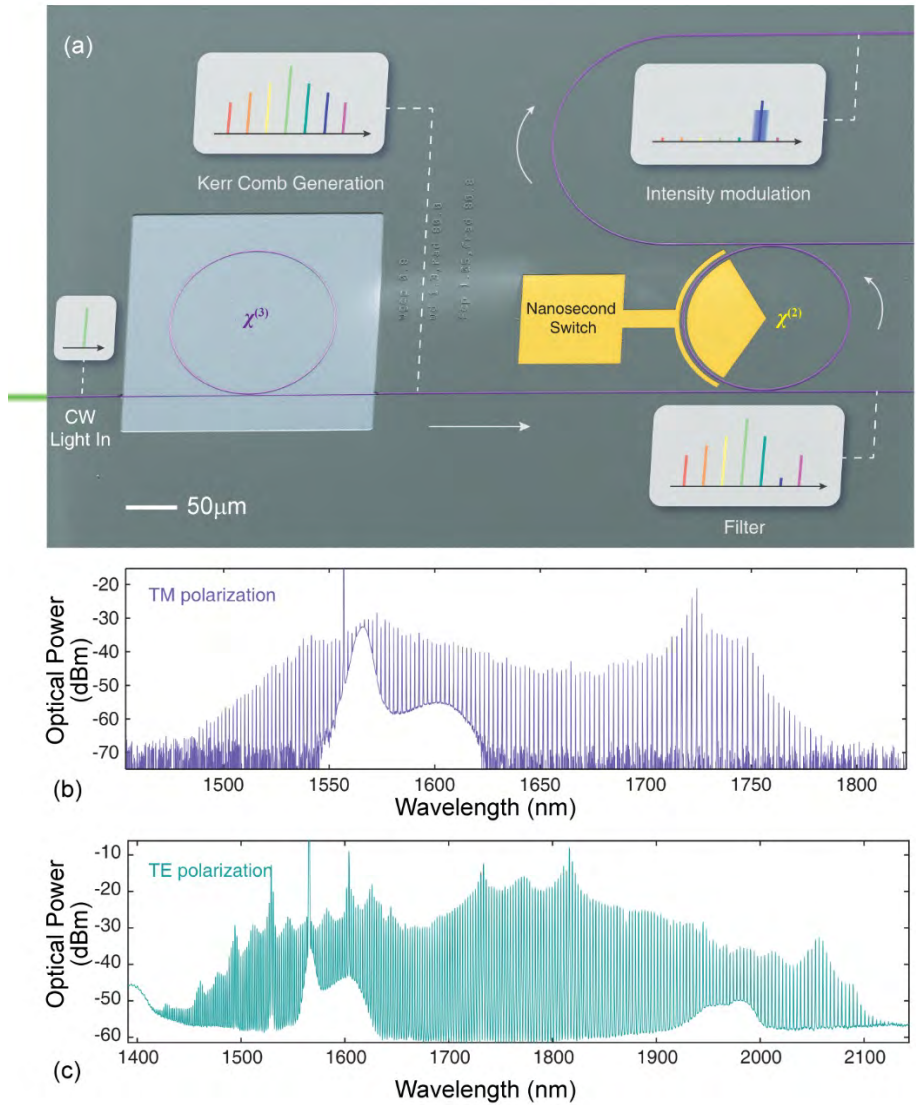

Fig. 28. Lithium niobate frequency comb with on-chip modulation. (a) A false-color SEM image highlighting the different resonant elements of the optical circuit: the $\chi^{(3)}$ frequency comb generator and an electro-optically ($\chi^{(2)}$) tunable add-drop filter. The comb is air-clad to achieve anomalous dispersion. (b)/(c) Broadband frequency comb generation when the input laser is tuned to a resonance with either TM or TE modes. Pump power of ~300 mW. Span is approximately 300 nm for TM and 700 nm for TE modes. Reprinted by permission from Nature [10]. (c) 2019

### 4.3.3 Aluminum Nitride

A similar material to $LiNbO_3$ is Aluminum Nitride (AlN). This noncentrosymmetric crystalline material exhibits both second- ($\chi^{(2)}$) and third-order ($\chi^{(3)}$) nonlinearities of similar magnitude to $LiNbO_3$ [388]. In addition, AlN has a high thermal conductivity (285 W/m·K) which improves the heat dissipation, enabling AlN devices to handle large optical loads. However, the optical transparency of AlN is not as broad as $LiNbO_3$. Therefore, the potential applications are primarily in the near-IR range. For example, in one demonstration, the dispersion of an AlN microring resonator is engineered to achieve a zero group-velocity dispersion in the near-IR region (~1550 nm). With the engineered dispersion, an AlN microring cavity with a Q of $8\times10^5$ generated a frequency comb that spanned approximately 200 nm (1450 ~ 1650 nm) with ~70 comb lines with 370 GHz spacing [380,389].

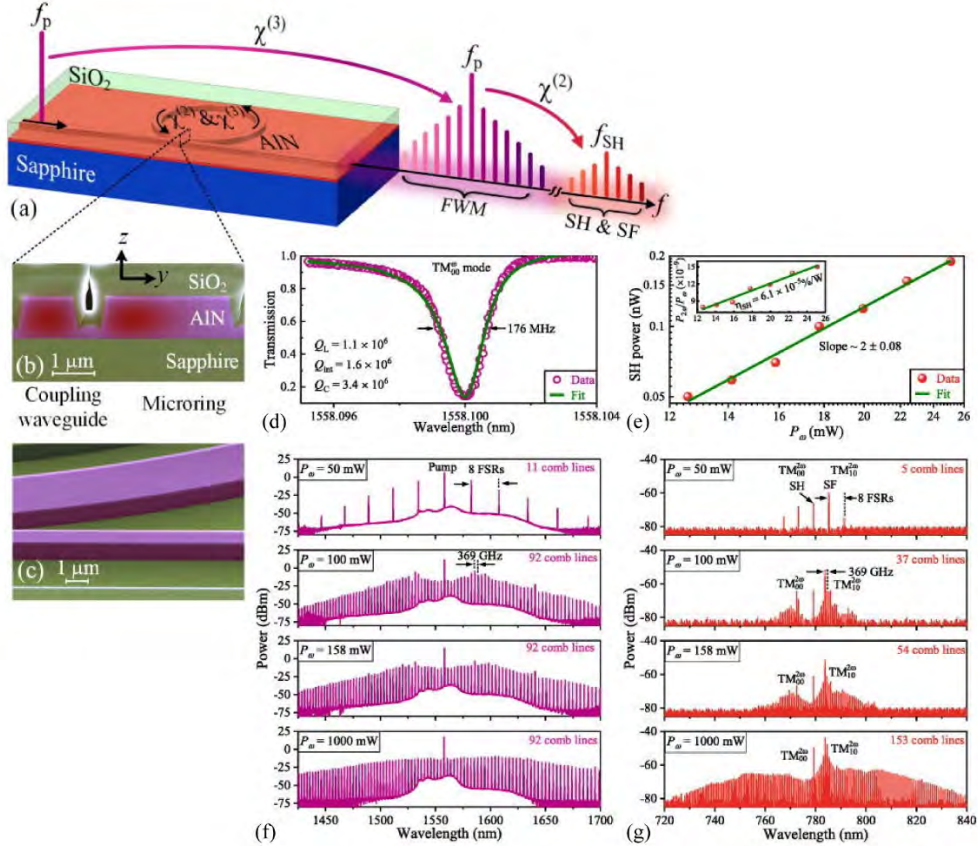

Fig. 29. AlN microring resonant cavity generating combs via multiple nonlinear behaviors. (a) Schematic of microcomb generation through four wave mixing, second harmonic (SH) generation, and sum frequency (SF) conversion. (b)/(c) False-colored SEM images of the AlN microcavity showing the cross-section of the coupling region and the device without the cladding layer. (d) Optical Q of AlN cavity. (e) Second harmonic power generated at a given input power. (f)/(g) Generation of combs via second harmonic conversion and sum frequency conversion at a series of different input powers. Reprinted with permission from [390]. Copyright 2018 by the American Physical Society.

An alternative approach for generating a broad-spanning comb is to combine the lines generated by second-harmonic and sum-frequency conversion with those generated by the Kerr nonlinearity. In this approach, both $\chi^{(2)}$ and $\chi^{(3)}$ nonlinearities contribute to and extend the comb generation. This complex approach was demonstrated using an AlN microring in the near-IR (Fig. 29). Using a 1550 nm pump, the AlN microring resonator generates comb lines in the range of 1400 ~ 1700 nm with single FSR spacing of ~369 GHz with input pump power of 1000 mW based on four wave mixing. These lines are complemented by emissions from 720-840 nm with spacing of ~369 GHz that are generated via second harmonic and sum frequency [389].

*4.3.4   High index materials*

While second and third order nonlinearities are clearly important, an alternative path to improving a cavity-based comb is to increase the refractive index, thereby increasing the circulating power. This design strategy motivated the use of high *n* (*n* > 3) materials, including gallium phosphide [383] and aluminum gallium arsenide [382]. One clear limit of this approach is its negative effect on the optical absorption. However, it does increase the circulating optical intensity by reducing the optical mode area, thus partially compensating for the Q degradation.

Gallium phosphide's (GaP) lattice is well-matched to silicon, making it ideal for deposition and integration with existing silicon-based components, and it is transparent from 0.55-11μm. In addition to a high refractive index, it also has relatively high second-order and third-order nonlinearities ($n_2 = 6\times10^{-18}$ m$^2$ W$^{-1}$). When combined, these characteristics enable tight confinement of the optical mode and efficient frequency comb generation and manipulation across a very wide optical spectrum range. However, challenges related to dispersion remain. In current devices, this issue is partially addressed by varying the width of the waveguide forming the ring. These dispersion-engineered cavities have achieved Q factors of $10^5$ with parametric oscillation thresholds as low as 3 mW. Broadband frequency combs (> 100 nm) have also been demonstrated. However, given the combination of material properties, GaP photonics has a bright future ahead.

Unlike the majority of materials discussed thus far, the transparency window of aluminum gallium arsenide ($Al_xGa_{1-x}As$) is limited to the near- to mid-infrared region. In addition, the refractive index is widely tunable. Even within this tunability, $Al_xGa_{1-x}As$ offers the highest refractive index ($n \approx 3.3$) and the highest third order nonlinear refractive index ($n_2$) of 2.6 x $10^{-17}$ m$^2$/W among the materials discussed thus far [382]. Additionally, AlGaAs also has strong second-order nonlinearity ($\chi^{(2)}$) due to the non-centrosymmetric crystal structure. Given the ability to control the index as well as geometry, the group velocity dispersion can be engineered from normal dispersion to anomalous dispersion in order to achieve phase matching and efficient four wave mixing. In one example, by varying the composition (x) of materials, the material was optimized for comb generation. Specifically, by using an aluminum fraction (x) of 17%, the bandgap was 1.63 eV and the refractive index (*n*) was 3.33. The bandgap was tailored to maximize the four-wave mixing efficiency over other nonlinear optical phenomena. Based on previous theoretical work, AlGaAs devices can exhibit anomalous dispersion in the near-IR and mid-IR, depending on the waveguide geometry.

In one interesting theoretical and experimental investigation, researchers studied how changes to the microring cross section and geometry effected the comb production [382]. Due to the large index of AlGaAs, light was efficiently confined, even in sub-micron waveguides (320 × 630 nm) which have an anomalous dispersion in the near-IR. Moreover, due to the small effective mode area ($A_{eff}$) of waveguide, the effective nonlinearity (γ) of the AlGaAs waveguide was ~660 W$^{-1}$m$^{-1}$, which is orders of magnitude higher than that of a typical $Si_3N_4$ waveguide. Given this performance improvement, an 810 μm long racetrack AlGaAs microresonator generated frequency comb spanning over ~350 nm (1400 ~ 1750 nm) with 72 mW of input pump power with ~0.82 nm (98 GHz) line spacing. This performance is notable given that the intrinsic Q of the cavity was ~2.0 x $10^5$. The threshold for four-wave mixing (FWM) in these AlGaAs devices was as low as ~ 6 mW.

*4.3.5   Applications*

While the development of integrated devices based on conducting materials happened faster than with dielectric devices, these integrated systems are relative newcomers to the field of frequency combs. As a result, applications using conducting materials are rare. Additionally, the generation of soliton combs is still challenging when using these materials. However, it is not anticipated that this is a fundamental limit, and with more development and optimization time, a route should be able to be developed.

Additionally, these materials provide a key advantage over dielectrics; namely, their high second-order nonlinearity, larger transparent window, and higher refractive index will allow for tunable and smaller combs to be developed.

One good example is using a conducting microresonator for dual-comb spectroscopy. Dual-comb spectroscopy has also been demonstrated from silica and silicon nitride systems as presented previously; however, frequency combs spanning from 2600 to 4100 nm can be achieved from silicon devices benefiting from its larger transparent window as well as larger nonlinear coefficients. [391] The mid-IR region combs offers higher sensitivity and selectivity for molecular spectroscopy because the absorption strength of molecular transitions in this region is typically 10 – 1000 times higher than in the visible or near-IR region. In addition to improved detection performance, silicon-chip based frequency combs, as well as other materials such as lithium niobate, show the potential of making a monolithic fully integrated chips to create tunable molecular spectroscopy systems.

*4.4 Hybrid device approaches*

Over the past few decades, there has been a revolution in nanomaterials. Initially, researchers focused efforts on nanoparticles, such as quantum dots and plasmonic particles, but recently, efforts have moved to 2D materials, such as graphene, phosphorene, and various disulfides. [392–397] In addition, machine learning algorithms are being developed and applied to material design and discovery, greatly accelerating innovation in organic materials. However, fabricating a resonant cavity device solely from these materials is challenging due to compatibility issues with conventional techniques as well as the fundamental length scales of the materials which impacts optical confinement. Therefore, there is interest in using these materials to enhance the optical properties of existing devices.

All the microcavity structures discussed thus far have been fabricated from a single material system. As such, given that the nonlinear parameters of the materials are fixed, the primary route available to researchers for improving comb performance is geometric structuring or design, such as tapering the ring. By integrating nonlinear optical nanomaterials with these conventional platforms, researchers can tune and improve the optical nonlinearity.

A wide range of approaches for fabricating hybrid devices is theoretically possible, and an overview of a few possible methods is shown in Fig. 30. As can be seen, there is typically a base or primary resonant device whose properties are modified by the additional layers. The location and structure of the additional functional layer or layers as well as the material properties of these layers governs the optical behavior of the hybrid cavity. Given their ease of fabrication and compatibility with nanofabrication and deposition methods, silica or silicon nitride are commonly used as the base cavity.

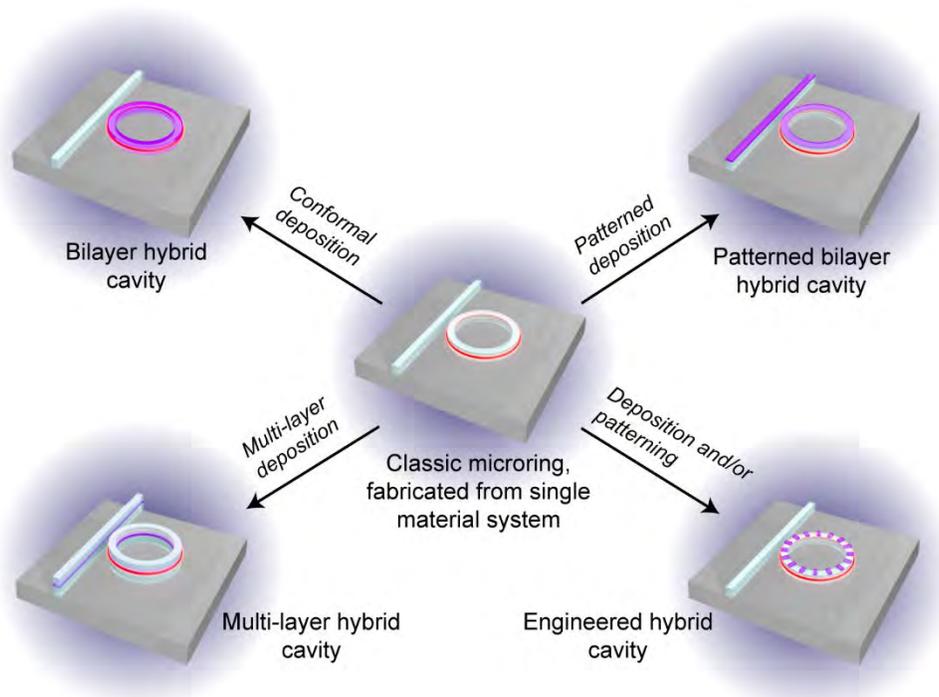

Fig. 30. Overview of different design strategies for fabricating hybrid optical devices that integrate disparate material systems.

Hybrid cavities can be divided by the type of functional layer (organic or inorganic). In the context of nonlinear optics, most of research has explored the fabrication of hybrid cavities for lasing applications, such as dye-doped polymer lasers, upconversion photon lasers, and Raman lasers to name a few. [69,151,398–403] The concept of using hybrid cavities for comb generation is an emerging area, and work in the field, particularly using organic layers, is just beginning. As a result, this field is truly in its infancy with all work to date focused on simply developing and optimizing the fabrication approaches to enable comb generation. Applications of these combs have yet to be explored. Therefore, of the methods discussed thus far, hybrid cavities have the largest potential, both for success and for failure, in enabling impactful applications and new directions for comb research.

### 4.4.1 *Inorganic optical materials*

There are two general methods for depositing thin layers of inorganic materials. The first relies on vapor deposition methods, such as atomic layer deposited (ALD) and LPCVD, and the second combines wet synthesis with spin or dip-coating. In the realm of hybrid devices for comb generation, both ALD and spin-coating have been demonstrated. The advantage of ALD over spin-coating is that high purity, crystalline layers can be deposited with nm-scale precision. However, for some materials, this precision is not required. Additionally, thicker layers which can be challenging to deposit using ALD are also desirable.

One hybrid device combined the $Si_3N_4$ microring cavity with $HfO_2$ deposited using ALD (Fig. 31). This combination was chosen based on the optical and chemical properties of both materials. [280] Specifically, as discussed previously, $Si_3N_4$ possesses a wide transparency window across the IR and relatively low refractive index. Along with its compatibility with standard CMOS fabrication processes, silicon nitride has also proven to be a good substrate for a wide range of ALD materials. [404,405]

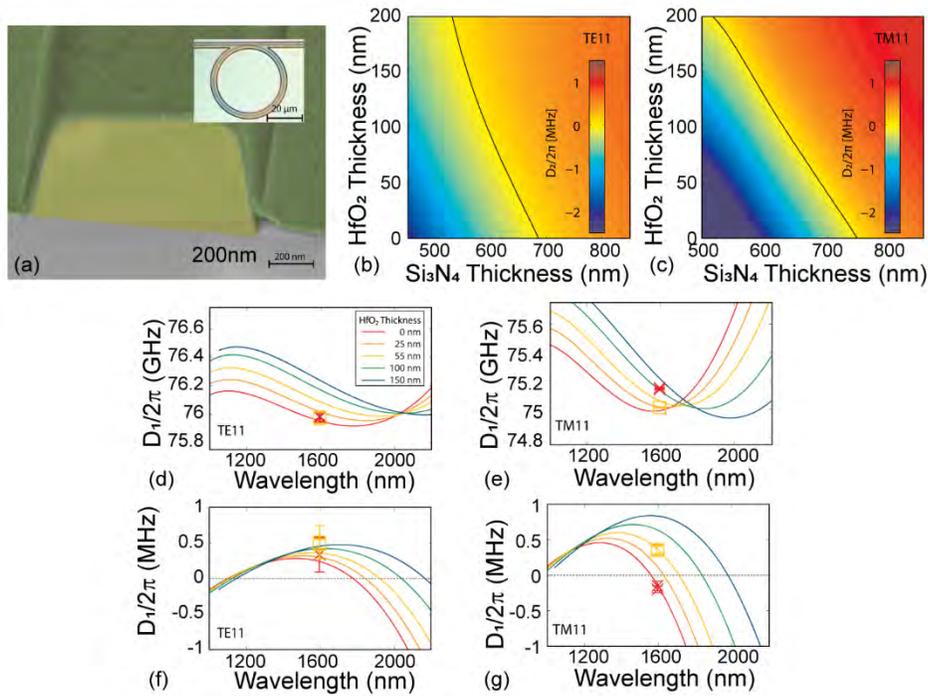

Fig. 31. Dispersion engineered $Si_3N_4$ microrings using ALD deposited $HfO_2$. (a) SEM image of device cross section. The 55 nm thick $HfO_2$ layer is clearly visible. Inset: Optical image of the device and coupling waveguide. (b)/(c) Dispersion at 1550 nm as a function of $HfO_2$ thickness for TE11 and TM11 mode. The black lines indicate the transition from normal to anomalous. (d)-(g) Simulations (lines) and measurements (symbols) of the FSR and dispersion as a function of wavelength of (d)/(f) TE11 and (e)/(g) TM11 for a series of different $HfO_2$ coating thicknesses. Adapted from [280].

The goal of the 55 nm coating was to introduce a material with large anomalous dispersion ($HfO_2$) to change the cavity dispersion of the $Si_3N_4$ which typically operates in the normal regime (Fig. 31). [273] The measured dispersion of the uncoated resonators ($D_2/2\pi$) at 1.55 µm possess normal dispersion (-180 kHz) whereas the coated devices possess a large anomalous dispersion (350 kHz). Although the dispersion could be effectively tailored, the ZDP changed only slightly ($TE_{11}$: 0.0021 nm/nm$^2$, $TM_{11}$: 0.0067 nm/nm$^2$). These overall results were validated with numerical results which showed a clear dependence of the cavity dispersion on the $HfO_2$ thickness (Fig. 31).

As mentioned, an alternative approach for fabricating a hybrid optical device is to deposit the functional layer using spin-coating. This method has been demonstrated using the silica toroidal optical cavity in combination with a metal-doped silica sol-gel thin film (Fig. 32). The silica sol-gel was synthesized using a hydrolysis condensation reaction, and the Zr was doped into the sol-gel during the reaction process. Approximately 400 nm layers were conformally deposited on the silica cavities. Due to the metal-doped silica layer, the Raman gain in these devices was significantly enhanced, resulting in higher efficiencies and lower thresholds for both Stokes and Anti-stokes emissions. As a result, like the $SrF_2$ and $BaF_2$ devices discussed in section 4.2.1, combs based on Raman-assisted FWM were demonstrated. Notably, because both Stokes and Anti-Stokes lines contributed to the Raman-assisted comb generation, combs spanning 300 nm using only 5 mW of input power were achieved with low Q values (~10 million range). Additionally, when compared to silica devices, the zero-dispersion point was shifted because of the inclusion of the Zr which modified the material dispersion.

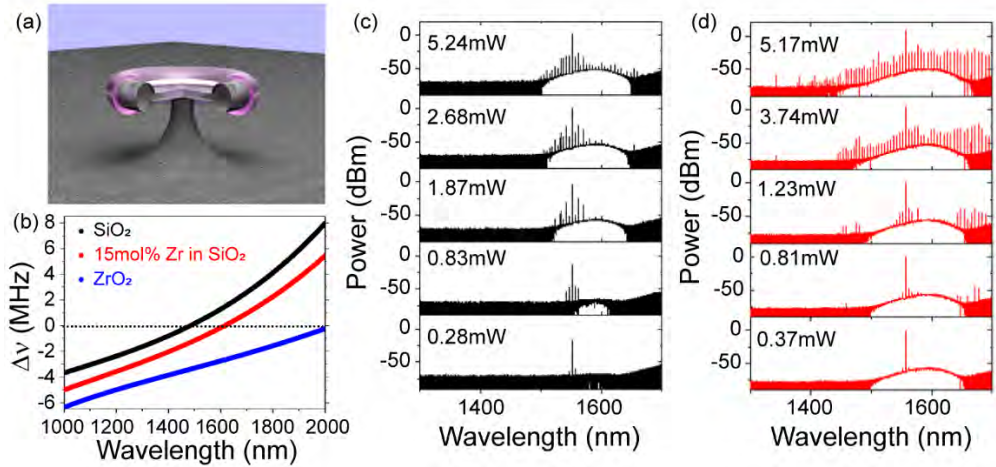

Fig. 32. Frequency comb based on silica toroidal cavity with 400 nm thick coating of 15mol% Zr-doped silica. (a) Rendering of the cavity design which consists of a silica toroidal cavity with a Zr-doped silica coating. (b) Calculated dispersion of plain silica, $ZrO_2$ and silica with Zr-doped silica coating. (c)/(d) Optical frequency comb generation from (c) silica and (d) hybrid cavities over a range of input powers. Notably, both Stokes and Anti-Stokes are clearly observed in the hybrid cavity. Adapted from [269].

Lastly, similar to the $HfO_2$ coated silicon nitride microring, one emerging platform combines a microring resonator with a 2D material with multiple nonlinearities: graphene. [173] In this system, a silicon nitride ring was partially coated with a single sheet of graphene which can be electrically modulated. The graphene was used because of its electrically tunable dispersion, as a result of its exceptional Fermi-Dirac tunability and ultrafast carrier mobility. The system demonstrated the control of the primary comb lines, the Cherenkov radiation, and soliton states by applying different gate voltages to the graphene layer. This study demonstrated the potential of using 2D materials for modifying monolithic systems to improve their performances.

### 4.4.2 Organic optical materials

While most reports on WGM resonators utilize inorganic materials to generate non-linear signals, there are a few reports of chemically modifying the surfaces to create hybrid cavities and to enhance the overall non-linear properties of the resonator. These reports include resonators with inorganic/organic hybrid films deposited on the surface, polymer coatings, and even vapor-deposited monolayers of organic molecules. The advantage of these materials lies in their high levels of optical non-linearity, especially with conjugated organic molecules. These organic molecules, based on the design, can possess second and third-order susceptibilities that are orders of magnitude higher than silica and other crystalline materials. In the past, these molecules have been used in other integrated optical devices such as all-optical switches, fast electro-optic modulators, and even two-photon absorbers. However, these materials require poling in order to align the molecules which is needed to achieve these high nonlinearities. Therefore, although reliable, their ultimate limitation in true integration is realized in the disparate fabrication methods which are largely incompatible with large-scale CMOS techniques and the ultimate lifetime of the poled state.

One initial demonstration of this concept relied on a slot waveguide device; specifically, a silicon-organic hybrid (SOH) phase modulator. [406] By coating the device with DLD164, an organic electro-optic chromophore, and then poling at elevated temperature with an applied voltage, an organic hybrid device was made. Previously, scientists determined that DLD164 is an excellent organic electro-optic chromophore due to its large EO coefficient. In bulk, the EO coefficient of DLD164 $r_{33}$ is 137 pm/V at 1310 nm, and when integrated into a device, the EO coefficient of DLD164 $r_{33,\text{ in-device}}$ is 160 or 180 pm/V at 1550 nm, depending on the slot width of the silicon waveguide. Also, DLD164 has a high second-order nonlinear optical coefficient $\chi^{(2)}_{333}$ = 1010 pm/V at 1550 nm in bulk. The SOH modulator generated a spectrally flat frequency comb of 7 lines which were within a difference of 2 dB, and the line spacing is 40 GHz. However, this device required high voltage poling, which is a fundamental limitation.

An alternative approach to poling for molecular alignment can be found in surface chemistry. Unlike spin-coating, which deposits molecules randomly on the surface, surface functionalization grafts or attaches molecules in an oriented manner to a surface. If the chemistry is correctly designed, single monolayers can be achieved, and the monolayers are intrinsically oriented without additional poling. A wide range of surface chemistries have been developed and can be leveraged in the fabrication of hybrid organic photonics, with adaption to consider the impact on the device optical performance.

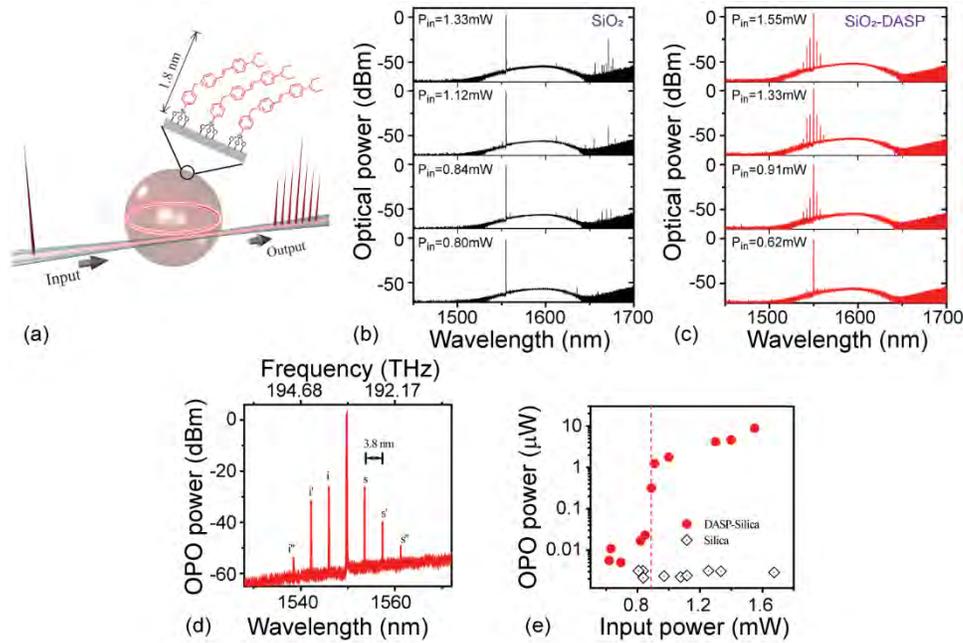

Fig. 33. Four-wave mixing enhanced by grafted organic small molecule monolayers. (a) Schematic of organic molecule (DASP) coating on silica optical microsphere cavity. The DASP molecule is anchored to the surface using surface silanization. (b)/(c) Spectra generated by silica and DASP functionalized silica cavities. The silica cavities generated Raman while the $SiO_2$-DASP cavities generated OPO due to the enhancement of the $\chi^{(3)}$. (d) Example results from an $SiO_2$-DASP cavity with the signal and idler lines identified. The line spacing is the cavity FSR. (e) Comparison of the OPO thresholds for the $SiO_2$ and $SiO_2$-DASP cavities. Reprinted from [400]. © The Authors, some rights reserved; exclusive licensee American Association for the Advancement of Science. Distributed under a Creative Commons Attribution NonCommercial License 4.0 (CC BY-NC) http://creativecommons.org/licenses/by-nc/4.0/"

For an initial demonstration, a silica microcavity was functionalized with 4-[4-diethylamino(styryl)]pyridinium (DASP), a highly nonlinear organic molecule ( Fig. 33). [400] Due to the conjugated nature of this molecule, the electrons are delocalized across the entire molecule, making it highly responsive to field perturbations. As a result, this molecule exhibits an $n_2$ of $2.54 \times 10^{-17}$ m$^2$/W, which is roughly three orders of magnitude higher than silica. The molecule was vapor-deposited on the surface and covalently attached to the hydroxylated silica surface, producing an oriented monolayer of about 1.8 nm thick. The circulating optical field evanescently decays outside of the silica sphere, interacting with the organic layer. The quality factor of the modified sphere was typically $7\times10^7$, representing about half an order of magnitude decreases in quality factor as compared to unmodified spheres. Despite the change in quality factor, significant non-linear phenomena were observed. Due to the increased non-linearity of the resonator/material system the threshold for hyperparametric oscillations was substantially decreased to about 880 µW compared to unmodified spheres. Furthermore, the addition of the organic molecule preferentially facilitated hyperparametric oscillations over stimulated Raman scattering, likely because of the increase in Kerr non-linearity. In silica spheres, depending on the size and coupling conditions, Raman scattering is typically observed first due to the large Raman gain of silica. However, with the DASP-modified spheres, and with similar geometric and coupling parameters, Kerr comb generation was preferred. This approach presents an interesting way to modify current WGM resonators to produce enhanced and coherent Kerr frequency combs. This method is also geometry independent and is only constrained to using devices with a moderately reactive surface.

In complementary work, an alternative immobilization and structuring strategy for organic molecules based on micropatterning was demonstrated using Crystal Violet (CV), an organic molecule. Silica microspheres were micropatterned, and CV was deposited on the micropatterned region by immersing the microsphere into the CV solution ($2.5\times10^{-5}$ M). [407] In this way, one can detect weak second harmonic generation (SHG) light signals on the CV-coated micropatterned surface because the central symmetry of the silica microsphere at the surface is not maintained. The CV-coated micropattern area facilitates quasi-phase matching in the nonlinear interaction modes. Owing to the high quality factor of the microsphere ($8.9\times10^6$), the number of CV molecules required is only 50-100, while most other studies need at least $4\times10^6$ to detect SHG signals. However, neither comb generation nor compatibility with an integrated device platform have been demonstrated using this approach, but it does provide an interesting potential path.

## 5 Remaining challenges and future prospects

### 5.1 Device stability

Although not discussed thus far, one limitation of ultra-high-Q cavities is their intrinsic sensitivity to environmental changes, in particular thermal effects. This challenge was originally highlighted in the initial papers on frequency comb generation in the 1980's and was re-iterated in the Nobel Prize lectures given in 2005.

Because the resonant wavelength is determined by both the refractive index and the device diameter, subtle change in the index will shift the resonant wavelength. This shift will result in changes in the comb, both the center frequency and the relative intensity of the lines. Therefore, while the line spacing is constant (because different frequencies are thermally shifted by almost the same values), other parameters will vary with environmental changes. And, if the changes are very large, the cavity can shift off-resonance, resulting in the loss of comb generation. Given that most comb applications are related to using the comb as a reference, this source of instability is not desirable. Therefore, significant efforts have been invested in developing methods to reduce the effects of thermal and electrical noise.

Initially, most methods were based on passive mutual locking of the resonator pumped mode and the laser using a thermal lock by leveraging the thermo-optic effect intrinsic to the cavity or based on actively utilizing a feedback-loop in the system. As broader combs spanning over an octave were developed, more advanced locking methods based on self-locking could be applied. These systems used the cavity itself as the feedback loop and allowed the demonstration of locked soliton combs which had long-term stability, offering the first glimpse of possible applications. More recently, to achieve even higher stability along with precision and accuracy, researchers are locking combs with each other. By using one broad (large FSR) and one narrow (small FSR) comb, the combs are extremely stable and calibrated with high precision.

*5.2 System integration and packaging*

One of the overarching challenges in the field and hurdles to translation of high-Q and ultra-high-Q resonant cavities is device-level and systems-level integration. Because the Q is very sensitive to coupling efficiency and stability, changes in coupling will greatly impact the overall device performance, including parameters like comb span. Therefore, the development of a fully integrated system, which includes the comb cavity, laser, detector, and on/off-chip interconnects is of great interest. Briefly, from a hardware perspective, the goal is a single chip with all components integrated that is battery powered. From the performance side, the goal is the generation of multi-octave combs. Analyzing these goals, to meet the performance and energy demands, it will be necessary to have Q's above 1-10 million with high nonlinearities, which will require low loss coupling and on-chip interconnects.

Currently, three different strategies based on single cavity platforms are being pursued, and they have had different levels of success, depending how success is defined Fig. 34). The strategies are based on the silica microdisk, silicon nitride microring, lithium niobate microring, and $MgF_2$ cavity. While the first three devices are fabricated on-chip, the $MgF_2$ cavity used in these systems is typically hand-polished. Additionally, all cavities meet or exceed the Q requirements, when operating in optimized conditions.

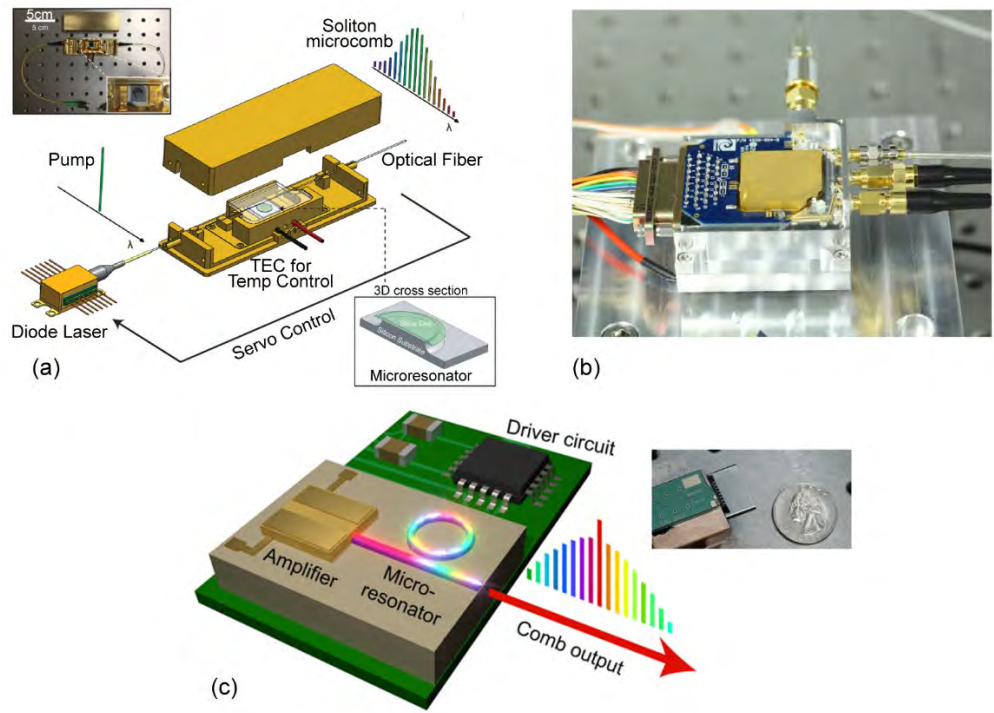

Fig. 34. Strategies for microcomb packaging and integration. (a) Modular approach for integrating a toroidal cavity with off-chip laser and detector. Adapted from [408]. (b) Packaged system with prism coupler for fluoride cavity microcombs. Image courtesy of OEWaves. (c) Battery-powered, integrated system for $Si_3N_4$ combs. Reprinted by permission from Nature [359]. © 2017

For silica cavities, two different strategies have been explored. Both methods are based on the wedge microdisk resonator and can maintain Q factors above 100 million. The first is the modular, plug-in-play approach shown in Fig. 34a, which relies on tapered optical fiber waveguide for device coupling and which connects to off-chip lasers and detectors. [408] This approach relies on using UV epoxy to mount and to stabilize the tapered optical fiber, an approach first shown using on-chip microdisks nearly a decade ago for high precision add-drop filters and devices [100]. Given its modular nature, it is easy to plug into existing systems, accelerating its use in a wide range of applications. However, compared to alternative methods, the footprint is large, and the electrical power can be high. To address these limitations, the second method is a more integrated approach which uses embedded $Si_3N_4$ waveguides and an on-chip laser and detector. [120] However, one limitation of silica devices is the requirement for hermetic sealing in order to achieve stable performance over long periods of time in non-controlled or ambient environments. This level of packaging for a comb device has not yet been demonstrated.

The packaged $MgF_2$ system was developed in parallel with the modular silica approach. Leveraging pick in place technology, this strategy combines micro-prism couplers, commercially available small-footprint lasers, and ultra-high-Q resonators to achieve a packaged system (Fig. 34b). The resonator and prism are cured and fixed in place leading to a robust device. These packaged systems have thus far been used for low-phase-noise laser and high-spectral purity RF signal generation but not for turn-key frequency comb generation.

Not surprisingly, given its compatibility with a wide range of substrates, combs based on $Si_3N_4$ have demonstrated the highest level of integration and power management, meeting all the previously defined metrics. By leveraging advances in heterogeneous fabrication and pick in place component methods, an entirely integrated optical comb photonic circuit that is battery powered has been demonstrated (Fig. 34c). Notably, the combs are generated by ultra-high-Q $Si_3N_4$ rings that maintain their Q factors in ambient environments, thus improving the lifetime stability of the devices and reducing the optical input power needed to the level that an on-chip amplifier is sufficient.

A fourth method which combines multiple cavity systems has also been demonstrated. [141] This system combines both a silicon nitride ring cavity and a silica wedge resonator with several on-chip lasers for pumping and locking. This approach requires integration of several different chips as the combs are fabricated on two different silicon dies and the other components are fabricated using a blend of III-V/Si processing. Despite this complexity, this dual-comb system has successfully demonstrated optical synthesis. [141]

*5.3 Turn-key generation of optical frequency combs*

Demonstration of mode-locked broadband Kerr microcombs, now commonly referred to as dissipative Kerr solitons (DKSs) in the literature, was a decisive milestone in the microresonator-based frequency comb generation research [211,212]. Many researchers were attracted to microcombs and their cumulative efforts paid handsome dividends through successful adoption of DKS pulse trains in significant proof of concept or even field applications such as optical communication [363,409], spectroscopy [6], exoplanet research [11,410], and optical clocks [411,412]. DKSs are coherent mode-locked frequency combs with smooth spectral envelope and small harmonic-to-harmonic power fluctuation, and repetition rates close to the host resonator FSR at the pumping frequency. Their formation in a CW-pumped microresonator occurs within a range of pump power and frequency values and is spontaneous, in the sense that it does not require a mode locking element such as a saturable absorber.

Despite these attractive and versatile properties, DKS microcomb formation is not deterministic: turning the pump laser on at the frequency and with the power corresponding to the desired DKS state will not simply lead to soliton generation [212,281]. The pump frequency must be swept over the pumped resonance at a tuning rate within a 'suitable' speed. During the laser sweep, comb generation is initiated in the modulational instability (MI) regime during which pump sidebands are created via parametric amplification of vacuum fluctuations, and the generated combs generally have sparse harmonics and are not as spectrally dense as a DKS comb. Further laser detuning carries the system through a chaotic state where the comb spectrum becomes dense, even though the comb is not stable. Finally, the desired stable solitonic states are achieved, although the ideal single soliton comb is hard to actualize. A theoretical framework explaining this suitable tuning speed and providing insight as to how it changes from one material platform, and resonator geometry and size, to another is missing and the proper range of tuning rate is found experimentally. We note that, while it is indeed possible to use steady-state approximate analytical models (see, e.g., the discussion surrounding Eq. 40) or numerical parametric sweeps based on the LLE (or, alternatively, coupled-wave equations) to find appropriate experimentally relevant parameters including the the resonator dispersion or pump power and detuning intervals for comb generation, the pump frequency tuning rate to reliably and repeatably achieve microcomb single soliton states is still an open problem in comb dynamics.

The non-deterministic nature of DKS formation is rooted in the multi-stability of the Kerr comb in the soliton formation regime [258]. From a nonlinear dynamical perspective, multiple soliton states, a single soliton state, or absence of solitons in the cavity are all possible outcomes of the laser sweep. From a practical perspective, this multi-stability means that even for the same experimental setup (i.e., same system parameters) a successive laser frequency sweep with the exact same tuning rate does not necessarily create the same solitonic state with the same number of soliton peaks and hence the same smooth power spectrum. This variability signifies a hindrance to large-scale production and industrial application adoption of DKS microcomb sources, because even if laser on-chip integration is properly addressed, laser frequency tuning with an essentially fabrication-dependent rate will entail system complexity and mass production difficulty. As the tuning problem is yet to be cracked theoretically, other approaches have been proposed and implemented to address this challenge to various levels of success.

One proposal based on chaos-avoiding trajectories in the power-detuning parameter plane has been suggested and studied theoretically [413]. The idea is to change the laser and power of the pump laser in such a way to avoid the chaotic middle stage of soliton formation. The challenge of theoretically or experimentally finding the optimal route has impeded the implementation of this proposal. On the other hand, it has been demonstrated that reverting the pump frequency tuning direction will provide a simpler route to stable single solitons [238].

An approach for deterministic soliton formation with potential for on-chip integration and obviating the need for laser sweep has been proposed [225] and experimentally demonstrated [227,414]. In this technique, phase modulation of the CW pump creates a spatially varying loss and gain profile in the cavity [275,415] and can create detuning ranges where the multi-stability is fully lifted and a single-peak DKS is the only possible microcomb state [414]. Pump phase or amplitude modulation at the comb repetition rate has additionally been shown to improve microcomb stability [225] and hence the phase noise performance of the detected RF signal [9]. Additionally, synchronous pumping, with a relatively small-bandwidth pulse, of a fiber Fabry-Perot resonators has been exploited for spontaneous excitation of DKS microcombs [233]. The driving pulse train was generated through a combination of CW pump (intensity and phase) modulation and delay compensation and offers enhanced comb nonlinear conversion efficiency. With the improvement of on-chip modulators, this technique has the potential of small-footprint integration.

It is worth noting that pump amplitude modulation for normal dispersion dark soliton generation has also been studied both theoretically and experimentally [226]. Furthermore, use of integrated micro-heaters with coupled auxiliary resonators to control avoided mode crossings for normal-dispersion Kerr microcomb generation with a fixed-frequency laser [416] and nesting the Kerr-nonlinear microresonator in a fiber loop with gain [239] have recently been reported. The latter demonstration relying on an external fiber loop required development of integrated low-loss material platforms including gain for on-chip integration. These demonstrations are promising for the turn-key generation of Kerr microcombs with improved nonlinear conversion efficiency.

Finally, we highlight that the interplay between thermal instabilities in Kerr resonators, discussed earlier in this review, and rapid intra-cavity power changes complicates DKS formation. Increased linear intra-cavity power (compared to when there is only one pumping frequency) has been shown to compensate for the power drop resulting from thermal shift of the pumped mode, thereby extending accessible range of soliton formation [417]. Compensating power coupled to the cavity could be supplied, for instance, by an accompanying pump component at the same frequency but orthogonally polarized (hence pumping a nearby mode family of different polarization, e.g., TM vs. TE) [292], or by a separate auxiliary pump [232,417].

The collection of the various approaches reviewed above, and possibly other novel techniques and combinations thereof, paints a bright horizon for the real-world applications of broad-band mode-locked microcombs and even their turn-key operation.

*5.4 Growth of novel material systems*

While initial efforts have focused on designing and fabricating broad spanning (multi-octave) combs, more recent efforts have discovered that short spans also have utility in fields like photon pair generation. While one simple approach for controlling the number of combs lines generated is to simply adjust the input power, this strategy does not give high precision. An alternative method is to control the dispersion of the cavity, which would effectively "quench" the comb at a precise width. In order to accomplish this feat, a rigorous understanding and a high degree of control over the material properties is required.

The basic science community has been working on developing this materials toolbox for several decades through numerous initiatives. The ongoing Materials Genome effort, whose goal is to create structure-function relationships for organic materials, as well as the transformative advances in 2D materials over the past several decades will greatly enable these types of blended devices. While many of the materials have been successfully integrated into nanoelectronics, they are just now beginning to appear in nonlinear optical systems.

For example, recent reports on novel low-dimensional materials interfaced with integrated microresonators show extremely large improvements in the overall device non-linearity with only moderate reductions in Q. $WS_2$ and $MoS_2$ both contain large thermo-optic coefficients and therefore allow for large bandwidth thermal modulation of ring resonators. [418] These materials coupled with integrated devices could allow for the fast modulation and tuning of coherent Kerr frequency combs, expanding device performance in the previously mentioned applications as well as enabling new applications. Similarly, the incorporation of phase change materials, such a vanadium dioxide, could provide a route for switching of combs with reduced power consumption. Most of these materials also possess large optical non-linearities, allowing for further enhancement of the comb operation itself. [395,419,420] Assuming minimal drops in resonator Q after deposition, increases in efficiency and span should be achievable with moderate input powers using hybrid cavity designs. By enhancing the operability of the Kerr comb source, we can likewise expect an overall enhancement in current and emergent applications. Given the infancy of these devices and the materials research, the future promise of these hybrid device platforms is very exciting.

# 6 Conclusions

In just the past couple years, the current packaged microcombs have realized many of the applications initially hypothesized nearly 30 years ago. For example, they have been used as an ultra-precise reference source for LIDAR, spectroscopy, and atomic clocks and as a frequency converter. [13,141,367,368] This rapid translation of the technology to multiple real-world applications is a very strong indicator of the long-term potential impact of frequency combs on the field.

In large part, these packaged systems have relied on frequency comb miniaturization which was accomplished through advances in device technologies and integration of multiple device components, and future research will undoubtably continue to emphasize both research directions in pursuit of higher efficiency and higher stability devices. The emerging material systems for comb generation simultaneously offer new strategies for broad comb generation via dispersion engineering and new paths for achieving reduction in power. By reducing power requirements while maintaining comb or even increasing comb span, smaller footprint, lighter weight devices can be fabricated, further increasing the portability of system.

While the materials and materials integration strategies presented here have successfully been leveraged to improve broad comb generation, there are nearly limitless potential material combinations still left to be explored. Additionally, with the recent realization that narrow combs have significant utility as references or rulers, there is renewed interest in the development of narrow combs that require minimal power and that are environmentally robust. These vastly different requirements can be achieved by varying the material system and the device architecture. Therefore, when coupled with recent innovations in materials design, the field of frequency combs has a very promising future.

**Funding**

We thank ARO (W911NF-18-1-0033) and ONR (N00014-17-1-2270) for supporting this work. AHD acknowledges a scholarship from the Deutscher Akademischer Austauschdienst.

**Acknowledgments**

We would like to thank researchers at EPFL (Kippenberg research group), Harvard (Loncar research group), Columbia (Lipson research group), and OEWaves for providing us with several of the figures and images.

**Disclosures**

The authors declare that there are no conflicts of interest related to this article.

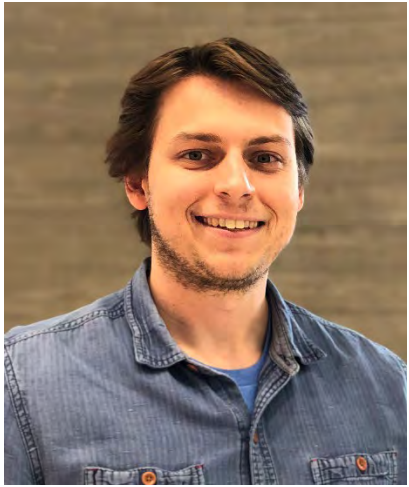

Andre Kovach received his BS in chemical engineering at the University of Southern California. He is currently a PhD candidate in the chemical engineering, also at the University of Southern California, under the advisement of Prof. Andrea M. Armani. His research focuses on the development and applications of hybrid whispering gallery mode microresonators. Specifically, he synthesizes novel optical nonlinear organic materials and combines them with integrated resonant cavities to investigate photoswitching and enhanced frequency comb and stimulated Raman generation.

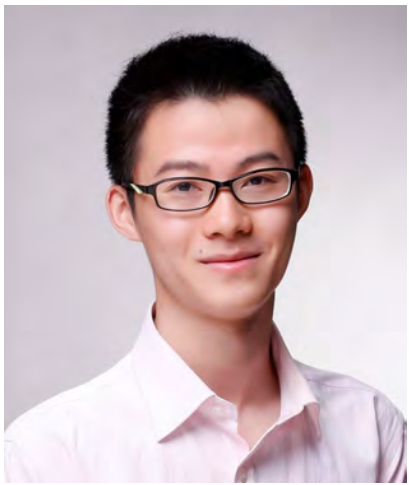

Dongyu Chen is currently a PhD student in Electrical Engineering-Electrophysics at the University of Southern California. He received his B.Sc. degree in physics in 2015 from Peking University in China. As a member of Prof. Andrea Armani's group, his research focuses on designing and fabricating silica and silicon oxynitride whispering gallery mode resonators and their applications in nonlinear optics, especially frequency comb generation and Raman emission. He is interested in interdisciplinary research, like exploring new materials and systems for these optical devices.

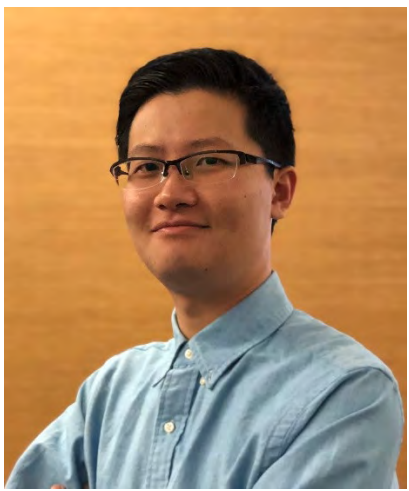

Jinghan He received his B.Eng. in polymer materials and engineering from Sichuan University in China in June 2014. Then, he continued his master study and received his M.Eng. in chemistry and biotechnology from the University of Tokyo in Japan in March 2017. He conducted his master research in Prof. Naoko Yoshie Group on the synthesis and characterization of nanopatterned polymer brushes by the "*grafting-to*" epitaxial crystallization method. Currently, he is a Ph.D. candidate in Chemistry at the University of Southern California working in Prof. Andrea Armani's laboratory, conducting interdisciplinary research into the synthesis and characterization of light-switching devices, self-healing materials and bio-imaging probes.

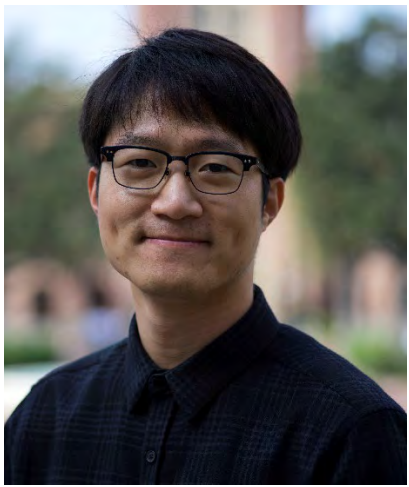

Hyungwoo Choi received his BS in chemical engineering and MS in energy engineering, both from Hanyang University in Seoul, South Korea. As a USC Provost's Fellow, he subsequently completed his PhD in chemical engineering at the University of Southern California under the supervision of Prof. Andrea Armani. His thesis focused on developing metal-doped high index silicon dioxides for improving nonlinear behaviors, such as stimulated Raman scattering and four-wave mixing, in whispering gallery mode resonant cavities. He is currently a senior transceiver product engineer at Lumentum.

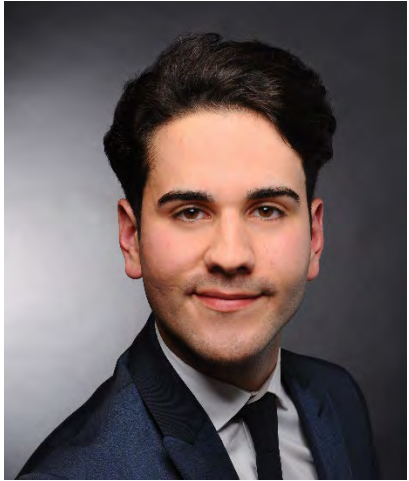

Adil Han Dogan is a visiting undergraduate researcher in the Department of Electrical and Computer Engineering at the University of California, Riverside (UCR). He is a senior undergraduate student in the Electrical Engineering and Information Technology program at the University of Stuttgart in Germany. He has worked as a technical assistant responsible for PCB design in various projects at the Institute of Electrical and Optical Communications Engineering at his home institution and has interned at the Electrical Drives Division of Bosch Corporation in Yokohama in Japan.

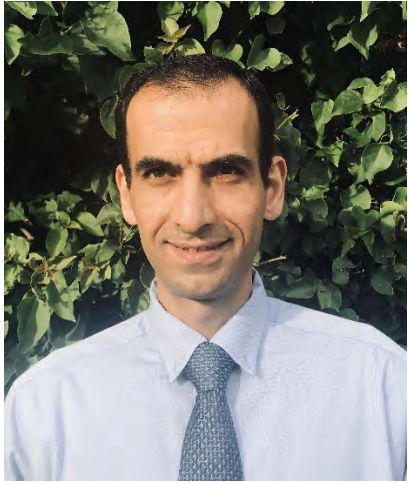

Mohammadreza Ghasemkhani is a project scientist in the Department of Electrical and Computer Engineering at the University of California, Riverside (UCR). Prior to his appointment at UCR, he was a member of technical staff at OEwaves Inc. in Pasadena CA. He earned his PhD in optical science and engineering at the University of New Mexico in Albuquerque NM, where he worked on cavity-enhanced optical refrigeration and spectroscopy.

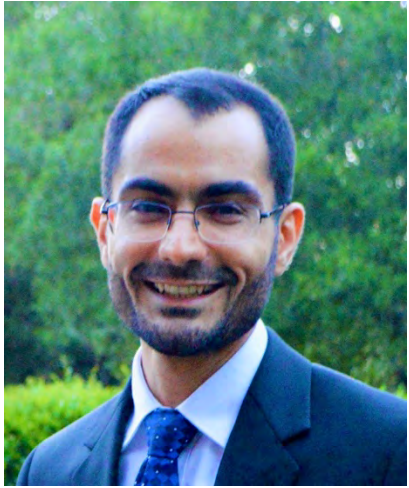

Hossein Taheri is an assistant research professor in the Department of Electrical and Computer Engineering at the University of California, Riverside (UCR). Prior to joining UCR, he was an R&D engineer at Foxconn Optical Interconnect Technology (now Broadcom Inc.) in San Jose CA, working on high-bit-rate datacenter fiber optic modules and optical transceivers. Since 2017, he has also been a consulting scientist with Bioxytech Retina, a startup based in the San Francisco Bay Area. In 2016, he was a research scientist at OEwaves Inc. in Pasadena CA, a JPL NASA spin-off company focused on microwave photonics products. He received his master's and Ph.D. degrees in electrical and computer engineering with minor in physics at the Georgia Institute of Technology in Atlanta GA, and his bachelor's degree in electrical and computer engineering at the University of Tehran.

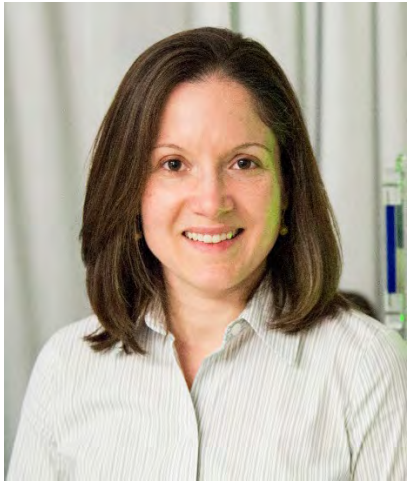

Andrea Armani received her BA in physics from the University of Chicago and her PhD in applied physics with a minor in biology from the California Institute of Technology. She is currently the Ray Irani Chair of Engineering and Materials Science and a Professor of Chemical Engineering and Materials Science at the University of Southern California. She is an elected member of Sigma Xi and the National Academy of Inventors (NAI) and a Fellow of OSA and SPIE. She leads a research group focused on the synthesis of polymeric and dielectric materials and the use of these materials in fabricating photonic technologies. Her research ranges from the molecular scale to instrumentation level and explores applications spanning from biodetection to non-linear optics, engaging researchers with diverse academic backgrounds.